\documentstyle[12pt,titlepage]{article}

\font\goth=eufm10

\newtheorem{example}{Example}[section]
\newtheorem{note}[example]{Note}
\newtheorem{theorem}[example]{Theorem}
\newtheorem{corollary}[example]{Corollary}
\newtheorem{definition}[example]{Definition}
\newtheorem{proposition}[example]{Proposition}

\newtheorem{lemma}[example]{Lemma}

\def\boxit#1#2{\setbox1=\hbox{\kern#1{#2}\kern#1}%
\dimen1=\ht1 \advance\dimen1 by #1 \dimen2=\dp1 \advance\dimen2 by #1
\setbox1=\hbox{\vrule height\dimen1 depth\dimen2\box1\vrule}%
\setbox1=\vbox{\hrule\box1\hrule}%
\advance\dimen1 by .4pt \ht1=\dimen1
\advance\dimen2 by .4pt \dp1=\dimen2 \box1\relax}
\def\bo#1{\boxit{1pt}{$#1$}}

\def\Proof{\medskip\noindent {\it Proof --- \ }}
\def\cqfd{\hfill $\Box$ \bigskip}
\def\adots{\mathinner{\mkern2mu\raise1pt\hbox{.}
\mkern3mu\raise4pt\hbox{.}\mkern1mu\raise7pt\hbox{.}}}
\def\<{\langle}
\def\>{\rangle}
\def\fsf{K\!\not< \! \bb{A}\!\not>}
\def\cf{{\it cf. }}
\def\eg{{\it e.g. }}
\def\mat{{\rm Mat}}
\def\limproj{\mathop{\oalign{lim\cr
\hidewidth$\longleftarrow$\hidewidth\cr}}}
\def\k{{\bf k}}
\def\ad{{\rm ad\,}}

\def\S{{\bf S}}
\def\Sym{{\bf Sym}}
\def\sym{{\sl Sym}}
\newfont{\bb}{msbm10}
\def\N{{\bf N}}
\def\Z{{\bf Z}}
\def\E{{\bf E}}
\def\Q{{\bf Q}}
\def\sgn{{\rm sgn\, }}
\def\D{{\rm Des \, }}
\def\Q{{\bf Q}}
\def\maj{{\rm maj\, }}
\def\C{{\bf C}}

\def\vt{\vrule height 0mm depth 2mm width 0mm}
\def\vtt{\vrule height 0mm depth 3mm width 0mm}
\def\vttt{\vrule height 0mm depth 13mm width 0mm}
\def\vtr#1{\vrule height 0mm depth #1mm width 0mm}
\vfill

\title{\sf NONCOMMUTATIVE SYMMETRIC FUNCTIONS}

\author{\sc I.M. Gelfand, D. Krob, A. Lascoux, \\
\sc B. Leclerc, V.S. Retakh and J.-Y. Thibon}
\date{}

\begin{document}

\newfont{\rml}{cmu10}
\newfont{\rmL}{cmr12}
\maketitle

\bigskip\bigskip
\bigskip\bigskip

\thispagestyle{empty}
\bigskip\bigskip
\noindent
Israel M. Gelfand\\
Department of Mathematics\\
Rutgers University\\
New Brunswick, N.J. 08903\\
U.S.A.

\bigskip\noindent
Daniel Krob\\
Laboratoire d'Informatique Th\'eorique et Programmation\\
Universit\'e Paris 7\\
2, place Jussieu\\
75251 Paris cedex 05\\
France

\bigskip\noindent
Alain Lascoux\\
Laboratoire d'Informatique Th\'eorique et Programmation\\
Universit\'e Paris 7\\
2, place Jussieu\\
75251 Paris cedex 05\\
France

\bigskip\noindent
Bernard Leclerc\\
Laboratoire d'Informatique Th\'eorique et Programmation\\
Universit\'e Paris 7\\
2, place Jussieu\\
75251 Paris cedex 05\\
France

\bigskip\noindent
Vladimir S. Retakh\\
DIMACS\\
Rutgers University\\
New Brunswick, N.J. 08903\\
U.S.A.

\bigskip\noindent
Jean-Yves Thibon\\
Institut Gaspard Monge\\
Universit\'e de Marne-la-Vall\'ee\\
2, rue de la Butte-Verte\\
93166 Noisy-le-Grand cedex\\
France

\newpage
\setcounter{page}{1}
\small
\tableofcontents
\normalsize
\newpage

\section{Introduction}

A large part of the classical theory of symmetric functions is fairly
independent of their interpretation as polynomials in some underlying
set of variables $X$. In fact, if $X$ is supposed infinite, the
elementary symmetric functions $\Lambda_k(X)$ are algebraically independent,
and what one really considers is just a polynomial algebra
$K[\Lambda_1,\Lambda_2,\ldots]$, graded by the weight function
$w(\Lambda_k)=k$ instead of the usual degree $d(\Lambda_k)=1$.

\smallskip
Such a construction still makes sense when the $\Lambda_k$ are interpreted
as noncommuting indeterminates and one can try to lift to
the free associative algebra $\Sym = K\<\Lambda_1,\Lambda_2,\ldots\>$ the
expressions of the other classical symmetric functions in terms of the
elementary ones, in order to define their noncommutative analogs
(Section \ref{ELEM}). Several algebraic and linear bases are obtained
in this way, including two families of ``power-sums", corresponding to
two different noncommutative analogs of the logarithmic derivative
of a power series. Moreover, most of the determinantal relations of the
classical theory remain valid, provided that determinants be replaced
by {\it quasi-determinants} (\cf \cite{GR1}, \cite{GR2} or \cite{KL}).

\smallskip
In the commutative theory, Schur functions constitute the fundamental
linear basis of the space of symmetric functions.
In the noncommutative case, it is possible to define a convenient notion
of {\it quasi-Schur function}
(for any skew Young diagram) using quasi-determinants, however most of
these functions are not polynomials in the generators $\Lambda_k$,
but elements of the skew field generated by the $\Lambda_k$. The only
quasi-Schur functions which remain polynomials in the generators are
those which are indexed by {\it ribbon} shapes (also called
{\it skew hooks}) (Section \ref{RIBBONS}).

\smallskip
A convenient property of these noncommutative ribbon Schur functions is
that they form a linear basis of $\Sym$. More importantly, perhaps,
they also suggest some kind of noncommutative analog of the fundamental
relationship between the commutative theory of symmetric functions
and the representation theory of the symmetric group. The role of
the character ring of
the symmetric group is here played by a certain subalgebra $\Sigma_n$
of its group algebra. This is the {\it descent algebra}, whose discovery
is due to L. Solomon (\cf \cite{So}). There is a close connection,
which has been known from the beginning, between
the product of the descent algebra, and the Kronecker product of
representations of the symmetric group. The fact that the homogeneous
components of $\Sym$ have the same dimensions as the
corresponding descent algebras allows us to transport the product
of the descent algebras, thus defining an analog of the usual {\it internal
product} of symmetric functions (Section \ref{DESCENT}).

\smallskip
Several Hopf algebra structures are classically defined on (commutative or
not) polynomial algebras. One possibility is to require that the generators
form an infinite sequence of divided powers. For commutative symmetric
functions, this is the usual structure, which is precisely compatible with the
internal product. The same is true in the noncommutative setting, and
the natural Hopf algebra structure of  noncommutative
symmetric functions provides an efficient tool for computing in the
descent algebra. This illustrates once more the importance of Hopf
algebras in Combinatorics, as advocated by Rota and his school (see \eg
\cite{JR}).

\smallskip
This can be explained by an interesting realization of  noncommutative
symmetric functions, as a certain subalgebra of the convolution algebra
of a free associative algebra (interpreted as a Hopf algebra in an appropriate
way). This algebra is discussed at length in the recent book \cite{Re}
by C. Reutenauer, where one finds many interesting results
which can be immediately translated in the language of noncommutative
symmetric functions. We illustrate this correspondence on
several examples. In particular, we show that the  Lie
idempotents in the descent algebra admit a simple interpretation
in terms of noncommutative symmetric functions.
We also discuss a certain recently discovered family of idempotents of
$\Sigma_n$ (the {\it Eulerian idempotents}), which arise quite
naturally in the context of noncommutative
symmetric functions, and explain to a large extent the combinatorics of
Eulerian polynomials.
All these considerations are strongly related to the combinatorics
of the Hausdorff series, more precisely of its continous analog,
which expresses the logarithm of the solution of a noncommutative
linear differential equation as a series of iterated integrals
(\cf \cite{Mag}\cite{Chen7}\cite{BMP}).
The theory of such series has been initiated and widely
developed by  K. T. Chen in his work
on homotopy theory (\cf \cite{Chen35}\cite{Chen42}), starting from his
fundamental observation that paths on a manifold can be conveniently
represented by noncommutative power series \cite{Chen5}.

\smallskip
The algebra of commutative symmetric functions  has a canonical
scalar product, for which it is self-dual as a Hopf algebra. In the
noncommutative theory, the algebra of symmetric functions differs
from its dual, which, as shown in \cite{MvR}, can be identified
with the algebra of quasi-symmetric functions (Section \ref{DUAL}).

\smallskip
Another classical subject in the commutative theory is the description of
the transition matrices between the various natural bases. This question
is considered in Section \ref{BASES}. It is worth noting that the rigidity
of the noncommutative theory leads to an explicit description of most
of these matrices.

\smallskip
We also investigate the general quasi-Schur functions.
As demonstrated in \cite{GR1} or \cite{GR2}, the natural object
replacing the determinant in noncommutative linear algebra is the
quasi-determinant, which is an analog of the ratio of two determinants.
Similarly, Schur functions will be replaced by quasi-Schur functions,
which are  analogs of the ratio of two ordinary Schur functions.
The various determinantal expressions of the classical Schur functions
can then be adapted to quasi-Schur functions (Section \ref{QUASI}).
This proves useful, for
example, when dealing with noncommutative continued fractions and
orthogonal polynomials. Indeed, the coefficients of the $S$-fraction or
$J$-fraction expansions of a noncommutative formal power series are
quasi-Schur functions of a special type, as well as the coefficients
of the typical three-term recurrence relation for noncommutative
orthogonal polynomials.

\smallskip
A rich field of applications of the classical theory is provided by
{\it specializations}. As pointed out by Littlewood, since the elementary
functions are algebraically independent, the process of specialization is
not restricted to the underlying variables $x_i$, but can be carried out
directly at the level of the
$\Lambda_k$, which can then be specialized in a totally arbitrary way, and
can still be formally considered as  symmetric functions of some fictitious
set of arguments. The same point of view can be adopted in the noncommutative
case, and we discuss several important examples of such specializations
(Section \ref{SPECIA}). The most natural question is whether
the formal symmetric functions can actually be interpreted as
functions of some set of noncommuting variables. Several  answers
can be proposed.

\smallskip
In Section \ref{RATSYM}, we take as generating function
$\lambda(t)=\sum_k \Lambda_k\, t^k$
of the elementary symmetric functions a quasi-determinant of the
Vandermonde matrix in the noncommutative indeterminates
$x_1,\ x_2,\ \ldots ,\ x_n$ and $x=t^{-1}$. This is a monic left polynomial
of degree $n$ in $x$, which is annihilated by the substitution
$x=x_i$ for every $i=1,\ \ldots \ n.$ Therefore  the so-defined functions
are noncommutative analogs of the ordinary symmetric
functions of $n$ commutative variables. They are actually
symmetric in the usual sense.
These functions are no longer polynomials but rational functions of the $x_i$.
We show that they can be expressed in terms of ratios of quasi-minors of
the Vandermonde matrix, as in the classical case. We also indicate in
Section \ref{SD} how to generalize these ideas in the context of skew
polynomial algebras.

\smallskip
In Section \ref{POLSYM}, we introduce another natural specialization, namely
$$
\lambda(t) = \overleftarrow{\prod_{1\le k\le n}}(1+tx_k)
=(1+tx_n)(1+tx_{n-1})(1+tx_{n-2})\cdots (1+tx_1) \ .
$$
This leads to
noncommutative polynomials which are symmetric for a special action
of the symmetric group on the free associative algebra.

\smallskip
In Section \ref{MATSYM}, we take $\lambda(t)$ to be a quasi-determinant of
$I+tA$, where $A=(a_{ij})$ is a matrix with noncommutative entries, and $I$
is the unit matrix. In this case, the usual families of symmetric functions
are polynomials in the $a_{ij}$ with integer coefficients, and admit a simple
combinatorial description in terms of paths in the
complete oriented graph labelled by the entries of $A$.
An interesting example, investigated in Section \ref{NCUGL},
is when $A=E_n= (e_{ij})$, the matrix
whose entries are the generators of the universal enveloping algebra
$U(gl_n)$. We obtain a description of the center of $U(gl_n)$ by means
of the symmetric functions associated with the matrices
$E_1,\ E_2-I,\ \ldots ,\ E_n-(n-1)I$. We also relate these functions to
Gelfand-Zetlin bases.

\smallskip
Finally, in Section \ref{SKEW}, other kinds of specializations in skew
polynomial algebras are considered.

\smallskip
The last section deals with some applications of quasi-Schur functions
to the study of rational power series with coefficients in a skew
field, and to some related points of noncommutative linear algebra. We
first discuss noncommutative continued fractions, orthogonal polynomials
and Pad\'e approximants. The results hereby found are
then applied to rational power series in one variable
over a skew field. One obtains in particular  a noncommutative
extension of the classical rationality criterion in terms
of Hankel determinants (Section \ref{RPS}).

\smallskip
The $n$ series $\lambda(t)$ associated with the generic matrix
of order $n$  (defined in Section \ref{MATSYM}) are examples
of rational series.
Their denominators appear as $n$ {\it pseudo-characteristic} polynomials,
for which a version of the
Cayley-Hamilton theorem can be established (Section \ref{CHAR}).
In particular, the generic matrix posesses $n$
{\it pseudo-determinants}, which are true noncommutative analogs of
the determinant. These pseudo-determinants reduce in the case of
$U(gl_n)$ to the Capelli determinant, and in the case of the quantum
group $GL_q(n)$, to the quantum determinant (up to a power of $q$).

\smallskip
The theory of noncommutative rational power series has been initiated
by M.P. Sch\"ut\-zen\-ber\-ger, in relation with problems in
formal languages and automata theory \cite{SCHUTZ}. This point of view is
briefly discussed in an Appendix.

\smallskip
The authors are grateful to C. Reutenauer for  interesting discussions
at various stages of the preparation of this paper.

\newpage
\section{Background} \label{BACKGR}

\subsection{Outline of the commutative theory}\label{OUTLINE}

Here is a brief review of the classical theory of symmetric functions.
A standard reference is Macdonald's book \cite{McD}. The notations
used here are those of \cite{LS}.

\smallskip
Denote by $X = \{x_1,\ x_2,\ \ldots \ \}$ an infinite set of {\it commutative}
indeterminates, which will be called a (commutative) {\it alphabet}. The
{\it elementary symmetric functions} $\Lambda_k(X)$ are then defined by
means of the generating series
\begin{equation}
\lambda(X,t) := \ \sum_{k\geq 0} \ t^k\, \Lambda_k(X) = \ \prod_{i\geq 1}
\ (1+x_it)\ .
\end{equation}
The {\it complete homogeneous symmetric functions} $S_k(X)$ are defined by
\begin{equation}
\sigma(X,t) := \ \sum_{k\geq 0} \ t^k\, S_k(X)
= \ \prod_{i\geq 1}\ (1-x_it)^{-1}\ ,
\end{equation}
so that the following fundamental relation holds
\begin{equation}\label{RSL}
\sigma(X,t)=\lambda(X,-t)^{-1}\ .
\end{equation}
The {\it power sums symmetric functions} $\psi_k(X)$ are defined by
\begin{equation}
\psi(X,t) := \ \sum_{k\geq 1}\ t^{k-1}\ \psi_k(X) = \ \sum_{i\geq 1}
\ x_i\, (1-x_it)^{-1}\ .
\end{equation}
These generating series satisfy the following relations
\begin{equation}\label{RPL}
\psi(X,t) = {d\over dt}\ {\rm log}\ \sigma(X,t) = -\, {d\over dt}\
{\rm log}\ \lambda(X,-t)\ ,
\end{equation}
\smallskip
\begin{equation}
{d\over dt}\ \sigma(X,t)=\sigma(X,t)\ \psi(X,t)\ ,
\end{equation}
\smallskip
\begin{equation}\label{NEW1}
-\, {d\over dt}\ \lambda(X,-t)=\psi(X,t)\ \lambda(X,-t)\ .
\end{equation}

\smallskip
Formula (\ref{NEW1}) is known as Newton's formula. The so-called
fundamental theorem of the theory of symmetric functions states that the
$\Lambda_k(X)$ are algebraically independent. Therefore, any formal power
series $f(t) = 1+\sum_{k\geq 1}\, a_k\, t^k$ may be considered as the
specialization of the series
$\lambda(X,t)$ to a virtual set of arguments $A$. The other families of
symmetric functions associated to $f(t)$ are then defined by
relations (\ref{RSL}) and (\ref{RPL}).
This point of view was introduced by Littlewood and extensively developped
in \cite{Li}. For example, the specialization $S_n=1/n!$
transforms the generating series $\sigma(X,t)$ into the exponential
function $e^t$. Thus, considering $\sigma(X,t)$ as a symmetric analog
of $e^t$, one can construct symmetric analogs for several related
functions, such as trigonometric functions, Eulerian polynomials or  Bessel
polynomials. This point of view allows to identify their coefficients
as the dimensions of certain representations of the symmetric
group \cite{F1}\cite{F2}. Also, any function admitting
a symmetric analog can be given a $q$-analog, for example
by means of the specialization $X = \{\, 1,q,q^2,\, \dots\, \}$.

\smallskip
We denote by $\sym$ the algebra of symmetric functions, {\it i.e.}  the
algebra generated over $\Q$ by the elementary functions. It is a graded
algebra for the weight function $w(\Lambda_k)=k$, and the dimension of its
homogeneous component of weight $n$, denoted by $\sym_n$, is equal to $p(n)$,
the number of partitions of $n$.  A {\it partition} is  a
finite non-decreasing sequence of positive integers,
$I = (i_1\leq i_2\leq\ldots \leq i_r)$. We shall also write
$I=(1^{\alpha_1}\, 2^{\alpha_2}\, \ldots)$, $\alpha_m$ being the
number of parts $i_k$ which
are equal to $m$. The \it weight \rm of $I$ is $|I|=\sum_k i_k$
and its \it length \rm is its
number of (nonzero) parts $\ell(I)=r$.

\smallskip
For a partition $I$, we set
$$
\psi^I = \psi_1^{\alpha_1}\, \psi_2^{\alpha_2}\, \cdots \ , \
\Lambda^I=\Lambda_1^{\alpha_1}\, \Lambda_2^{\alpha_2}\, \cdots \ , \
S^I = S_1^{\alpha_1}\, S_2^{\alpha_2}\, \cdots \ .
$$
For $I\in\Z^r$, not necessarily a partition,  the {\it Schur function} $S_I$
is defined by
\begin{equation}
S_I = \det\, \bigl( S_{i_k+k-h} \bigr)_{1\leq h,k\leq r}
\end{equation}
where $S_j=0$ for $j<0$. The Schur functions indexed by partitions form a
$\Z$-basis of $\sym$, and one usually endows $\sym$
with a scalar product $(\cdot ,\cdot )$
for which this basis is orthonormal. The $\psi^I$ form then an orthogonal
$\Q$-basis of $\sym$, with $(\psi^I,\psi^I) = 1^{\alpha_1}\, \alpha_1!\,
2^{\alpha_2}\, \alpha_2!\, \cdots$ Thus, for a
partition $I$ of weigth $n$, $n!/(\psi^I,\psi^I)$ is the cardinality of the
conjugacy class of ${\goth{S}}_n$ whose elements have $\alpha_k$
cycles of length $k$ for all $k\in [1,n]$. A permutation $\sigma$ in this
class will be said {\it of type} $I$, and we shall write $T(\sigma)=I$.

\smallskip
These definitions are motivated by the following classical results
of Frobenius. Let $CF($\S$_n)$ be the ring of central functions
of the symmetric group $\S_n$. The {\it Frobenius characteristic map}
${\cal{F}}\, :\, CF(\S_n) \longrightarrow \sym_n$
associates with any central function $\xi$ the symmetric function
$$
{\cal{F}}(\xi)
=
{{1}\over{n!}} \ \sum_{\sigma\in\S_n} \xi(\sigma)\ \psi^{T(\sigma)}
= \ \sum_{|I|=n} \ \xi(I)\ {{\psi^I} \over {(\psi^I,\psi^I)}}
$$
where $\xi(I)$ is the common value of the $\xi(\sigma)$ for all $\sigma$
such that $T(\sigma)=I$. We can also consider ${\cal{F}}$ as a map from
the representation ring $R(\S_n)$ to $\sym_n$ by setting
${\cal{F}}([\rho])={\cal{F}}(\chi_\rho)$, where $[\rho]$ denotes
the equivalence class of a representation $\rho$
(we use the same letter ${\cal{F}}$ for the two maps since this does
not lead to ambiguities). Glueing these maps together, one has a linear
map
$$
{\cal{F}}\, :\, R:=\bigoplus_{n\ge 0}R(\S_n) \longrightarrow \sym \ ,
$$
which turns out to be an isomorphism of graded rings (see for instance
\cite{McD} or \cite{Zel}). We denote by $[I]$ the class of the irreducible
representation of ${\goth{S}}_n$ associated with the partition $I$, and by
$\chi_I$ its character. We have then ${\cal{F}}(\chi_I)=S_I$ (see \it e.g.
\rm \cite{McD} p. 62).

\smallskip
The product $*$, defined on the homogeneous component $\sym_n$ by
\begin{equation}
{\cal{F}}([\rho]\otimes[\eta]) = {\cal{F}}(\chi_\rho \chi_\eta)
= {\cal{F}}([\rho])*{\cal{F}}([\eta]) \ ,
\end{equation}
and extended to $\sym$ by defining the product of two homogeneous functions
of different weights to be zero, is called the {\it internal product}.

\smallskip
One can identify the tensor product $\sym\otimes\sym$ with the algebra
$\sym(X,Y)$ of polynomials which are separately symmetric in two infinite
disjoint sets of indeterminates $X$ and  $Y$, the correspondence
being given by $F\otimes G \longmapsto F(X)\, G(Y)$. Denoting by $X+Y$
the disjoint union of $X$ and $Y$, one can then define a comultiplication
$\Delta$ on $\sym$ by setting $\Delta (F) = F(X+Y)$. This comultiplication
endows $\sym$ with the structure of a self-dual Hopf algebra, which is very
useful for calculations involving characters of symmetric groups
(see for instance \cite{Gei}, \cite{Zel}, \cite{ST} or \cite{Th}). The
basic formulas are
\begin{equation}
(F_1F_2\cdots F_r)*G = \mu_r[(F_1\otimes F_2\otimes\cdots\otimes F_r) *
\Delta^r G] \ ,
\end{equation}
where $\mu_r$ denotes the $r$-fold ordinary multiplication and
$\Delta^r$ the iterated coproduct, and
\begin{equation}
\Delta^r(F*G) =\Delta^r(F) *\Delta^r(G) \ .
\end{equation}
It will be shown in the sequel that both of these formulas admit
natural noncommutative analogs. The antipode of the Hopf algebra
$\sym$ is given, in $\lambda$-ring notation, by
\begin{equation}
\tilde\omega (F(X)) = F(-X) \ .
\end{equation}
The symmetric functions of $(-X)$ are defined by the generating series
\begin{equation}
\lambda(-X,t):=\sigma(X,-t) = [\lambda(X,t)]^{-1}
\end{equation}
and one can more generally consider {\it differences} of alphabets.
The symmetric functions of $X-Y$ are given by the generating series
\begin{equation}
\lambda(X-Y,t) := \lambda(X,t)\ \lambda(-Y,t) = \lambda(X,t)\ \sigma(Y,-t) \ .
\end{equation}
In particular, $\psi_k(X-Y)=\psi_k(X)-\psi_k(Y)$.

\smallskip
There is another coproduct $\delta$ on $\sym$, which is obtained by considering
products instead of sums, that is
\begin{equation}
\delta (F) = F(XY) \ .
\end{equation}
One can check that its adjoint is the internal product :
\begin{equation}
(\delta F\, ,\, P\otimes Q) = (F\, ,\, P*Q) \ .
\end{equation}
This equation can be used to give an intrinsic definition of the internal
product, {\it i.e.} without any reference to characters of the
symmetric group. Also, one can see that two bases $(U_I), (V_J)$ of $\sym$
are adjoint to each other iff
\begin{equation}
\sigma(XY,1) = \ \sum_I \ U_I(X)\, V_I(Y) \ .
\end{equation}
For example, writing $\sigma(XY,1) = \prod_i\ \sigma(Y,x_i)$ and expanding
the product, one obtains that the adjoint basis of $S^I$ is formed by the
{\it monomial functions} $\psi_I$.


\subsection{Quasi-determinants}\label{QUASIDET}

Quasi-determinants have been defined in \cite{GR1} and further
developed in \cite{GR2} and \cite{KL}. In this section we briefly survey
their main properties, the reader being referred to these papers for a
more detailed account.

\smallskip
Let $K$ be a field, $n$ an integer and
$\bb{A} = \{\, a_{ij},\, 1 \leq i,j \leq n\, \}$ an {\it alphabet} of
order $n^2$, {\it i.e.} a set of $n^2$ noncommutative indeterminates.
Let $\fsf$ be the {\it free field} constructed on $K$
and generated by $\bb{A}$. This is the universal field of fractions
of the free associative algebra $K\<\bb{A}\>$  (\cf \cite{Co}). The matrix
$A = ( a_{ij} )_{1\leq i,j\leq n}$ is  called the {\it generic matrix}
of order $n$. This matrix is invertible over $\fsf$.

\smallskip
Let $A^{pq}$ denote the matrix obtained from the generic matrix $A$ by
deleting the $p$-th row and the $q$-th column. Let also
$\xi_{pq}=(a_{p1},\ldots ,\hat a_{pq},\ldots ,a_{pn})$ and
$\eta_{pq}=(a_{1q},\ldots ,\hat a_{pq},\ldots ,a_{nq})$.

\begin{definition}\label{DEFQUASIDET}
The quasi-determinant $|A|_{pq}$ of order $pq$
of the generic matrix $A$ is the element of $\fsf$ defined by

\medskip
\centerline{$
|A|_{pq} = a_{pq} - \xi_{pq}\, (A^{pq})^{-1}\, \eta_{pq}\ =
a_{pq} - \displaystyle\sum_{i\not=p, j\not=q} \
a_{pj}\, ((A^{pq})^{-1})_{ji} \, a_{iq} \ .
$}
\end{definition}

\noindent
It is sometimes convenient to adopt the following more explicit notation

\medskip
\centerline{$
|A|_{pq} = \left|\matrix{
a_{11} & \ldots & a_{1q} & \ldots & a_{1n} \cr
\vdots &        & \vdots &        & \vdots \cr
a_{p1} & \ldots & \bo{a_{pq}} & \ldots & a_{pn} \cr
\vdots &        & \vdots &        & \vdots \cr
a_{n1} & \ldots & a_{nq} & \ldots & a_{nn} \cr
}\right| \ .
$}

\bigskip
\noindent
Quasi-determinants are here only defined
for generic matrices. However, using substitutions, this
definition can be applied to matrices with
entries in an arbitrary skew field. In fact, one can
even work in a noncommutative ring,
provided that  $A^{pq}$ be an invertible matrix.

\begin{example} \label{QD2}
{\rm For $n = 2$, there are four quasi-determinants :}

\medskip
\vskip 1mm
\centerline{$
\left|
\matrix{
\bo{a_{11}} & a_{12} \cr
a_{21} & a_{22} \cr
}
\right|
= a_{11} - a_{12}\, a_{22}^{-1}\, a_{21}\ , \ \
\left|
\matrix{
a_{11} & \bo{a_{12}} \cr
a_{21} & a_{22} \cr
}
\right|
= a_{12} - a_{11}\, a_{21}^{-1}\, a_{22}\ ,
$}

\medskip
\vskip 1mm
\centerline{$
\left|
\matrix{
a_{11} & a_{12} \cr
\bo{a_{21}} & a_{22} \cr
}
\right|
= a_{21} - a_{22}\, a_{12}^{-1}\, a_{11}\ , \ \
\left|
\matrix{
a_{11} & a_{12} \cr
a_{21} & \bo{a_{22}} \cr
}
\right|
= a_{22} - a_{21}\, a_{11}^{-1}\, a_{12}\ .
$}
\end{example}

\smallskip
The next result can be considered as another definition
of quasi-determinants (for generic matrices).

\begin{proposition}\label{INVM}
Let $A$ be the generic matrix of order $n$ and let
$B = A^{-1}=(b_{pq})_{1\leq p,q\leq n}$ be its inverse. Then one has
$|A|_{pq} = b_{qp}^{-1}$ for every $1\leq p,q \leq n$.
\end{proposition}

It follows from Proposition~\ref{INVM}  that
$|A|_{pq} = (-1)^{p+q}\, {\rm det}\, A/{\rm det}\, A^{pq}$ when the $a_{ij}$
are commutative variables. Thus quasi-determinants are noncommutative analogs
of the ratio of a determinant to one of its principal minors. If $A$ is an
invertible matrix with entries in a arbitrary skew field, the
above relation still holds for every $p,q$ such that $b_{qp} \not= 0$.
Another consequence of Proposition~\ref{INVM} is that

\medskip
\vskip 1mm
\centerline{$
\hfill
|A|_{pq} = a_{pq} - \displaystyle\sum_{i\not=p, j\not=q} \
a_{pj}\, |A^{pq}|^{-1}_{ij} \, a_{iq} \ , \hfill
$}

\medskip
\noindent
which provides a recursive definition of quasi-determinants.

\medskip
Let $I$ be the unit matrix of order $n$. The expansions of the
quasi-determinants of $I-A$ into formal power series are conveniently
described in terms of paths in a graph. Let  ${\cal A}_n$ denote
the complete oriented graph with $n$
vertices $\{\, 1,2,\dots,n\,\}$, the arrow from $i$ to $j$ being
labelled by $a_{ij}$.
We denote by ${\cal P}_{ij}$ the set of words labelling a path in
${\cal A}_n$ going from $i$ to $j$, {\it i.e.} the set of words of the
form $w = a_{ik_1}\, a_{k_1k_2}\, a_{k_2k_3}\, \ldots\, a_{k_{r-1}j}$.
A {\it simple path} is a path such that $k_s \not = i,\ j$ for every $s$.
We denote by ${\cal SP}_{ij}$ the set of words labelling simple paths from
$i$ to $j$.

\begin{proposition}\label{AUTO}
Let $i,\ j$ be two distinct integers between $1$ and $n$. Then,
\begin{equation}
|I-A|_{ii} = 1 - \sum_{{\cal SP}_{ii}} \ w \ , \quad
|I-A|_{ij}^{-1} =  \sum_{{\cal P}_{ji}} \ w  \ \ .
\end{equation}
\end{proposition}

\begin{example}
{\rm For $n=2$,
$$
\left|
\matrix{
\bo{1-a_{11}} & -a_{12} \cr
-a_{21} & 1-a_{22} \cr
}
\right|
= 1 - a_{11} -\ \sum_{p\ge 0} \ a_{12}\, a_{22}^p\, a_{21}\ .
$$
}
\end{example}

\medskip

As a general rule, quasi-determinants are not polynomials in their entries,
except in the following special case, which is particularly important for
noncommutative symmetric functions
(a graphical interpretation of this formula can be found in the Appendix).

\begin{proposition} \label{SSTRI}
The following quasi-determinant is a polynomial in its entries :
\begin{equation}\label{POLENT}
\left|\matrix{
a_{11} & a_{12} & a_{13}  & \ldots & \bo{a_{1n}} \cr
-1     & a_{22} & a_{23}  & \ldots & a_{2n}      \cr
0      & -1     & a_{33}  & \ddots &  \vdots     \cr
\vdots & \ddots & \ddots  & \ddots & a_{n-1\,n}  \cr
0      & \ldots &   0     & -1     & a_{nn}      \cr
}\right|
=
a_{1n} +
\, \sum_{1\le j_1 <j_2 < \ldots <j_k <n} \,
a_{1j_1}\, a_{j_1+1\,j_2}\, a_{j_2+1\,j_3} \, \ldots\, a_{j_k+1\,n} \ .
\end{equation}
\end{proposition}

\medskip
We now recall some useful properties of quasi-determinants. Quasi-determinants
behave well with respect to permutation of rows and columns.

\begin{proposition}\label{PERMUT}
A permutation of the rows or columns of a quasi-determinant does not
change its value.
\end{proposition}

\begin{example}
$$
\left|\matrix{
a_{11} & a_{12} & a_{13} \cr
a_{21} & a_{22} & a_{23} \cr
\bo{a_{31}} & a_{32} & a_{33} \cr
}\right|
=
\left|\matrix{
a_{21} & a_{22} & a_{23} \cr
a_{11} & a_{12} & a_{13} \cr
\bo{a_{31}} & a_{32} & a_{33} \cr
}\right|
=
\left|\matrix{
a_{22} &a_{21} & a_{23} \cr
a_{12} &a_{11} & a_{13} \cr
a_{32} &\bo{a_{31}} & a_{33} \cr
}\right|\ .
$$
\end{example}

\noindent
One can also give the following analogs of the classical behaviour of a
determinant with respect to linear combinations of rows and columns.

\begin{proposition}
If the matrix $B$ is obtained from the matrix $A$ by multiplying the $p$-th
row {\rm on the left} by $\lambda$, then

\medskip
\centerline{$
|B|_{kq}
=
\left\{\matrix{
\lambda\, |A|_{pq} & {\rm for} \ \ k=p \ , \cr
|A|_{kq}  & {\rm for} \ \ k\not =p \ .\cr
}\right.
$}

\medskip
\noindent
Similarly, if the matrix $C$ is obtained from the matrix $A$ by multiplying
the $q$-th column {\rm on the right} by $\mu$, then

\centerline{$
|C|_{pl}
=
\left\{\matrix{
|A|_{pq}\, \mu & {\rm for} \ \ l=q\ , \cr
|A|_{pl}  & {\rm for} \ \ l\not =q\ .\cr
}\right.
$}

\medskip
\noindent
Finally, if the matrix $D$ is obtained from $A$ by adding to some row
(resp. column) of $A$ its $k$-th row (resp. column), then
$|D|_{pq} = |A|_{pq}$ for every $p \not=k$ (resp. $q \not= k$).
\end{proposition}

The following proposition gives important identities which are called
{\it homological relations}.

\begin{proposition} The quasi-minors of the generic matrix $A$ are
related by :

\medskip
\centerline{$
|A|_{ij} \, (|A^{il}|_{kj})^{-1} = - |A|_{il}\, (|A^{ij}|_{kl})^{-1}\ ,
$}

\medskip
\centerline{$
(|A^{kj}|_{il})^{-1} \, |A|_{ij} = - (|A^{ij}|_{kl})^{-1}\, |A|_{kj}\ .
$}
\end{proposition}

\begin{example}
$$
\left|\matrix{
a_{11} & a_{12} \cr
a_{31} & \bo{a_{32}} \cr
}\right|^{-1}
\left|\matrix{
a_{11} & a_{12} & a_{13} \cr
a_{21} & a_{22} & a_{23} \cr
a_{31} & a_{32} & \bo{a_{33}} \cr
}\right|
=
-
\left|\matrix{
a_{11} & a_{12} \cr
a_{21} & \bo{a_{22}} \cr
}\right|^{-1}
\left|\matrix{
a_{11} & a_{12} & a_{13} \cr
a_{21} & a_{22} & \bo{a_{23}} \cr
a_{31} & a_{32} & a_{33} \cr
}\right|
\ .
$$
\end{example}

The classical expansion of
a determinant by one of its rows or columns is replaced by
the following property.

\begin{proposition} \label{DEVQDET}
For quasi-determinants, there holds :

\medskip
\vskip 1mm
\centerline{$
|A|_{pq} = a_{pq}\, -\
\displaystyle\sum_{j\not = q}\ a_{pj}\,(|A^{pq}|_{kj})^{-1}\, |A^{pj}|_{kq}
\ ,
$}

\medskip
\centerline{$
|A|_{pq} = a_{pq}\, -\
\displaystyle\sum_{i\not = p}\ |A^{iq}|_{pl}\,(|A^{pq}|_{il})^{-1}\, a_{iq}
\ ,
$}

\smallskip
\vskip 0.5mm
\noindent
for every $k \not =p$ and $l \not =q$.
\end{proposition}

\begin{example}
{\rm Let $n=p=q=4$. Then,}
$$
\left|\matrix{
a_{11} & a_{12} & a_{13} & a_{14} \cr
a_{21} & a_{22} & a_{23} & a_{24} \cr
a_{31} & a_{32} & a_{33} & a_{34} \cr
a_{41} & a_{42} & a_{43} & \bo{a_{44}} \cr
}\right|
=
a_{44}-
a_{43}
\left|\matrix{
a_{11} & a_{12} & a_{13} \cr
a_{21} & a_{22} & a_{23} \cr
a_{31} & a_{32} & \bo{a_{33}} \cr
}\right|^{-1}
\left|\matrix{
a_{11} & a_{12} & a_{14} \cr
a_{21} & a_{22} & a_{24} \cr
a_{31} & a_{32} & \bo{a_{34}} \cr
}\right|
$$
\medskip
$$
-\, a_{42}
\left|\matrix{
a_{11} & a_{12} & a_{13} \cr
a_{21} & a_{22} & a_{23} \cr
a_{31} & \bo{a_{32}} & a_{33} \cr
}\right|^{-1}
\left|\matrix{
a_{11} & a_{13} & a_{14} \cr
a_{21} & a_{23} & a_{24} \cr
a_{31} & a_{33} & \bo{a_{34}} \cr
}\right|
-
a_{41}
\left|\matrix{
a_{11} & a_{12} & a_{13} \cr
a_{21} & a_{22} & a_{23} \cr
\bo{a_{31}} & a_{32} & a_{33} \cr
}\right|^{-1}
\left|\matrix{
a_{12} & a_{13} & a_{14} \cr
a_{22} & a_{23} & a_{24} \cr
a_{32} & a_{33} & \bo{a_{34}} \cr
}\right|
\ .$$
\end{example}

\medskip
Let $P,\ Q$ be subsets of $\{1,\ldots ,n\}$ of the same cardinality.
We denote by $A^{PQ}$ the matrix obtained by removing from $A$ the rows whose
indices belong to $P$ and the columns whose indices belong to $Q$.
Also, we denote by $A_{PQ}$ the submatrix of $A$ whose row
indices belong to $P$ and column indices to $Q$.
Finally, if $a_{ij}$ is an entry of some submatrix $A^{PQ}$ or $A_{PQ}$, we
denote by $|A^{PQ}|_{ij}$ or $|A_{PQ}|_{ij}$ the corresponding
quasi-minor.

\medskip
Here is a noncommutative version of Jacobi's ratio theorem which
relates the quasi-minors of a matrix with those of its inverse.
This identity will be frequently used in the
sequel.

\begin{theorem}\label{JACOBI}
Let $A$ be the generic matrix of order $n$, let $B$ be its inverse and let
$(\{i\},L,P)$ and $(\{j\},M,Q)$ be two partitions of $\{1,2,\ldots, n\}$
such that $|L| = |M|$ and $|P| = |Q|$. Then, one has

\medskip
\vskip 0.5mm
\centerline{$
|B_{M\cup\{j\},L\cup\{i\}}|_{ji}
=
|A_{P\cup\{i\},Q\cup\{j\}}|_{ij}^{-1}
\ \, .
$}
\end{theorem}

\begin{example} {\rm Take $n=5$, $i=3$, $j=4$, $L=\{1,2\}$,
$M=\{1,3\}$, $P=\{4,5\}$ and $Q=\{2,5\}$. There holds}

\medskip
\centerline{$
\left|\matrix{
a_{32} & \bo{a_{34}} & a_{35} \cr
a_{42} & a_{44} & a_{45} \cr
a_{52} & a_{54} & a_{55} \cr
}\right|
=
\left|\matrix{
b_{11} & b_{12} & b_{13} \cr
b_{31} & b_{32} & b_{33} \cr
b_{41} & b_{42} & \bo{b_{43}} \cr
}\right|^{-1}\ .
$}
\end{example}

\medskip
We conclude with noncommutative versions of Sylvester's  and
Bazin's theorems.

\begin{theorem} \label{SYLVESTER}
Let $A$ be the generic matrix of order $n$ and let $P,Q$ be two subsets of
$[1,n]$ of cardinality $k$. For $i\notin P$ and $j\notin Q$, we set
$b_{ij} = |A_{P\cup\{i\},Q\cup\{j\}}|_{ij}$ and form the matrix
$B = (b_{ij})_{i\notin P,j\notin Q}$ of order $n-k$. Then,
$$
|A|_{lm} = |B|_{lm}
$$
for every $l \notin P$ and every $m \notin Q$.
\end{theorem}

\begin{example}
{\rm Let us take $n=3$, $P = Q = \{3\}$ and $l =m = 1$. Then,}
$$
\left|
\matrix{
\bo{a_{11}} & a_{12} & a_{13} \cr
a_{21} & a_{22} & a_{23} \cr
a_{31} & a_{32} & a_{33}
}
\right|
=
\left|
\matrix{
\bo{\left|\matrix{ \bo{a_{11}} & a_{13} \cr a_{31} & a_{33} } \right|} &
\left|\matrix{ \bo{a_{12}} & a_{13} \cr a_{32} & a_{33} } \right| \cr
\left|\matrix{ \bo{a_{22}} & a_{23} \cr a_{32} & a_{33} } \right| &
\left|\matrix{ \bo{a_{21}} & a_{23} \cr a_{31} & a_{33} } \right|
}
\right| \ .
$$
\end{example}

\smallskip
Let $A$ be a matrix of order $n\!\times\! p$, where $p\ge n$. Then, for every
subset
$P$ of cardinality $n$ of $\{1,\ldots ,p\}$, we denote by $A_P$ the
square submatrix of $A$ whose columns are indexed by $P$.

\begin{theorem} \label{BAZIN}
Let $A$ be the generic matrix of order
$n\!\times\! 2n$ and let $m$ be an integer in $\{1,\ldots,n\}$. For
$1\leq i,j\leq n$, we set
$b_{ij} = |A_{\{j,n+1,\ldots ,n+i-1, n+i+1,\ldots ,2n\}}|_{mj}$ and form
the matrix $B = (b_{ij})_{1\leq i,j \leq n}$. Then we have

\medskip
\vskip 1mm
\centerline{$
|B|_{kl} = |A_{\{n+1,\ldots ,2n\}}|_{m,n+k} \
|A_{\{1,\ldots,l-1,l+1,\ldots ,n,n+k\}}|_{m,n+k}^{-1} \
|A_{\{1,\ldots ,n\}}|_{ml} \
$}

\medskip
\noindent
for every integers $k,l$ in $\{ 1,\ldots, n\}$.
\end{theorem}

\begin{example}\label{EXBAZ}
{\rm Let $n = 3$ and $k = l = m =1$. Let us
adopt more appropriate notations, writing for example $|2\bo{4}5|$ instead
of $|M_{\{2,4,5\}}|_{14}$. Bazin's identity reads}

\medskip
\vskip 1mm
\centerline{$
\left|\matrix{
\bo{|\bo{1}56|} & |\bo{2}56| & |\bo{3}56| \cr
|\bo{1}46| & |\bo{2}46| & |\bo{3}46| \cr
|\bo{1}45| & |\bo{2}45| & |\bo{3}45| \cr
}\right|
=
|\bo{4}56|\ |23\bo{4}|^{-1} \ |\bo{1}23|
\ .
$}
\end{example}

\medskip
We shall also need the following variant of Bazin's theorem~\ref{BAZIN}.

\begin{theorem} \label{BAZINPLUS}
Let $A$ be the generic matrix of order
$n\!\times\! (3n-2)$ and let $m$ be an integer in $\{1,\ldots,n\}$. For
$1\leq i,j\leq n$, we set
$c_{ij} = |A_{\{j,n+i,n+i+1, \ldots ,2n+i-2\}}|_{mj}$ and form
the matrix $C = (c_{ij})_{1\leq i,j \leq n}$. Then we have

\medskip
\vskip 1mm
\centerline{$
|C|_{11} = |A_{\{n+1,\ldots ,2n\}}|_{m,2n} \
|A_{\{2,3 \ldots,n,2n\}}|_{m,2n}^{-1} \
|A_{\{1,\ldots ,n\}}|_{m1} \
$ \ .}

\end{theorem}

\begin{example} {\rm Let $n = 3$, $m =1$, and keep the notations of
\ref{EXBAZ}.
Theorem~\ref{BAZINPLUS} reads}

\medskip
\vskip 1mm
\begin{equation}\label{EX3}
\left|\matrix{
\bo{|\bo{1}45|} & |\bo{2}45| & |\bo{3}45| \cr
|\bo{1}56| & |\bo{2}56| & |\bo{3}56| \cr
|\bo{1}67| & |\bo{2}67| & |\bo{3}67| \cr
}\right|
=
|45\bo{6}|\ |23\bo{6}|^{-1} \ |\bo{1}23|
\ .
\end{equation}
\end{example}

\Proof We use an induction on $n$. For $n=2$, Theorem~\ref{BAZINPLUS}
reduces to Theorem~\ref{BAZIN}. For $n=3$, we have to prove (\ref{EX3}).
To this end, we note that the specialization $7 \rightarrow 4$ gives
back Bazin's theorem \ref{BAZIN}. Since the right-hand side
does not contain 7, it is therefore enough to show that the left-hand
side does not depend on 7. But by Sylvester's theorem \ref{SYLVESTER},
the left-hand side is equal to
$$
\left|\matrix{
\bo{|\bo{1}45|} & |\bo{2}45| \cr
|\bo{1}56|      & |\bo{2}56| \cr
}\right|
-
\left|\matrix{
|\bo{2}45| & \bo{|\bo{3}45|} \cr
|\bo{2}56|      & |\bo{3}56| \cr
}\right|
\,
\left|\matrix{
|\bo{2}56| & |\bo{3}56| \cr
|\bo{2}67| & \bo{|\bo{3}67|} \cr
}\right|^{-1}
\,
\left|\matrix{
|\bo{1}56| & |\bo{2}56| \cr
|\bo{\bo{1}67|} & |\bo{2}67| \cr
}\right|
$$
$$
=|45\bo{6}|\,|25\bo{6}|^{-1}\,|\bo{1}25| -
|45\bo{6}|\,|25\bo{6}|^{-1}\,|2\bo{3}5|
|2\bo{3}6|^{-1}|\bo{1}26|\ ,
$$
which is independent of 7. Here, the second expression is derived by means of
\ref{BAZIN} and Muir's law of extensible minors for quasi-determinants
\cite{KL}.
The general induction step is similar, and we shall not detail it. \cqfd

\newpage
\section{Formal noncommutative symmetric functions}\label{FORMAL}

In this section, we introduce the algebra $\Sym$ of formal noncommutative
symmetric functions which is just the free associative algebra
$K\<\Lambda_1,\Lambda_2,\ldots\>$ generated by an infinite sequence of
indeterminates $(\Lambda_k)_{k\ge 1}$ over some fixed commutative field $K$
of characteristic zero. This algebra will be graded by the weight function
$w(\Lambda_k)=k$. The homogeneous component of weight $n$ will be denoted by
$\Sym_n$. The $\Lambda_k$ will be informally regarded as the elementary
symmetric functions of some virtual set of arguments.
When we need several copies of $\Sym$, we can give them labels
$A,B,\ldots$ and use the different sets of indeterminates
$\Lambda_k(A),\Lambda_k(B),\ldots$ together with their generating series
$\lambda(A,t), \lambda(B,t),\ldots$, and so on. The corresponding algebras
will be denoted $\Sym (A)$, $\Sym(B)$, {\it etc.}

\smallskip
We recall that a {\it composition} is a vector $I = (i_1,\dots,i_k)$
of nonnegative integers, called the {\it parts} of $I$.
The {\it length} ${\em l}(I)$ of the composition $I$
is the number $k$ of its parts and the weigth of $I$ is the sum
$|I| = \sum \, i_j\, $ of its parts.


\subsection{Elementary, complete and power sums functions}\label{ELEM}

Let $t$ be another indeterminate, commuting with all the $\Lambda_k$.
It will be convenient in the sequel to set $\Lambda_0=1$.

\begin{definition}
The {\rm elementary} symmetric functions are the $\Lambda_k$ themselves,
and their generating series is denoted by
\begin{equation}
\lambda(t) := \ \sum_{k\geq 0} \ t^k\, \Lambda_k
= 1 + \ \sum_{k\geq 1} \ t^k \, \Lambda_k \ .
\end{equation}
The {\rm complete homogeneous} symmetric functions $S_k$ are defined by
\begin{equation}\label{DEFS}
\sigma(t) := \ \sum_{k\geq 0}\ t^k\, S_k  = \lambda(-t)^{-1} \ .
\end{equation}
The {\rm power sums} symmetric functions of the {\rm first kind} $\Psi_k$ are
defined by
\begin{equation}
\psi(t) := \ \sum_{k\geq 1}\ t^{k-1}\, \Psi_k\ ,
\end{equation}
\begin{equation}\label{DEFPSI1}
{d\over dt}\ \sigma(t) = \sigma(t)\ \psi(t)\ .
\end{equation}
\smallskip
The {\rm power sums} symmetric functions of the {\rm second kind} $\Phi_k$
are defined by
\begin{equation} \label{DEFPHII}
\sigma (t) = \exp\, (\ \sum_{k\geq 1} \ t^k\, {\Phi_k \over k}\ ) \ ,
\end{equation}
or equivalently by one of the series
\begin{equation}\label{DEFPHI}
\Phi(t) :=
\ \sum_{k\geq 1} \ t^k\, {\Phi_k \over k} =
\log\, ( \, 1 + \sum_{k \geq 1} \ S_k\, t^k \ )
\end{equation}
or
\begin{equation}
\phi(t) :=
\ \sum_{k\ge 1} \ t^{k-1}\, \Phi_k
= {d\over dt} \, \Phi(t) = {d\over dt}\, \log \sigma(t)
\ .
\end{equation}
\end{definition}

\medskip

Although the two kinds of power sums  coincide
in the commutative case, they are quite different at the
noncommutative level.
For instance,
\begin{equation}
\Phi_3 = \Psi_3 + {1 \over 4} \, ( \Psi_1\, \Psi_2 - \Psi_2\, \Psi_1) \ .
\end{equation}
The appearance of two families of power sums is due to the fact
that one does not have a unique notion of logarithmic derivative
for power series with noncommutative coefficients. The two
families selected here correspond to the most natural
noncommutative analogs. We shall see below that both of them
admit interesting interpretations in terms of Lie algebras.
Moreover, they may be related to each other via appropriate specializations
(\cf Note~\ref{PHIPSI}).
\smallskip

One might also introduce a third family of power sums by replacing
(\ref{DEFPSI1}) by
$$
{d\over dt}\ \sigma(t) = \psi(t)\ \sigma(t)\ \ ,
$$
but this would lead to essentially the same functions. Indeed, $\Sym$
is equiped with several natural involutions, among which the
\it anti\rm -automorphism which leaves invariant the $\Lambda_k$. Denote this
involution by $F \longrightarrow F^*$. It follows from (\ref{DEFS}) and
(\ref{DEFPHI}) that one also  has $S_k^* = S_k,\ \Phi_k^* = \Phi_k$,
and,
$$
{d\over dt}\ \sigma(t) = \psi(t)^*\ \sigma(t)\ \ ,
$$
with $\psi(t)^* = \sum_{k\geq 1}\ t^{k-1}\, \Psi_k^*$.

\medskip

Other involutions, to be defined below, send $\Lambda_k$ on $\pm S_k$.
This is another way to handle the left-right symmetry in $\Sym$, as shown
by the following proposition which expresses $\psi(t)$ in terms of
$\lambda(t)$.

\begin{proposition}
One has
\begin{equation}\label{DEFPSI2}
-\, {d\over dt}\ \lambda(-t) = \psi(t)\ \lambda(-t)\ .
\end{equation}
\end{proposition}

\Proof Multiplying from left and right equation (\ref{DEFPSI1}) by
$\lambda(-t)$, we obtain
$$ \lambda(-t)\ \left({d\over dt}\ \sigma(t)\right)\ \lambda(-t) =
\sigma(t)^{-1}\ \left({d\over dt}\ \sigma(t)\right)\ \sigma(t)^{-1}
=\psi(t)\ \lambda(-t)\ .
$$
But one also has
$$ \sigma(t)^{-1}\ \left({d\over dt}\ \sigma(t)\right)\ \sigma(t)^{-1}
= -\, {d\over dt}\ \sigma(t)^{-1} = -\, {d\over dt}\ \lambda(-t)\ .
$$
\cqfd

\smallskip
These identities between formal power series imply the following relations
between their coefficients.

\begin{proposition} \label{FORMIN}
For $n \geq 1$, one has
\begin{equation}\label{INV}
\sum_{k=0}^n \ (-1)^{n-k}\, S_k\, \Lambda_{n-k} =
\ \sum_{k=0}^n \ (-1)^{n-k}\, \Lambda_{k}\, S_{n-k} = 0
\ ,
\end{equation}
\smallskip
\begin{equation}\label{NEWTON}
\sum_{k=0}^{n-1} \ S_k\, \Psi_{n-k} = n \, S_n\ ,\quad
\sum_{k=0}^{n-1} \ (-1)^{n-k-1}\, \Psi_{n-k} \, \Lambda_k
= n\, \Lambda_n\ .
\end{equation}
\end{proposition}

\Proof Relation (\ref{INV}) is obtained by considering the coefficient of
$t^n$ in the defining equalities
$\sigma(t)\, \lambda(-t) = \lambda(-t)\, \sigma(t) = 1$. The other identity
is proved similarly from relations (\ref{DEFPSI1}) and (\ref{DEFPSI2}). \cqfd

In the commutative case, formulas (\ref{INV}) and (\ref{NEWTON}) are
respectively called Wronski and Newton formulas.

\smallskip
It follows from relation (\ref{DEFPHI}) that
for $n \geq 1$
\begin{equation}\label{RATCONS}
\Phi_n = n \, S_n + \ \sum_{i_1+\dots+i_m=n,m>1} \ c_{i_1,\dots,i_m} \,
S_{i_1} \, \dots\, S_{i_m}
\end{equation}
where $c_{i_1,\dots,i_m}$ are some rational constants,
so that Proposition \ref{FORMIN} implies
in particular that $\Sym$ is freely generated by any of the families
$(S_k)$, $(\Psi_k)$ or $(\Phi_k)$. This observation leads to the following
definitions.

\begin{definition}
Let $I = (i_1,\dots,i_n) \in (\N^*)^n$ be a composition. One  defines the
products of complete symmetric functions
\begin{equation}
S^I = S_{i_1} \, S_{i_2} \, \dots \, S_{i_n} \ .
\end{equation}
Similarly, one has the products of elementary symmetric functions
\begin{equation}
\Lambda^I = \Lambda_{i_1} \, \Lambda_{i_2} \, \dots \, \Lambda_{i_n} \ ,
\end{equation}
the products of power sums of the first kind
\begin{equation}
\Psi^I = \Psi_{i_1} \, \Psi_{i_2} \, \dots \, \Psi_{i_n} \ ,
\end{equation}
and the products of power sums of the second kind
\begin{equation}
\Phi^I = \Phi_{i_1} \, \Phi_{i_2} \, \dots \, \Phi_{i_n} \ .
\end{equation}
\end{definition}

\begin{note}
{\rm As in the classical case, the complete and elementary symmetric
functions form $\Z$-bases of $\Sym$ ({\it i.e.} bases of the algebra
generated over $\Z$ by the $\Lambda_k$)
while the power sums of first and second kind are just $\Q$-bases.}
\end{note}

The systems of linear equations given by
Proposition \ref{FORMIN} can be solved by means of quasi-determinants. This
leads to the following quasi-determinantal formulas.

\begin{corollary}\label{DETFORM}
For every $n \geq 1$, one has
\begin{equation}\label{SN}
S_n = (-1)^{n-1}
\left|\matrix{
\Lambda_1 & \Lambda_2 & \ldots & \Lambda_{n-1} &\bo{\Lambda_n}\cr
\Lambda_0 & \Lambda_1 & \ldots & \Lambda_{n-2} &\Lambda_{n-1} \cr
0         & \Lambda_0 & \ldots & \Lambda_{n-3} &\Lambda_{n-2} \cr
\vdots    & \vdots    & \ddots & \vdots        &\vdots \cr
0         & 0         & \ldots & \Lambda_0     &\Lambda_1     \cr
}\right|\ ,
\end{equation}
\smallskip
\begin{equation}\label{LN}
\Lambda_n = (-1)^{n-1}
\left|\matrix{
S_1      & S_0     & 0       & \dots  & 0 \cr
S_2      & S_1     & S_0     & \dots  & 0 \cr
S_3      & S_2     & S_1     & \dots  & 0 \cr
\vdots   & \vdots  & \vdots  & \ddots & \vdots \cr
\bo{S_n} & S_{n-1} & S_{n-2} & \dots  & S_1
}\right| \ ,
\end{equation}
\smallskip
\begin{equation} \label{QDSPSI}
n\, S_n =
\left|\matrix{
\Psi_1       & \Psi_2       & \ldots & \Psi_{n-1} &\bo{\Psi_n}\cr
-1           & \Psi_1       & \ldots & \Psi_{n-2} &\Psi_{n-1} \cr
0            & -2           & \ldots & \Psi_{n-3} &\Psi_{n-2} \cr
\vdots       & \vdots       & \ddots & \vdots     &\vdots     \cr
0            & 0            & \ldots & -n+1       &\Psi_1     \cr
}\right|\ ,
\end{equation}
\smallskip
\begin{equation}
n\, \Lambda_n =
\left|\matrix{
\Psi_1       &  1        &   0       &  \ldots  &0 \cr
\Psi_2       &\Psi_1     &   2       &  \ldots  &0 \cr
\Psi_3       &\Psi_2     &\Psi_1     &  \ldots  &0 \cr
\vdots       &\vdots     &\vdots     &\ddots    &\vdots\cr
\bo{\Psi_n}  &\Psi_{n-1} &\Psi_{n-2} &\ldots    &\Psi_1\cr
}\right|\ ,
\end{equation}
\smallskip
\begin{equation} \label{QDPSIS}
\Psi_n =
\left|\matrix{
S_1        & S_0     &  0      &  \ldots  & 0 \cr
2S_2       & S_1     & S_0     &  \ldots  & 0 \cr
3S_3       & S_2     & S_1     &  \ldots  & 0 \cr
\vdots     & \vdots  & \vdots  & \ddots   & \vdots\cr
\bo{nS_n}  & S_{n-1} & S_{n-2} & \ldots   & S_1\cr
}\right|=
\left|\matrix{
\Lambda_1  & 2\Lambda_2 & \ldots & (n-1)\Lambda_{n-1} & \bo{n\Lambda_n}\cr
\Lambda_0  & \Lambda_1  & \ldots & \Lambda_{n-2}      & \Lambda_{n-1} \cr
0          & \Lambda_0  & \ldots & \Lambda_{n-3}      & \Lambda_{n-2} \cr
\vdots     & \vdots     & \ddots & \vdots             & \vdots     \cr
0          & 0          & \ldots & \Lambda_0          & \Lambda_1     \cr
}\right|\ .
\end{equation}
\end{corollary}

\Proof Formulas (\ref{SN}) and (\ref{LN}) follow from
relations (\ref{INV}) by means of Theorem 1.8 of \cite{GR1}. The other
formulas are proved in the same way. \cqfd

\begin{note}
{\rm In the commutative theory, many determinantal formulas can be
obtained from Newton's relations. Most of them can be lifted
to the noncommutative case, as illustrated on the following
relations, due to Mangeot in the commutative case (see \cite{LS}).}
\begin{equation}\label{MAN}
(-1)^{n-1} \,n\, S_n =
\left|
\matrix{
2\,\Lambda_1 & 1            & 0            & \dots  & 0 \cr
4\,\Lambda_2 & 3\,\Lambda_1 & 2\,\Lambda_0 & \dots  & 0 \cr
\vdots       & \vdots       & \vdots       &  & \vdots \cr
(2n-2)\,\Lambda_{n-1} & (2n-3)\,\Lambda_{n-2} & (2n-4)\,\Lambda_{n-3} &
\dots & (n-1)\,\Lambda_0 \cr
\bo{n\,\Lambda_{n}} & (n-1)\,\Lambda_{n-1} & (n-2)\,\Lambda_{n-2} &
\dots & \Lambda_1
}
\right| \ .
\end{equation}
{\rm To prove this formula, it suffices to note that, using relations
(\ref{NEWTON}), one can show as in the commutative case that the matrix
involved in (\ref{MAN}) is the product of the matrix}
$$
{\bf \Psi} =
\left(
\matrix{
\Psi_1  & 1       & 0 & \dots & 0 \cr
-\Psi_2 & \Psi_1  & 2 & \dots & 0 \cr
\Psi_3  & -\Psi_2 & \Psi_1 & \dots & 0 \cr
\vdots  & \vdots  & \vdots &  & \vdots \cr
(-1)^{n-1}\,\Psi_n & (-1)^{n-2}\,\Psi_{n-1} & (-1)^{n-3}\,\Psi_{n-2} & \dots
& \Psi_1
}
\right)
$$
{\rm by the matrix ${\bf \Lambda} = (\Lambda_{j-i})_{0\leq i,j\leq n-1}$.
Formula (\ref{MAN}) then follows from Theorem 1.7 of \cite{GR1}.}
\end{note}

Proposition~\ref{FORMIN} can also be interpreted in the Hopf algebra
formalism. Let us consider the comultiplication $\Delta$ defined in
$\Sym$ by
\begin{equation}
\Delta ( \Psi_k ) = 1 \otimes \Psi_k + \Psi_k \otimes 1
\end{equation}
for  $k \geq 1$. Then, as in the commutative case (\cf \cite{Gei}
for instance), the elementary and complete functions form infinite
sequences of divided powers.

\begin{proposition} \label{COPROD} For every $k \geq 1$, one has
$$
\Delta (S_k) = \sum_{i=0}^k \ S_i \otimes S_{k-i} \ , \quad
\Delta (\Lambda_k) = \sum_{i=0}^k \ \Lambda_i \otimes \Lambda_{k-i} \ .
$$
\end{proposition}

\Proof We prove the claim  for $\Delta(S_k)$, the other proof being similar.
Note first that there is nothing to prove for $k = 0$ and $k =1$. Using now
induction on $k$ and relation (\ref{NEWTON}), we obtain
$$
k\, \Delta(S_k) = \ \sum_{i=0}^{k-1} \ \sum_{j=0}^i \
(S_{j} \otimes S_{i-j} \Psi_{k-i} + S_{i-j} \Psi_{k-i} \otimes S_{j})
$$
$$
=  \ \sum_{j=0}^{k-1} \ S_j \otimes ( \sum_{i=j}^{k-1} S_{i-j} \Psi_{k-i} )
\ + \ \sum_{j=0}^{k-1} \ ( \sum_{i=j}^{k-1} S_{i-j} \Psi_{k-i}) \otimes S_j
$$
$$
= \ \sum_{j=0}^{k-1} \ ( S_j \otimes (k-j)\, S_{k-j} ) \
+ \ \sum_{j=0}^{k-1} \ (k-j)\, S_{k-j} \otimes S_j
= \ k\ \sum_{j=0}^k \ (S_j \otimes S_{k-j}) \ .
$$
\cqfd

We define an anti-automorphism $\omega$ of $\Sym$ by setting
\begin{equation}
\omega(S_k) = \Lambda_k
\end{equation}
for  $k \geq 0$. The different formulas of Corollary \ref{DETFORM}
show that
$$
\omega(\Lambda_k) = S_k \ , \quad \omega(\Psi_k) = (-1)^{k-1}\, \Psi_k \ .
$$
In particular we see that $\omega$ is an involution.
To summarize:

\begin{proposition}
The comultiplication $\Delta$ and the antipode $\tilde{\omega}$, where
$\tilde{\omega}$ is the anti-au\-to\-mor\-phism of $\Sym$ defined by
\begin{equation}
\tilde{\omega}(S_k) = (-1)^{k}\, \Lambda_k,\ \ k\geq 0\ ,
\end{equation}
endow $\Sym$ with the structure of a Hopf algebra.
\end{proposition}

The following property,
which is equivalent to the Continuous Baker-Campbell-Hausdorff theorem
of Magnus \cite{Mag} and Chen \cite{Chen7},
 has some interesting consequences.

\begin{proposition}\label{LIE}
The Lie algebra generated by the family $(\Phi_k)$
coincides with the Lie algebra generated by the family $(\Psi_k)$. We shall
denote it by $L(\Psi)$. Moreover the difference $\Phi_k - \Psi_k$ lies in
the Lie ideal $L^{2}(\Psi)$ for every $k \geq 1$.
\end{proposition}

\Proof In order to see that the two Lie algebras $L(\Psi)$ and $L(\Phi)$
coincide, it is sufficient to show that $\Phi$ lies in $L(\Psi)$.
According to Friedrichs' criterion (\cf \cite{Re} for instance),
we just have  to check that $\Phi$ is primitive
for the comultiplication $\Delta$. Using (\ref{DEFPHI}) and
Proposition \ref{COPROD}, we can now write
\begin{equation}
\sum_{k\geq 1}\ t^k \, {\Delta (\Phi_k) \over k} =
\ \log \, (\, \sum_{k\geq 0} \ \Delta(S_k) \, t^k \, )
= \log \, (\, U V \,) \ ,
\end{equation}
where we respectively set $U = \sum_{k\geq 0} \ (1\otimes S_k)\, t^k$ and
$V = \sum_{k\geq 0} \ (S_k \otimes 1)\, t^k$. Since all coefficients of $U$
commute with all coefficients of $V$, we have $\log (UV) = \log(U) + \log(V)$
from which, applying again (\ref{DEFPHI}), we  obtain
\begin{equation}
\sum_{k\geq 1}\ t^k \, {\Delta (\Phi_k) \over k} =
\sum_{k\geq 1}\ t^k \, {{1\otimes \Phi_k}\over k} \ + \
\sum_{k\geq 1}\ t^k \, {{\Phi_k \otimes 1}\over k} \ ,
\end{equation}
as required. The second point follows from the fact that $\Phi_k - \Psi_k$
is of order at least 2 in the $\Psi_i$, which is itself a consequence of
(\ref{RATCONS}).
\cqfd

\begin{example}
{\rm The first Lie relations between $\Phi_k$ and $\Psi_k$ are given below.}
$$
\Phi_1 = \Psi_1, \quad \Phi_2 = \Psi_2, \quad \Phi_3 = \Psi_3 + {1\over 4}
\, [\Psi_1,\Psi_2], \quad \Phi_4 = \Psi_4 + {1\over 3}\, [\Psi_1,\Psi_3],
$$
$$
\Phi_5 = \Psi_5 + {3\over 8}\, [\Psi_1,\Psi_4] +
{1\over 12}\, [\Psi_2,\Psi_3] +
{1\over 72}\, [\Psi_1,[\Psi_1,\Psi_3]]
$$
$$
+ {1\over 48}\, [[\Psi_1,\Psi_2],\Psi_2] +
{1\over 144}\, [[[\Psi_2,\Psi_1],\Psi_1],\Psi_1] \ .
$$
\end{example}

\smallskip
It is worth noting that, according to Proposition~\ref{LIE}, the Hopf
structures defined by requiring the $\Phi_k$ or the $\Psi_k$ to be primitive
are the same. Note also that the antipodal property of $\tilde{\omega}$
and the fact that
\begin{equation}
\Delta(\Phi_k) = 1 \otimes \Phi_k + \Phi_k \otimes 1
\end{equation}
show that we have
\begin{equation}
\omega(\Phi_k) = (-1)^{k-1}\, \Phi_k \ .
\end{equation}


\subsection{Ribbon Schur functions} \label{RIBBONS}

A general notion of {\it quasi-Schur function} can be obtained by replacing
the classical Jacobi-Trudi determinant by an appropriate quasi-determinant
(see Section \ref{QUASI}). However, as mentioned in the introduction, these
expressions are no longer polynomials in the ge\-ne\-rators of $\Sym$, except
in one case. This is when all the subdiagonal elements $a_{i+1,i}$ of their
defining quasi-determinant are scalars, which corresponds to Schur functions
indexed by {\it ribbon shaped} diagrams. We recall that a {\it ribbon} or
{\it skew-hook} is a skew Young diagram containing no $2\times 2$
block of boxes. For instance, the skew diagram $\Theta = I/J$ with
$I = (3,3,3,6,7)$ and $J = (2,2,2,5)$

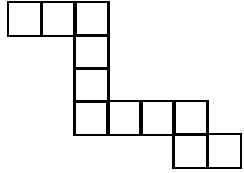
\centerline{\setlength{\unitlength}{0.25pt}
\begin{picture}(300 ,300 )
\put(0,200){\framebox(50,50){}}
\put(50,200){\framebox(50,50){}}
\put(100,200){\framebox(50,50){}}
\put(100,150){\framebox(50,50){}}
\put(100,100){\framebox(50,50){}}
\put(100,50){\framebox(50,50){}}
\put(150,50){\framebox(50,50){}}
\put(200,50){\framebox(50,50){}}
\put(250,50){\framebox(50,50){}}
\put(250,0){\framebox(50,50){}}
\put(300,0){\framebox(50,50){}}
\end{picture}
}

\bigskip
\noindent
is a ribbon. A ribbon $\Theta$ with $n$ boxes is naturally encoded by a
{\it composition} $I = (i_1,\ldots,i_r)$ of $n$, whose parts are the lengths
of its rows (starting from the top). In the above example, we have
$I=(3,1,1,4,2)$. Following \cite{MM}, we also define the
{\it conjugate composition} $I^\sim$ of $I$ as the one whose associated
ribbon is the conjugate (in the sense of skew diagrams) of the ribbon of
$I$. For example, with $I= (3,1,1,4,2)$, we have $I^\sim=(1,2,1,1,4,1,1)$.

\smallskip
In the commutative case, the skew Schur functions indexed by ribbon
diagrams possess interesting properties and are strongly related to
the combinatorics of compositions and descents of permutations
(\cf \cite{Ge}, \cite{GeR} or \cite{Re} for instance). They have been defined
and investigated by MacMahon (\cf \cite{MM}). Although the commutative ribbon
functions are not linearly independent, {\it e.g.} $R_{12} = R_{21}$,
it will be shown in Section
\ref{BASES} that the noncommutative ones form a linear basis of $\Sym$.

\begin{definition}
Let $I = (i_1,\dots,i_n) \in (\N^*)^n$ be a composition. Then the {\rm ribbon}
Schur function $R_I$ is defined by
\begin{equation}
R_I = (-1)^{n-1}\
\left|
\matrix{
S_{i_1} & S_{i_1+i_2} & S_{i_1+i_2+i_3} & \dots  & \bo{S_{i_1+\dots+i_n}} \cr
S_0     & S_{i_2}     & S_{i_2+i_3}     & \dots  & S_{i_2+\dots+i_n} \cr
0       & S_0         & S_{i_3}         & \dots  & S_{i_3+\dots+i_n} \cr
\vdots  & \vdots      & \vdots          & \ddots & \vdots \cr
0       & 0           & 0               & \dots  & S_{i_n}
}
\right|
\end{equation}
\end{definition}

\smallskip
As in the commutative case, we have $S_n = R_n$ and $\Lambda_n = R_{1^n}$
according to Corollary~\ref{DETFORM}. Moreover, the ribbon functions
$R_{(1^k,n-k)}$ play a particular role. In the commutative case, they
are the {\it hook Schur functions}, denoted $S_{(1^k,n-k)}$. We shall also
use this notation for the noncommutative ones.

We have for the multiplication of ribbon Schur functions the following
formula, whose commutative version is due to MacMahon (\cf \cite{MM}).

\begin{proposition}\label{MR} Let $I=(i_1,\ldots,i_r)$ and
$J=(j_1,\ldots,j_s)$ be two compositions. Then,
$$
R_I \,  R_J = R_{I\triangleright J} + R_{I\cdot J}
$$
where $I\triangleright J$ denotes the composition
$(i_1,\ldots,i_{r-1},i_r+j_1,j_2,\ldots,j_s)$
and $I\cdot J$ the com\-po\-si\-tion $(i_1,\ldots,i_r,j_1,\ldots,j_s)$.
\end{proposition}

\Proof Using the above notations, we can write that $R_I$ is equal to
\smallskip
$$
(-1)^{k-1} \,
\left|
\matrix{
S_{i_1} & \ldots     & \bo{S_{i_1+\dots+i_{k-1}}} \cr
S_0     & \ldots     & S_{i_2+\dots+i_{k-1}} \cr
\vdots  &            &\vdots  \cr
0       & \ldots     & S_{i_{k-1}} \cr
}
\right|
\left|
\matrix{
S_0    & \ldots    & S_{i_2+\dots+i_{k-1}} \cr
0      & \ldots    & S_{i_3+\dots+i_{k-1}} \cr
\vdots &           & \vdots \cr
0      &   \ldots  & \bo{S_0}    \cr
}
\right|^{-1}
\left|
\matrix{
S_{i_1}   & S_{i_1+i_2}  & \ldots & S_{i_1+\dots+i_k} \cr
S_0       & S_{i_2}      & \ldots & S_{i_2+\dots+i_k} \cr
\vdots    & \vdots       & \ddots & \vdots  \cr
0         & 0            & \ldots & \bo{S_{i_k}} \cr
}
\right|\ ,
$$
$$
= R_{(i_1,\dots,i_{k-1})}\, (S_{i_k} - R_{(i_1,\dots,i_{k-1})}^{-1}
\, R_{(i_1,\dots,i_{k-2},i_{k-1}+i_k)} )\ ,
$$
the first equality following from the homological relations between
quasi-determinants of the same matrix, and the second one from the expansion
of the last quasi-determinant by its last row together with the fact
that the second quasi-determinant involved in this relation is equal to $1$.
Hence,
\begin{equation}\label{RS}
R_I \, R_n = R_{I \triangleright n} + R_{I\cdot n} \ .
\end{equation}
Thus, the proposition holds for $\ell(J) = 1$. Let
now $J = (j_1,\dots,j_n)$ and $J' = (j_1,\dots,j_{n-1})$. Then
relation (\ref{RS}) shows that
$$
R_J = R_{J'}\, R_{j_n} - R_{J' \triangleright j_n} \ .
$$
Using this last identity together with  induction on $\ell(J)$, we get
$$
R_I \, R_J = R_{I \triangleright J'} \, R_{j_n} + R_{I \cdot J'} \, R_{j_n}
- R_{I \triangleright (J' \triangleright j_n)} -
R_{I \cdot (J' \triangleright j_n)} \ ,
$$
and the conclusion follows  by means of relation (\ref{RS})
applied two times. \cqfd

Proposition~\ref{MR} shows in particular that the product of
an elementary by a complete function is the sum of two hook functions,
\begin{equation}
\Lambda_k\ S_l = R_{1^k l} + R_{1^{k-1} (l+1)} \ .\label{PRODLS}
\end{equation}

\medskip
We also note the following expression of the
power sums $\Psi_n$ in terms of hook Schur functions.

\begin{corollary}\label{psihook}
For every $n \geq 1$, one has
$$
\Psi_n = \ \sum_{k=0}^{n-1} \ (-1)^k\, R_{1^k(n-k)} \ .
$$
\end{corollary}

\Proof The identity
$\lambda(-t)\ {\displaystyle {d\over dt}}\ \sigma(t) = \psi(t)$ implies that
\begin{equation}
\Psi_n = \ \sum_{k=0}^{n-1} \ (-1)^k\, (n-k)\, \Lambda_k\, S_{n-k}\ ,
\end{equation}
and the result follows from (\ref{PRODLS}). \cqfd

\smallskip
Let us introduce the two infinite Toeplitz matrices
\begin{equation}
{\bf S} = \left( S_{j-i} \right)_{i,j\geq 0}\ ,\quad
{\bf \Lambda} = \left( (-1)^{j-i}\,\Lambda_{j-i} \right)_{i,j\geq 0}\ ,
\label{TOEPLITZ}
\end{equation}
where we set $S_k = \Lambda_k = 0$ for $k<0$. The identity
$\lambda(-t)\ \sigma(t)=\sigma(t)\ \lambda(-t) = 1$ is equivalent to the
fact that ${\bf \Lambda \ S}={\bf S\ \Lambda} = I$. We know that the
quasi-minors of the inverse matrix $A^{-1}$ are connected with those of $A$
by Jacobi's ratio theorem for quasi-determinants (Theorem \ref{JACOBI}).
Thus the ribbon functions may also be expressed
in terms of the $\Lambda_k$, as shown by the next proposition.

\begin{proposition}\label{LRIB} Let $I \in \N^n$ be a composition and let
$I^{\sim} = (j_1,\dots,j_m)$ be the conjugate composition. Then one has
the relation
$$
R_I = (-1)^{m-1} \
\left|
\matrix{
\Lambda_{j_m} & \Lambda_{j_{m-1}+j_m} & \Lambda_{j_{m-2}+j_{m-1}+j_m} & \ldots
& \bo{\Lambda_{j_1+\dots+j_m}} \cr
\Lambda_0 & \Lambda_{j_{m-1}} & \Lambda_{j_{m-2}+j_{m-1}} & \ldots
& \Lambda_{j_1+\dots+j_{m-1}} \cr
0         & \Lambda_0     & \Lambda_{j_{m-2}} & \ldots
&\Lambda_{j_1+\dots+j_{m-
2}} \cr
\vdots    & \vdots        & \vdots & \ddots    & \vdots \cr
0         & 0             & 0      & \ldots    & \Lambda_{j_1} \cr
}\right|\ .
$$
\end{proposition}

\Proof This formula is obtained by applying Jacobi's theorem for
quasi-determinants (Theorem \ref{JACOBI}) to the definition of $R_I$. \cqfd

\begin{corollary}
For any composition $I$, one has
$$
\omega(R_I) = R_{I^{\sim}} .
$$
\end{corollary}


\subsection{Quasi-Schur functions}\label{QUASI}

We shall now consider general quasi-minors of the matrices ${\bf S}$ and
${\bf \Lambda}$ and proceed to the definition of quasi-Schur functions.
As already mentioned, they are no longer elements of $\Sym$, but of the
free field $K\!\not< \! S_1,\,S_2,\, \ldots \!\not>$ generated by the $S_i$.

\begin{definition}\label{QSF}
Let $I = (i_1,i_2,\ldots,i_n)$ be a {\rm partition}, i.e. a weakly increasing
sequence of nonnegative integers. We define the {\rm quasi-Schur function}
$\check{S}_I$ by setting
\begin{equation} \label{QSFF}
\check{S}_I = (-1)^{n-1} \,
\left|
\matrix{
S_{i_1}     & S_{i_2+1}   & \ldots & \bo{S_{i_n +n-1}} \cr
S_{i_1-1}   & S_{i_2}     & \ldots & S_{i_n +n-2} \cr
\vdots      & \vdots      & \ddots & \vdots      \cr
S_{i_1-n+1} & S_{i_2-n+2} & \ldots & S_{i_n} \cr
}
\right|\ .
\end{equation}
\end{definition}

\medskip
In particular we have $\check{S}_i = S_i$, $\check{S}_{1^i} = \Lambda_i$
and $\check{S}_{1^i(n-i)} = R_{1^i(n-i)}$. However it must be noted that
for a general partition $I$, the quasi-Schur function $\check{S}_I$ reduces
in the commutative case to the {\it ratio} of two Schur functions
$S_I/S_J$, where $J = (i_1-1,i_2-1,\ldots ,i_{n-1}-1)$.
One can also define in the same way {\it skew Schur functions}. One
has to take the same minor as in the commutative case, with the
box in the upper right corner and the sign $(-1)^{n-1}$, where
$n$ is the order of the quasi-minor. The ribbon Schur functions
are the only quasi-Schur functions which are polynomials in the
$S_k$, and they reduce to the ordinary ribbon functions in the
commutative case. To emphasize this point, we shall also denote the
quasi-Schur function $\check{S}_{I/J}$ by $S_{I/J}$
when $I/J$ is a ribbon.

\medskip
Quasi-Schur functions are indexed by {\it partitions}.
It would have been possible to define more general functions indexed by
compositions, but the homological relations imply that such functions
can always be expressed as  noncommutative
rational fractions in the quasi-Schur functions.
For instance
$$
\check{S}_{42}
=
-
\left|
\matrix{
S_4 & \bo{S_3} \cr
S_3 &  S_2
}
\right|
=
-
\left|
\matrix{
\bo{S_3} & S_4  \cr
S_2      & S_3
}
\right|
=
\left|
\matrix{
S_3 & \bo{S_4} \cr
S_2 & S_3
}
\right|
\
S_3^{-1} \, S_2
=
\check{S}_{33} \, S_3^{-1} \, S_2 \ .
$$

\medskip

Definition \ref{QSF} is a noncommutative analog of the so-called Jacobi-Trudi
formula. Using Jacobi's theorem for the quasi-minors of the inverse matrix
as in the proof of Proposition~\ref{LRIB}, we derive the following analog
of Naegelbasch's formula.

\begin{proposition}\label{NAE}
Let $I$ be a partition and let $I^{\sim} = (j_1,\ldots ,j_p)$ be its conjugate
partition, i.e. the partition whose diagram is obtained by interchanging the
rows and columns of the diagram of $I$. Then,
$$
\check{S}_I = (-1)^{p-1} \
\left|
\matrix{
\Lambda_{j_p}       & \Lambda_{j_p+1}    & \ldots & \bo{\Lambda_{j_p+p-1}} \cr
\Lambda_{j_{p-1}-1} & \Lambda_{j_{p-1}}  & \ldots & \Lambda_{j_{p-1}+p-2} \cr
\vdots              & \vdots             & \ddots & \vdots  \cr
\Lambda_{j_1-p+1}   & \Lambda_{j_1-p+2}  & \ldots &\Lambda_{j_1} \cr
}
\right| \ .
$$
\end{proposition}

Let us extend $\omega$ to the skew field generated by the $S_k$. Then we
have.

\begin{proposition}
Let $I$ be a partition and let $I^{\sim}$ be its conjugate partition. There
holds
$$
\omega(\check{S}_I)=\check{S}_{I^{\sim}} \ .
$$
\end{proposition}

\Proof This is a consequence of Proposition~\ref{NAE}. \cqfd

\medskip
We can also extend the $*$ involution to the division ring generated  by the
$S_i$.
It follows from Definition \ref{DEFQUASIDET} that $\check S_I^*$ is equal to
the
transposed quasi-determinant
$$
\check{S}_I^* = (-1)^{n-1} \
\left|
\matrix{
S_{i_n}       & S_{i_n+1}    & \ldots & \bo{S_{i_n+n-1}} \cr
S_{i_{n-1}-1} & S_{i_{n-1}}  & \ldots & S_{i_{n-1}+n-2} \cr
\vdots              & \vdots             & \ddots & \vdots  \cr
S_{i_1-n+1}   & S_{i_1-n+2}  & \ldots &S_{i_1} \cr
}
\right| \ .
$$
In particular, if $I=n^k$ is a rectangular partition, the quasi-Schur function
indexed by $I$ is invariant under $*$.

\medskip
In the commutative case, Schur functions can also be expressed as determinants
of hook Schur functions, according to a formula of Giambelli. This formula is
stated more conveniently, using Frobenius' notation for partitions. In this
notation, the hook

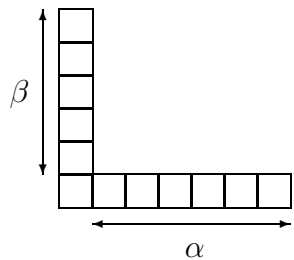
\setlength{\unitlength}{0.25pt}
\centerline{
\begin{picture}(450,450)(-50,-75)
\put(0,0){\framebox(50,50){}}
\put(50,0){\framebox(50,50){}}
\put(100,0){\framebox(50,50){}}
\put(150,0){\framebox(50,50){}}
\put(200,0){\framebox(50,50){}}
\put(250,0){\framebox(50,50){}}
\put(300,0){\framebox(50,50){}}
\put(0,50){\framebox(50,50){}}
\put(0,100){\framebox(50,50){}}
\put(0,150){\framebox(50,50){}}
\put(0,200){\framebox(50,50){}}
\put(0,250){\framebox(50,50){}}
\put(-25,100){\vector(0,1){200}}
\put(-25,100){\vector(0,-1){50}}
\put(150,-25){\vector(1,0){200}}
\put(150,-25){\vector(-1,0){100}}
\put(-75,160){$\beta$}
\put(190,-75){$\alpha$}
\end{picture}
}

\noindent
is written $(\beta |\alpha) := 1^{\beta}\, (\alpha +1)$. A general partition is
decomposed into diagonal hooks. Thus, for the partition $I=(2,3,5,7,7)$ for
instance, we have the following decomposition, where the different hooks are
distinguished by means of the symbols $\star,\, \bullet$ and $\diamond$.

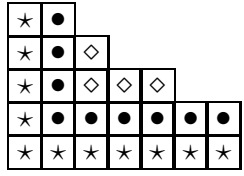
\setlength{\unitlength}{0.25pt}
\centerline{
\begin{picture}(400,400)(0,-50)
\put(0,0){\framebox(50,50){$\star$}}
\put(50,0){\framebox(50,50){$\star$}}
\put(100,0){\framebox(50,50){$\star$}}
\put(150,0){\framebox(50,50){$\star$}}
\put(200,0){\framebox(50,50){$\star$}}
\put(250,0){\framebox(50,50){$\star$}}
\put(300,0){\framebox(50,50){$\star$}}
\put(0,50){\framebox(50,50){$\star$}}
\put(0,100){\framebox(50,50){$\star$}}
\put(0,150){\framebox(50,50){$\star$}}
\put(0,200){\framebox(50,50){$\star$}}
\put(50,50){\framebox(50,50){$\bullet$}}
\put(100,50){\framebox(50,50){$\bullet$}}
\put(150,50){\framebox(50,50){$\bullet$}}
\put(200,50){\framebox(50,50){$\bullet$}}
\put(250,50){\framebox(50,50){$\bullet$}}
\put(300,50){\framebox(50,50){$\bullet$}}
\put(50,100){\framebox(50,50){$\bullet$}}
\put(50,150){\framebox(50,50){$\bullet$}}
\put(50,200){\framebox(50,50){$\bullet$}}
\put(100,100){\framebox(50,50){$\diamond$}}
\put(150,100){\framebox(50,50){$\diamond$}}
\put(200,100){\framebox(50,50){$\diamond$}}
\put(100,150){\framebox(50,50){$\diamond$}}
\end{picture}
}

\noindent
It is denoted $(134\ | \ 256)$ in Frobenius' notation. We can now state.

\begin{proposition} {\rm (Giambelli's formula for quasi-Schur functions)}
Let $I$ be a par\-tition represented by
$(\beta_1 \ldots \beta_k\ | \ \alpha_1 \ldots \alpha_k)$ in Frobenius'
notation. One has
$$
\check{S}_I =
\left|
\matrix{
\check{S}_{(\beta_1|\alpha_1)} & \check{S}_{(\beta_1|\alpha_2)} & \ldots &
\check{S}_{(\beta_1|\alpha_k)}\cr
\check{S}_{(\beta_2|\alpha_1)} & \check{S}_{(\beta_2|\alpha_2)} & \ldots &
\check{S}_{(\beta_2|\alpha_k)}\cr
\vdots     &  \vdots  & \ddots & \vdots      \cr
\check{S}_{(\beta_k|\alpha_1)} & \check{S}_{(\beta_k|\alpha_2)} & \ldots &
\bo{\check{S}_{(\beta_k|\alpha_k)}}\cr
}
\right| \ .
$$
\end{proposition}

\Proof This is an example of relation between the quasi-minors of the matrix
{\bf S}. The proposition is obtained by means of Bazin's theorem for
quasi-determinants \ref{BAZIN}. It is sufficient to illustrate the
computation in the case of $I=(2,3,5,7,7) = (134\ |\ 256)$. We denote by
$|i_1i_2i_3i_4\bo{i_5}|$ the quasi-minor of {\bf S} defined by
$$
|i_1i_2i_3i_4\bo{i_5}| =
\left|
\matrix{
S_{i_1}    &  S_{i_2}    & \ldots & \bo{S_{i_5}}\cr
S_{i_1-1}  &  S_{i_2-1}  & \ldots & S_{i_5-1}\cr
\vdots     &  \vdots     & \ddots & \vdots      \cr
S_{i_1-4}  &  S_{i_2-4}  & \ldots & S_{i_5-4}\cr
}
\right|\ .
$$
Using this notation, we have
$$
\left|
\matrix{
\check{S}_{(1|2)}  &  \check{S}_{(1|5)}  & \check{S}_{(1|6)}\cr
\check{S}_{(3|2)}  &  \check{S}_{(3|5)}  & \check{S}_{(3|6)}\cr
\check{S}_{(4|2)}  &  \check{S}_{(4|5)}  & \bo{\check{S}_{(4|6)}}\cr
}\right|\
=
\left|\matrix{
|0124\bo{7}| & |0124\bo{10}| &|0124\bo{11}|\cr
|0234\bo{7}| & |0234\bo{10}| &|0234\bo{11}|\cr
|1234\bo{7}| & |1234\bo{10}| &\bo{|1234\bo{11}|}\cr
}\right|\ ,
$$
$$
= |\bo{0}1234|\ |\bo{0}247\ 10|^{-1} \
|247\ 10\ \bo{11}| = |247\ 10\ \bo{11}| = \check{S}_{23577}\ ,
$$
the second  equality following from Bazin's theorem. \cqfd

\medskip

There is also an expression of quasi-Schur functions as quasi-determinants
of ribbon Schur functions, which extends to the noncommutative case a
formula given in \cite{LP}. To state it, we
introduce some notations. A ribbon $\Theta$ can be seen as the
outer strip of a Young diagram $D$. The unique box of $\Theta$ which is
on the diagonal of $D$ is called the {\it diagonal box} of $\Theta$.
The boxes of $\Theta$ which are strictly above the diagonal form a
ribbon denoted $\Theta^+$. Similarly those strictly under the diagonal
form a ribbon denoted $\Theta^-$. Given two ribbons $\Theta$ and $\Xi$,
we shall denote by $\Theta^+ \&\, \Xi^-$ the ribbon obtained from $\Theta$
by replacing $\Theta^-$ by $\Xi^-$. For example, with $\Theta$ and
$\Xi$ corresponding to the compositions $I=(2,1,1,3,2,4)$, $J=(1,3,3,1,1,2)$,
$\Theta^+$ corresponds to $(2,1,1,1)$, $\Xi^-$ to $(1,1,1,2)$ and
$\Theta^+ \&\, \Xi^-$ to $(2,1,1,3,1,1,2)$. Given a partition $I$, we can peel
off its diagram into successive ribbons $\Theta_p,\ldots,\Theta_1$, the
outer one being $\Theta_p$ (see \cite{LP}). Using these notations, we have
the following identity.

\begin{proposition}\label{LASPRAG}
Let $I$ be a partition and $(\Theta_p,\ldots,\Theta_1)$ its
decomposition into ribbons. Then, we have
$$
\check{S}_I=
\left|\matrix{
\check{S}_{\Theta_1} & \check{S}_{\Theta_1^+\&\Theta_2^-} & \ldots &
\check{S}_{\Theta_1^+\&\Theta_p^-}
\vtr{3} \cr
\check{S}_{\Theta_2^+\&\Theta_1^-} & \check{S}_{\Theta_2} & \ldots &
\check{S}_{\Theta_2^+\&\Theta_p^-}
\vtr{3} \cr
\vdots & \vdots & \ddots & \vdots \vtr{3} \cr
\check{S}_{\Theta_p^+\&\Theta_1^-} & \check{S}_{\Theta_p^+\&\Theta_2^-}
&\ldots & \bo{\check{S}_{\Theta_p}} \cr
}\right| \ .
$$
\end{proposition}

\Proof The two proofs proposed in \cite{LP} can be adapted to the
noncommutative case. The first one, which rests upon Bazin's
theorem, is similar to the proof of Giambelli's formula given
above. The second one takes Giambelli's formula as a starting
point, and proceeds by a step by step deformation of hooks
into ribbons, using at each stage the multiplication formula for
ribbons and subtraction to a row or a column of a multiple
of an other one. To see how it works, let us consider for
example the quasi-Schur function $\check S_{235}$. Its expression
in terms of hooks is
$$
\check S_{235}=
\left|\matrix{
R_{12}  &  R_{15}  \cr
R_{112} & \bo{R_{115}}\cr
}\right| \ .
$$
Subtracting to the second row the first one multiplied to
the left by $R_1$ (which does not change the value of the
quasi-determinant) and using the multiplication formula
for ribbons, we arrive at a second expression
$$
\check S_{235}=
\left|\matrix{
R_{12}  &  R_{15}  \cr
R_{22} & \bo{R_{25}}\cr
}\right|\ .
$$
Now, subtracting to the second column the first one multiplied
to the right by $R_3$ we finally obtain
$$
\check S_{235}=
\left|\matrix{
R_{12}  &  R_{123}  \cr
R_{22} & \bo{R_{223}}\cr
}\right|\ ,
$$
which is the required expression. \cqfd

\newpage
\section{Transition matrices}\label{BASES}

This section is devoted to the study of
the transition matrices between the previously
introduced bases of $\Sym$. As we shall see, their description is
rather simpler than in the commutative case. We recall that $\Sym$ is a
graded algebra
$$
\Sym = \bigoplus_{n\ge0} \ \Sym_n
$$
$\Sym_n$ being the subspace of dimension $2^{n-1}$  generated
by the symmetric functions $S^I$, for all compositions $I$ of $n$.

\smallskip
The description of the transition matrices
can be abridged by using  the involution $\omega$.
Note that the action of $\omega$ on the usual bases is given by
$$
\omega(S^I) = \Lambda^{\overline{I}} \, , \quad
\omega(\Lambda^I) = S^{\overline{I}} \, , \quad
\omega(\Psi^I) = (-1)^{|I|-\ell(I)} \, \Psi^{\overline{I}} \, , \quad
\omega(\Phi^I) = (-1)^{|I|-\ell(I)} \, \Phi^{\overline{I}} \, ,
$$
where we denote by $\overline{I}$ the mirror image of the composition $I$,
i.e. the new composition obtained by reading $I$ from right to left.

\medskip
We shall consider two orderings on the set of all compositions of
an integer $n$. The first one is the {\it reverse refinement order}, denoted
$\preceq$, and defined by $I\preceq J$ iff $J$ is finer than $I$.
For example, $(326) \preceq (212312)$.
The second one is the reverse lexicographic ordering, denoted $\le$.
For example, $(6)\le(51)\le(42)\le(411)$. This ordering is well suited
for the indexation of matrices, thanks to the following property.

\begin{lemma}\label{LEX}
Let $C_n$ denote the sequence of all compositions of $n$ sorted
by reverse lexicographic order. Then,
$$
C_n = ( \, 1 \triangleright C_{n-1}  \ , \ 1 \, . \, C_{n-1} \, )
$$
where $1 \triangleright C_{n-1} $ and $1\, .\, C_{n-1}$ denote respectively the
compositions
obtained from the compositions of $C_{n-1}$ by adding $1$ to their first part
and by considering $1$ as their new first part, the other parts being
unchanged.
\end{lemma}

\begin{note}\label{REGRUB}{\rm
Writing $S_1^n = S_1 \, S_1^{n-1}$, Lemma \ref{LEX} proves in particular
that
\begin{equation}
(S_1)^n = \sum_{|I|=n} R_I \ ,
\end{equation}
a noncommutative analogue of a classical formula which is relevant
in the representation theory of the symmetric group (the left hand
side is the characteristic of the regular representation, which is thus
decomposed as the direct sum of all the ribbon representations). This
decomposition appears when the regular representation is realized
in the space ${\cal H}_n$ of $\S_n$-harmonic polynomials, or, which
amounts to the same, in the cohomology of the variety of complete
flags (more details will be given in Section \ref{LIEID}).}
\end{note}

\medskip
For every pair $(F_I)$, $(G_I)$ of graded
bases of $\Sym$ indexed by compositions, we denote by $M(F,G)_n$ the
transition matrix from the basis $(F_I)$ with $|I| = n$ to the basis $(G_I)$
with
$|I| = n$ indexed by compositions sorted in reverse lexicographic order. Our
convention is that the {\it row} indexed by $I$ contains the components of
$F_I$ in the basis $(G_J)$.


\subsection{$S$ and $\Lambda$}

The matrices $M(S,\Lambda)_n$ and $M(\Lambda,S)_n$ are easily described.
With the choosen indexation, they appear as Kronecker powers of a simple
$2\times 2$ matrix.

\begin{proposition} \label{MATRANLS}
For every $n \geq 1$, we have
\begin{equation}
M(S,\Lambda)_n = M(\Lambda,S)_n
=
\left(\matrix{
-1 & 1 \cr
0 & 1 \cr
}\right)^{\otimes (n-1)} \ .
\end{equation}
\end{proposition}

\Proof The defining relation $\sigma(-t)\, \lambda(t) = 1$ shows that
$$
\sigma(-t)
=
(\, 1 - \ \sum_{i\geq 1} \ \Lambda_i \, t^i \, )^{-1}
=
1 + \ \sum_{k\geq 1} \ (-1)^k \, (\ \sum_{i\geq 1} \ \Lambda_i \, t^i \, )^k
\ .
$$
Identifying the coefficients, we get
$$
S_k
=
\ \sum_{|J|=k} \ (-1)^{{\em l}(J)-k} \, \Lambda^J \ ,
$$
so that
$$
S^I
=
\ \sum_{J\succeq I} \ (-1)^{{\em l}(J)-|I|} \, \Lambda^J \ ,
$$
for every composition $I$.  The conclusion follows then from Lemma \ref{LEX}.
Applying $\omega$ to this relation, we see
that the same relation holds when $S$ and $\Lambda$ are interchanged.  \cqfd

\begin{example}
{\rm For $n = 2$ and $n = 3$, we have}
$$
M(S,\Lambda)_2 = M(\Lambda,S)_2 =
\matrix{
&\ 2 \ \ \ \ 11 \vt \cr
\matrix{ 2 \cr 11 } \!\!\!\!\! &
\left(\matrix{
-1 & 1 \cr
0 & 1 \cr
}\right) \vt
}
\ ,
$$
\smallskip
$$
M(S,\Lambda)_3 = M(\Lambda,S)_3 =
\matrix{
& 3 \ \ \ 21 \ \ \ 12 \ \ \ 111 \vt \cr
\matrix{ 3 \cr 21 \cr 12 \cr 111} \!\!\!\!\! &
\left(\matrix{
1  & -1 & -1 & 1 \cr
0  & -1 & 0  & 1 \cr
0  & 0  & -1 & 1 \cr
0  & 0  & 0  & 1
}\right)
}\ .
$$
\end{example}

\medskip
As shown in the proof of   Proposition \ref{MATRANLS}, we have
$$
S^I = \sum_{J\succeq I} \ (-1)^{\ell(J)-|I|} \, \Lambda^J \ , \quad
\Lambda^I = \sum_{J\succeq I} \ (-1)^{\ell(J)-|I|} \, S^J \ ,
$$
for every composition $I$. Hence the matrices
$M(S,\Lambda)_n = M(\Lambda,S)_n$ are upper triangular matrices whose
entries are only $0$, $1$ or $-1$ and such that moreover the non-zero
entries of each column are all equal.


\subsection{$S$ and $\Psi$}

Before describing the matrices $M(S,\Psi)_n$ and $M(\Psi,S)_n$, let us
introduce some notations. Let $I = (i_1,\dots,i_m)$ be a composition. We
define $\pi_u(I)$ as follows
$$
\pi_u(I) = i_1 \, (i_1+i_2) \, \dots \, (i_1+i_2+\dots+i_m) \ .
$$
In other words, $\pi_u(I)$ is the product of the successive partial sums of
the entries of the composition $I$. We also use a special notation for the
last part of the composition by setting $lp(I) = i_m$. Let now $J$ be a
composition which is finer than $I$. Let then $J = (J_1,\dots,J_m)$ be the
unique decomposition of $J$ into compositions $(J_i)_{i=1,m}$ such that
$|J_p|=i_p$, $p=1,\ldots,m$. We now define
$$
\pi_u(J,I) = \ \prod_{i=1}^m \ \pi_u(J_i) \ .
$$
Similarly, we set
$$
lp(J,I) = \ \prod_{i=1}^m \ lp(J_i) \ ,
$$
which is just the product of the last parts of all compositions $J_i$.

\begin{proposition} \label{MATRSPSI}
For every composition $I$, we have
$$
S^I = \sum_{J \succeq I} \ {1 \over \pi_u(J,I)}\, \Psi^J \ ,
$$
$$
\Psi^I = \sum_{J \succeq I} \ (-1)^{\ell(J)-\ell(I)}\ lp(J,I) \ S^J \ .
$$
\end{proposition}

\Proof The two formulas of this proposition are consequences
of the quasi-determinantal relations given in Corollary \ref{DETFORM}.
Let us establish the first one. According to relation (\ref{QDSPSI}) and
to basic properties of quasi-determinants, we can write
$$
n\, S_n =
\left|\matrix{
\Psi_1  & \Psi_2 & \ldots & \Psi_{n-1}          &\bo{\Psi_n}\vtr{3} \cr
-1      & \Psi_1 & \ldots & \Psi_{n-2}          &\Psi_{n-1} \vtr{3} \cr
0       & -1     & \ldots & \displaystyle {\Psi_{n-3}\over 2} &
\displaystyle {\Psi_{n-2}\over 2} \cr
\vdots  & \vdots & \ddots & \vdots              &\vdots  \vtr{3}     \cr
0       & 0      & \ldots & -1                  &
\displaystyle {\Psi_1\over n+1} \cr
}\right|\ .
$$
But this quasi-determinant can be explicitely expanded by means of
Proposition \ref{SSTRI}
\begin{equation}\label{SnPsi}
S_n = \ \sum_{|J|=n} \ \displaystyle {1\over \pi_u(J)} \, \Psi^J \ .
\end{equation}
Taking the product of these identities
for $n=i_1,i_2,\ldots, i_m$, one obtains the first formula
of Proposition \ref{MATRSPSI}. The second one is established
in the same way, starting from relation (\ref{QDPSIS}) which
expresses $\Psi_n$ as a quasi-determinant in the $S_i$. \cqfd

\begin{note}
{\rm One can also prove (\ref{SnPsi}) by solving the differential
equation $\sigma'(t)= \sigma(t)\, \psi(t)$ in terms of iterated
integrals. This yields}
\begin{equation}\label{INTIT}
\sigma(t)= 1 + \int_0^t \! dt_1\, \psi(t_1) +
\int_0^t \! dt_1 \int_0^{t_1} \! dt_2 \, \psi(t_2)\psi(t_1) +
\int_0^t \! dt_1 \int_0^{t_1} \! dt_2 \int_0^{t_2} \! dt_3\,
\psi(t_3)\psi(t_2)\psi(t_1) + \cdots
\end{equation}
{\rm and one obtains relation (\ref{SnPsi}) by equating the coefficients of
$t^n$ in both sides of (\ref{INTIT}).}
\end{note}

\bigskip
The matrices $M(\Psi,S)_n$ have a simple block structure.

\begin{proposition} \label{BLOCPSIS}
For every $n \geq 0$, we have
$$
M(\Psi,S)_n = \left(
\matrix{
M(\Psi,S)_{n-1} + A_{n-1} & -M(\Psi,S)_{n-1} \cr
0 & M(\Psi,S)_{n-1}
}\right)
$$
where $A_n$ is the matrix of size $2^{n-1}$ defined by
$$
A_n = \left(
\matrix{
I_{2^{n-2}} & 0 \cr
0 & M(\Psi,S)_{n-1}
}\right) \ ,
$$
$M(\Psi,S)_0$ denoting the empty matrix.
\end{proposition}

\Proof This follows from Proposition \ref{MATRSPSI} and Lemma \ref{LEX}. \cqfd

\begin{example}
{\rm For $n = 2$ and $n = 3$, we have}
$$
M(S,\Psi)_2 =
\matrix{
& 2 \ \ \ \ 11 \vt \cr
\matrix{ 2 \cr 11 } \!\!\!\!\! &
\left(\matrix{
1/2 & 1/2 \cr
0   & 1 \cr
}\right) \vt
}
\ , \quad
M(\Psi,S)_2 =
\matrix{
& 2 \ \ \ 11 \vt \cr
\matrix{ 2 \cr 11 } \!\!\!\!\! &
\left(\matrix{
2  & -1 \cr
0  & 1 \cr
}\right) \vt
}
\ ,
$$
\medskip
$$
M(S,\Psi)_3 =
\matrix{
& 3 \ \ \ \ 21 \ \ \ \ 12 \ \ \ 111 \vt \cr
\matrix{ 3 \cr 21 \cr 12 \cr 111} \!\!\!\!\! &
\left(\matrix{
1/3 & 1/6 & 1/3 & 1/6 \cr
0   & 1/2 & 0   & 1/2 \cr
0   & 0   & 1/2 & 1/2 \cr
0   & 0   & 0   & 1
}\right)
}\ , \quad
M(\Psi,S)_3 =
\matrix{
& 3 \ \ \ 21 \ \ \ \ 12 \ \ \ 111 \vt \cr
\matrix{ 3 \cr 21 \cr 12 \cr 111} \!\!\!\!\! &
\left(\matrix{
3 & -1  & -2  & 1 \cr
0 & 2   & 0   & -1 \cr
0 & 0   & 2   & -1 \cr
0 & 0   & 0   & 1
}\right)
}\ .
$$
\end{example}


\subsection{$S$ and $\Phi$}

We first introduce some notations. Let $I=(i_1,\ldots,i_m)$ be a composition
and let $J$ be a finer composition. As in the preceding section, decompose
$J = (J_1,\dots,J_m)$ into compositions with respect to the refinement
relation $J\succeq I$, {\it i.e.} such that $|J_p| = i_p$ for every $p$.
Then, we set
$$
\ell(J,I) = \ \prod_{i=1}^m \ \ell(J_i) \ ,
$$
which is just the product of the lengths of the different compositions $J_i$.
We also denote by $\pi(I)$ the product of all parts of $I$. We also set
$$
sp(I) = \ell(I)!\ \pi(I) = m! \ i_1\, \dots \, i_m \ .
$$
and
$$
sp(J,I) = \ \prod_{i=1}^m \ sp(J_i) \ .
$$

The following proposition can be found in \cite{GaR} and in \cite{Re}.

\begin{proposition}\label{SPh}
For every composition $I$, we have
$$
\Phi^I = \ \sum_{J \succeq I} \ (-1)^{\ell(J)-\ell(I)} \,
{\pi(I) \over \ell(J,I)} \, S^J \ ,
$$
$$
S^I = \ \sum_{J \succeq I} \ {1 \over sp(J,I)} \, \Phi^J \ .
$$
\end{proposition}

\Proof Using the series expansion  of the logarithm
and the defining relation (\ref{DEFPHI}) yields
$$
\Phi_n = \ \sum_{|J|=n} \ (-1)^{\ell(J)-1} \ \displaystyle {n\over \ell(J)} \
S^J \ ,
$$
from which the first part of the proposition follows.
On the other hand,  the series expansion of the exponential
and the defining relation (\ref{DEFPHII}) gives
$$
S_n = \ \sum_{|J|=n} \ \displaystyle {1\over sp(J)} \ \Phi^J \ ,
$$
which  implies the second equality.  \cqfd

\smallskip
\begin{note}
{\rm It follows from  Proposition \ref{SPh} that the coefficients of
the expansion of $S_n$ on the basis $\Phi^I$ only depend on the
partition $\sigma(I) = (1^{m_1} \, 2^{m_2} \, \dots \, n^{m_n})$
associated with $I$. In fact, one has}
$$
sp(I) = \left( \matrix{ m_1 + \dots + m_n \cr m_1,\ \dots\ ,\, m_n } \right)
\ 1^{m_1} \, \dots \, n^{m_n} \ m_1 ! \ \dots \ m_n ! \ ,
$$
{\rm which shows that $1/sp(I)$ is  equal to the usual commutative
scalar product $(S_n\, ,\, \psi_{\sigma(I)})$ divided by the number of
compositions with associated partition $\sigma(I)$.}
\end{note}

\smallskip
\begin{example}
{\rm For $n = 2$ and $n = 3$, we have}
$$
M(S,\Phi)_2 =
\matrix{
& 2 \ \, \ \ \ 11 \vt \cr
\matrix{ 2 \cr 11 } \!\!\!\!\! &
\left(\matrix{
1/2 & 1/2 \cr
0   & 1 \cr
}\right) \vt
}
\ , \quad
M(\Phi,S)_2 =
\matrix{
& 2 \ \ \ 11 \vt \cr
\matrix{ 2 \cr 11 } \!\!\!\!\! &
\left(\matrix{
2  & -1 \cr
0  & 1 \cr
}\right) \vt
}
\ ,
$$
\medskip
$$
M(S,\Phi)_3 = \!\!
\matrix{
& 3 \ \ \ \ \ 21 \ \ \ \ 12 \ \ \ \ 111 \vt \cr
\matrix{ 3 \cr 21 \cr 12 \cr 111} \!\!\!\!\!\! &
\left(\matrix{
1/3 & 1/4 & 1/4 & 1/6 \cr
0   & 1/2 & 0   & 1/2 \cr
0   & 0   & 1/2 & 1/2 \cr
0   & 0   & 0   & 1
}\right)
}\ , \ \
M(\Phi,S)_3 = \!\!
\matrix{
& 3\ \ \ \ \ 21\ \ \ \ \ 12\ \ \ \ \ 111 \vt \cr
\matrix{ 3 \cr 21 \cr 12 \cr 111} \!\!\!\!\!\! &
\left(\matrix{
3 & -3/2 & -3/2 & 1 \cr
0 & 2    & 0    & -1 \cr
0 & 0    & 2    & -1 \cr
0 & 0    & 0    & 1
}\right)
}\ .
$$
\end{example}

\begin{note}{\rm
One can produce combinatorial identities by taking the commutative
images of the relations between the various bases of noncommutative
symmetric functions. For example, taking into account the fact that $\Psi$
and $\Phi$ reduce in the commutative case to the same functions, one deduces
from the description of the matrices $M(S,\Psi)$ and $M(S,\Phi)$ that
$$
\sum_{\sigma(J) = I} \ \displaystyle{1 \over \pi_u(J)} =
\displaystyle {1\over \pi(I)} \ ,
$$
the sum in the left-hand side being taken over all compositions
corresponding to the same partition $I$.}
\end{note}


\subsection{$S$ and $R$}

The matrices $M(S,R)_n$ and $M(R,S)_n$ are given by Kronecker powers
of $2\times 2$ matrices.

\begin{proposition} \label{MATRANSR}
For every $n \geq 1$, we have
\begin{equation}
M(S,R)_n =
\left(\matrix{
1 & 0 \cr
1 & 1 \cr
}\right)^{\otimes (n-1)} \ ,
\end{equation}
\begin{equation}
M(R,S)_n =
\left(\matrix{
1 & 0 \cr
-1 & 1 \cr
}\right)^{\otimes (n-1)} \ .
\end{equation}
\end{proposition}

\Proof It is sufficient to establish the second relation,
since one has
$$
\left(\matrix{
1 & 0 \cr
1 & 1 \cr
}\right)^{-1}
=
\left(\matrix{
1 & 0 \cr
-1 & 1 \cr
}\right) \ .
$$
Going back to the definition of a ribbon Schur function and using the
same technique as in the proof of Proposition \ref{MATRSPSI}, one
arrives at
$$
R_I = \ \sum_{I\succeq J} \ (-1)^{\ell(J)-\ell(I)} \ S^J \ ,
$$
for every composition $I$. The conclusion follows again from
Lemma \ref{LEX}.  \cqfd

\begin{example}
{\rm For $n = 2$ and $n = 3$, we have}
$$
M(S,R)_2 =
\matrix{
& 2 \ \ \ 11 \vt \cr
\matrix{ 2 \cr 11 } \!\!\!\!\! &
\left(\matrix{
1 & 0 \cr
1 & 1 \cr
}\right) \vt
}
\ , \quad
M(R,S)_2 =
\matrix{
& \ 2 \ \ \ 11 \vt \cr
\matrix{ 2 \cr 11 } \!\!\!\!\! &
\left(\matrix{
1 & 0 \cr
-1 & 1 \cr
}\right) \vt
}
\ ,
$$
\smallskip
$$
M(S,R)_3 =
\matrix{
& 3 \ \, 21 \, \ 12 \, \ 111 \vt \cr
\matrix{ 3 \cr 21 \cr 12 \cr 111} \!\!\!\!\! &
\left(\matrix{
1  & 0  & 0  & 0 \cr
1  & 1  & 0  & 0 \cr
1  & 0  & 1  & 0 \cr
1  & 1  & 1  & 1
}\right)
}\ , \quad
M(R,S)_3 =
\matrix{
& \ 3 \ \ \ \ 21 \ \ \ 12 \ \ \ 111 \vt \cr
\matrix{ 3 \cr 21 \cr 12 \cr 111} \!\!\!\!\! &
\left(\matrix{
1   & 0   & 0 & 0 \cr
-1  & 1   & 0  & 0 \cr
-1  & 0   & 1  & 0 \cr
1   & -1  & -1 & 1
}\right)
}\ .
$$
\end{example}

\medskip
It follows from the proof of Proposition \ref{MATRANSR} that
\begin{equation}\label{R2S}
S^I = \sum_{I\succeq J} \ R_J \ , \quad
R_I = \sum_{I\succeq J} \ (-1)^{\ell(I)-\ell(J)} \, S^J \ ,
\end{equation}
for every composition $I$. This is the noncommutative analog of a
formula of MacMahon (\cf \cite{MM}). These formulas
are equivalent to the well-known fact
that the M\"obius function of the order $\preceq$ on compositions is equal to
$\mu_{\preceq}(I,J) = (-1)^{\ell(I)-\ell(J)}$ if $I \succeq J$ and
to $\mu_{\preceq}(I,J) = 0$ if $I \prec J$.


\subsection{$\Lambda$ and $\Psi$}

The matrices relating the two bases $\Lambda$ and $\Psi$ can be described
similarly to the matrices relating $S$ and $\Psi$ with the only difference
that the combinatorial descriptions reverse left and right.

\begin{proposition}
For every composition $I$, we have
$$
\Lambda^I = \sum_{J \succeq I} \ (-1)^{|I|-\ell(J)} \,
{1 \over \pi_u(\overline{J},\overline{I})}\, \Psi^J \ ,
$$
$$
\Psi^I = (-1)^{|I|}\sum_{J \succeq I} \ (-1)^{\ell(J)} \,
lp(\overline{J},\overline{I}) \, \Lambda^J \ .
$$
\end{proposition}

\Proof It suffices to apply $\omega$ to the formulas given by Proposition
\ref{MATRSPSI}. \cqfd

The block structure of $M(\Psi,\Lambda)_n$ is also simple.

\begin{proposition}
For  $n \geq 2$,
$$
M(\Psi,\Lambda)_n = \left(
\matrix{
-A_{n-1} - M(\Psi,\Lambda)_{n-1} & A_{n-1} \cr
0 & M(\Psi,\Lambda)_{n-1}
}\right)
$$
where $A_n$ is the matrix of size $2^{n-1}$ defined by
$$
A_n = \left(
\matrix{
A_{n-1} & A_{n-1} \cr
0 & M(\Psi,\Lambda)_{n-1}
}\right) \ ,
$$
with $A_1=(1)$ and $M_1=(1)$.
\end{proposition}

\Proof This follows from Propositions \ref{BLOCPSIS} and \ref{MATRANLS}
using the fact that
$$
M(\Psi,\Lambda)_n = M(\Psi,S)_n \, M(S,\Lambda)_n \ .
$$
\cqfd

\begin{example}
{\rm For $n = 2$ and $n = 3$, one has}
$$
M(\Psi,\Lambda)_2 =
\matrix{
& 2 \ \ \ 11 \vt \cr
\matrix{ 2 \cr 11 } \!\!\!\!\! &
\left(\matrix{
-2 & 1 \cr
0  & 1 \cr
}\right) \vt
}
\ , \quad
M(\Lambda,\Psi)_2 =
\matrix{
& 2 \ \ \ \ \ \ 11 \vt \cr
\matrix{ 2 \cr 11 } \!\!\!\!\! &
\left(\matrix{
-1/2 & 1/2 \cr
0    & 1 \cr
}\right) \vt
}
\ ,
$$
\medskip
$$
M(\Psi,\Lambda)_3 = \!\!
\matrix{
& 3 \ \ \ 21 \ \ \ 12 \ \ \ 111 \vt \cr
\matrix{ 3 \cr 21 \cr 12 \cr 111} \!\!\!\!\!\! &
\left(\matrix{
3 & -2 & -1 & 1 \cr
0 & -2 & 0  & 1 \cr
0 & 0  & -2 & 1 \cr
0 & 0  & 0  & 1
}\right)
}\ , \ \
M(\Lambda,\Psi)_3 = \!\!
\matrix{
& 3 \ \ \ \ \ \ 21 \ \ \ \ \ 12 \ \ \ \ \ 111 \vt \cr
\matrix{ 3 \cr 21 \cr 12 \cr 111} \!\!\!\!\!\! &
\left(\matrix{
1/3 & -1/3  & -1/6  & 1/6 \cr
0   & -1/2  & 0     & 1/2 \cr
0   & 0     & -1/2  & 1/2 \cr
0   & 0     & 0     & 1
}\right)
}\ .
$$
\end{example}


\subsection{$\Lambda$ and $\Phi$}

The matrices relating $\Lambda$ and $\Phi$ are again, due to the action
of $\omega$, essentially the same as the matrices relating $S$ and
$\Phi$.

\begin{proposition}\label{LAMPHI}
For every composition $I$, we have
$$
\Phi^I = \ \sum_{J \succeq I} \ (-1)^{|I|-\ell(J)} \,
{\pi(I) \over \ell(J,I)} \, \Lambda^J \ ,
$$
$$
\Lambda^I = \ \sum_{J \succeq I} \ (-1)^{|I|-\ell(J)} \,
{1 \over sp(J,I)} \, \Phi^J \ .
$$
\end{proposition}

\Proof It suffices again to apply $\omega$ to the relations given by
Proposition \ref{SPh}. \cqfd

\begin{example}
{\rm For $n = 2$ and $n = 3$, we have}
$$
M(\Lambda,\Phi)_2 =
\matrix{
& 2 \ \ \ \ \ 11 \vt \cr
\matrix{ 2 \cr 11 } \!\!\!\!\! &
\left(\matrix{
-1/2 & 1/2 \cr
0    & 1 \cr
}\right) \vt
}
\ , \quad
M(\Phi,\Lambda)_2 =
\matrix{
& 2 \ \ \ 11 \vt \cr
\matrix{ 2 \cr 11 } \!\!\!\!\! &
\left(\matrix{
-2 & 1 \cr
0  & 1 \cr
}\right) \vt
}
\ ,
$$
\medskip
$$
M(\Phi,\Lambda)_3 = \!\!\!
\matrix{
& 3 \ \ \ \ 21 \ \ \ \ 12 \ \ \ \ 111 \vt \cr
\matrix{ 3 \cr 21 \cr 12 \cr 111} \!\!\!\!\!\! &
\left(\matrix{
3 & -3/2 & -3/2 & 1 \cr
0 & -2   & 0    & 1 \cr
0 & 0    & -2   & 1 \cr
0 & 0    & 0    & 1
}\right)
}\, , \
M(\Lambda,\Phi)_3 = \!\!\!
\matrix{
& 3\ \ \ \ \ 21\ \ \ \ \ 12\ \ \ \ \ 111 \vt \cr
\matrix{ 3 \cr 21 \cr 12 \cr 111} \!\!\!\!\!\! &
\left(\matrix{
1/3 & -1/4  & -1/4  & 1/6 \cr
0   & -1/2  & 0     & 1/2 \cr
0   & 0     & -1/2  & 1/2 \cr
0   & 0     & 0     & 1
}\right)
}\, .
$$
\end{example}


\subsection{$\Lambda$ and $R$}

The matrices $M(\Lambda,R)$ and $M(R,\Lambda)$ are again Kronecker
powers of simple $2\times 2$ matrices.

\begin{proposition}\label{MATRANSLR}
For every $n \geq 1$, we have
$$
M(\Lambda,R)_n =
\left(
\matrix{
0 & 1 \cr
1 & 1 }
\right) ^{\otimes (n-1)} \ , \quad
M(R,\Lambda)_n =
\left(
\matrix{
-1 & 1 \cr
1 & 0 }
\right) ^{\otimes (n-1)} \ .
$$
\end{proposition}

\Proof This follows from Propositions \ref{MATRANSR} and \ref{MATRANLS},
since
$$
M(\Lambda,R)_n = M(\Lambda,S)_n \, M(\S,R)_n \quad \hbox{and} \quad
M(R,\Lambda)_n = M(R,S)_n \, M(S,\Lambda)_n \ .
$$
Another possibility is to apply $\omega$ to the formulas (\ref{R2S}).  \cqfd

\begin{note}
{\rm Proposition \ref{MATRANSLR} is equivalent to the following formulas
\begin{equation}\label{R2LA}
\Lambda^I = \sum_{\overline{I}\succeq J^{\sim}} \ R_J \ , \quad
R_I = \sum_{I^{\sim}\succeq \overline{J}} \ (-1)^{\ell(I^{\sim})-\ell(J)}
\, \Lambda^J \ .
\end{equation}
}
\end{note}

\begin{example}
{\rm For $n = 2$ and $n = 3$, we have}
$$
M(\Lambda,R)_2 =
\matrix{
& 2 \ \ 11 \vt \cr
\matrix{ 2 \cr 11 } \!\!\!\!\! &
\left(\matrix{
0 & 1 \cr
1 & 1 \cr
}\right) \vt
}
\ , \quad
M(R,\Lambda)_2 =
\matrix{
& 2 \ \ \ 11 \vt \cr
\matrix{ 2 \cr 11 } \!\!\!\!\! &
\left(\matrix{
-1 & 1 \cr
1  & 0 \cr
}\right) \vt
}
\ ,
$$
\medskip
$$
M(\Lambda,R)_3 = \!\!
\matrix{
& 3 \ \ 21 \ \ 12 \ \ 111 \vt \cr
\matrix{ 3 \cr 21 \cr 12 \cr 111} \!\!\!\!\!\! &
\left(\matrix{
0 & 0 & 0 & 1 \cr
0 & 0 & 1 & 1 \cr
0 & 1 & 0 & 1 \cr
1 & 1 & 1 & 1
}\right)
}\ , \quad
M(R,\Lambda)_3 = \!\!
\matrix{
& 3 \ \ \ \ 21 \ \ \ \ 12 \ \ \ 111 \vt \cr
\matrix{ 3 \cr 21 \cr 12 \cr 111} \!\!\!\!\!\! &
\left(\matrix{
1   & -1  & -1 & 1 \cr
-1  & 0   & 1  & 0 \cr
-1  & 1   & 0  & 0 \cr
1   & 0   & 0  & 0
}\right)
}\ .
$$
\end{example}


\subsection{$\Psi$ and $R$}

Let $I = (i_1,\dots,i_r)$
and $J = (j_1,\dots,j_s)$ be two compositions of $n$. The ribbon
decomposition of $I$ relatively to $J$ is the unique decomposition
of $I$ as
\begin{equation} \label{DECOMPIJ}
I = I_1 \bullet I_2 \bullet \, \dots\, \bullet I_s \ ,
\end{equation}
where $I_i$ denotes a composition of length $j_i$ for every $i$ and where
$\bullet$ stands for $\cdot$ or $\triangleright$ as defined in the statement
of Proposition \ref{MR}.
For example,
if $I = (3,1,1,3,1)$ and $J = (4,3,2)$, we have
$$
I = (31) \cdot (12) \triangleright (11) \ ,
$$
which can be read on the ribbon diagram associated with $I$:

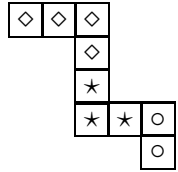
\centerline{\setlength{\unitlength}{0.25pt}
\begin{picture}(250,300)
\put(0,200){\framebox(50,50){$\diamond$}}
\put(50,200){\framebox(50,50){$\diamond$}}
\put(100,200){\framebox(50,50){$\diamond$}}
\put(100,150){\framebox(50,50){$\diamond$}}
\put(100,100){\framebox(50,50){$\star$}}
\put(100,50){\framebox(50,50){$\star$}}
\put(150,50){\framebox(50,50){$\star$}}
\put(200,50){\framebox(50,50){$\circ$}}
\put(200,0){\framebox(50,50){$\circ$}}
\end{picture}
}

\bigskip
For a pair $I,J$ of compositions of the same integer $n$, let
$(I_1,\dots,I_s)$ be the ribbon decomposition of $I$ relatively to $J$
as given by (\ref{DECOMPIJ}). We define then $psr(I,J)$ by
$$
psr(I,J) = \left\{ \
\matrix{
(-1)^{l(I_1)+\dots+\l(I_s)-s} & \ \hbox{if every} \ I_j \ \hbox{is a hook}
\hfill \hbox{ } \cr
0 & \ \hbox{otherwise} \hfill \hbox{ }
}
\right. \ \ .
$$
We can now give the expression of $\Psi^I$ in the basis of ribbon Schur
functions.

\begin{proposition} \label{PsiRib}
For any composition $I$,
\begin{equation}
\Psi^I = \ \sum_{|J| = n} \ psr(J,I) \, R_J \ .
\end{equation}
\end{proposition}

\Proof This follows from Corollary \ref{psihook}
and Proposition \ref{MR}. \cqfd

The block structure of the matrix $M(\Psi,R)_n$ can also be described.

\begin{proposition}
$$
M(\Psi,R)_n = \left(
\matrix{
A_{n-1} & - M(\Psi,R)_{n-1} \cr
M(\Psi,R)_{n-1} & M(\Psi,R)_{n-1}
}\right) \ ,
$$
where $A_n$ is a matrix of order $2^{n-1}$, itself  given  by the block
decomposition
$$
A_n = \left(
\matrix{
A_{n-1} & 0 \cr
M(\Psi,R)_{n-1} & M(\Psi,R)_{n-1}
}
\right) \ .
$$
\end{proposition}

\Proof The result follows from Proposition \ref{PsiRib} and from Lemma
\ref{LEX}. \cqfd

The structure of the matrix $M(R,\Psi)_n$ is
more complicated. To describe it, we introduce some notations. For a
vector $v = (v_1,\ldots,v_n)\in\Q^n$, we denote by $v[i,j]$ the vector
$(v_i,v_{i+1},\ldots,v_j)\in\Q^{j-i}$. We also denote by $\overline{v}$
the vector obtained by reading the entries of $v$ from right to left and
by $v\, .\, w$ the vector obtained by concatenating the entries of $v$
and $w$.

\begin{proposition} \label{RPsi}
1) One can write $M(R,\Psi)_n$ in the following way
$$
M(R,\Psi)_n
=
{1 \over n!} \
\left( \
\matrix{
A(R,\Psi)_n   & A(R,\Psi)_n \cr
-A(R,\Psi)_n  & B(R,\Psi)_n
}
\ \right)
$$
where $A(R,\Psi)_n$ and $B(R,\Psi)_n$ are matrices of order $2^{n-2}$ whose
all entries are integers and that satisfy to the relation
$$
A(R,\Psi)_n + B(R,\Psi)_n = n \, M(R,\Psi)_{n-1} \ .
$$
Hence $M(R,\Psi)_n$ is in particular completely determined
by the structure of $B(R,\Psi)_n$.

\smallskip
2) The matrix $B(R,\Phi)_{n}$ can be block decomposed as follows
\smallskip
\begin{equation} \label{BlocRPsi}
B(R,\Phi)_{n} = \left( \
\matrix{
-B_n^{(0)} &
\matrix{
-B_{n}^{(00)} & \matrix{ -B_n^{(000)} & \dots \vtt\cr B_n^{(000)}
& \dots \vtt } \cr
B_{n}^{(00)} & \matrix{ -B_n^{(001)} & \dots \vtt\cr B_n^{(001)}
& \dots \vtt }
} \vttt \cr
B_n^{(0)} &
\matrix{
- B_{n}^{(01)} & \matrix{ -B_n^{(010)} & \dots \vtt \cr B_n^{(010)}
& \dots \vtt }\cr
 B_{n}^{(01)} & \matrix{ -B_n^{(011)} & \dots \vtt \cr B_n^{(011)}
& \dots \vtt }
}}
\ \ \ \right)
\end{equation}
\medskip
where $(B_n^{(i_1,\dots,i_r)})_{r=1,\dots,n-1}$ are square blocks of order
$2^{n-1-r}$.

\smallskip
3) Every block $B^{(i_1,\dots,i_r)}_n$ has itself a block structure
of the form given by (\ref{BlocRPsi}) with blocks denoted here
$B_{(i_1,\dots,i_n)}^{(j_1,\dots,j_s)}$. These blocks statisfy the
two following properties.

\medskip
\hspace{5mm}
-- Every block $B_{(i_1,\dots,i_n)}^{(j_1,\dots,j_s)}$ of order
$2\times 2$ has the structure
$$
B_{(i_1,\dots,i_n)}^{(j_1,\dots,j_s)}
=
\left( \
\matrix{
p  & q \cr
-p & p
}
\ \right) \ ,
$$
for some integers $p,q \in \Z$.

\smallskip
\hspace{5mm}
-- For every $(i_1,\dots,i_n)$ and $(j_1,\dots,j_s)$, let us consider the
rectangular matrix $C_{(i_1,\dots,i_r)}^{(j_1,\dots,j_s)}$ defined as
follows
$$
C_{(i_1,\dots,i_r)}^{(j_1,\dots,j_s)}
=
\left( \
\matrix{
B_{(i_1,\dots,i_n)}^{(j_1,\dots,j_s)}
&
\matrix{
- B_{(i_1,\dots,i_n)}^{(j_1,\dots,j_s,0)} & \dots \vtt \cr
B_{(i_1,\dots,i_n)}^{(j_1,\dots,j_s,0)}   & \dots \vtt
}
}
\ \right) \ .
$$
If we denote by $LC(M)$ the last column of a matrix $M$, we then have
$$
LC(B_{(i_1,\dots,i_r)}^{(j_1,\dots,j_s)})
=
(\, LC(-B_{(i_1,\dots,i_r)}^{(j_1,\dots,j_s,0)}) \cdot
LC(B_{(i_1,\dots,i_r)}^{(j_1,\dots,j_s,0)}) \, )
\, - \,
LC(C_{(i_1,\dots,i_r)}^{(j_1,\dots,j_s)}) \ .
$$

\medskip
Note that these recursive properties allow to recover all the block structure
of $B(R,\Psi)_n$ from the last column of this matrix.

\smallskip
4) Let $V_n$ be the vector of order $2^{n-2}$ corresponding to the last
column of $B(R,\Psi)_n$ read from {\rm bottom to top}. The vector $V_n$ is
then determined by the recursive relations
$$
V_2 = ( \, 1\, ) \ , \ \
$$
$$
V_n [1] = 1 \ , \ \ V_n [2] = n-1 \ ,
$$
$$
V_n [1,  2^k] + V_n [2^{k}+1,  2^{k+1}]
=
\left( \, \matrix{ n \cr k+1 } \, \right) \
V_{k} [1,  2^{k-1}] \, . \, \overline{V_{k} [1,  2^{k-1}]}
\ ,
$$
for $k\in\{1,\dots,n-4\}$, and
$$
V_n [1,  2^{n-3}] +
\overline{V_n [2^{n-3}+1,  2^{n-2}]}
= n \ V_{n-1} \ .
$$
\end{proposition}

\begin{example}
{\rm Here are the first vectors $V_n$ from which the matrix
$M(R,\Psi)_n$ can be recovered: }
$$
V_2 = (\, 1\, ) \ , \ V_3 = (\, 1 \ \, 2 \, ) \ , \
V_4 = (\, 1 \ \, 3 \ \, 5 \ \, 3\, ) \ , \
V_5 = (\, 1 \ \, 4 \ \, 9 \ \, 6 \ \, 9 \ \, 16 \ \, 11 \ \, 4 \, ) \ ,
$$
$$
V_6 = (\, 1 \ \, 5 \ \, 14 \ \, 10 \ \, 19 \ \, 35 \ \, 26 \ \, 10 \ \,
14 \ \, 40 \ \, 61 \ \, 35 \ \, 26 \ \, 40 \ \, 19 \ \, 5 \, ) \ .
$$
{\rm For $n=2,3,4$ the  matrices relating $\Psi^I$
and $R_I$.}
$$
M(\Psi,R)_2 =
\matrix{
& 2 \ \ \ 11 \vt \cr
\matrix{ 2 \cr 11 } \!\!\!\!\! &
\left(\matrix{
1 & -1 \cr
1 & 1 \cr
}\right) \vt
}
\ , \quad
M(R,\Psi)_2 =
\matrix{
& 2 \ \ \ \ \ 11 \vt \cr
\matrix{ 2 \cr 11 } \!\!\!\!\! &
\left(\matrix{
1/2 & 1/2 \cr
-1/2 & 1/2 \cr
}\right) \vt
}
\ ,
$$
\medskip
$$
M(\Psi,R)_3 = \!\!\!
\matrix{
& 3 \ \ \ 21 \ \ \ 12 \ \ \ 111 \vt \cr
\matrix{ 3 \cr 21 \cr 12 \cr 111} \!\!\!\!\!\! &
\left(\matrix{
1 & 0  & -1 & 1 \cr
1 & 1  & -1 & -1 \cr
1 & -1 & 1  & -1 \cr
1 & 1  & 1  & 1
}\right)
}\, , \ \,
M(R,\Psi)_3 = \!\!\!
\matrix{
& 3\ \ \ \ \ \ 21 \ \ \ \ \ \ 12 \ \ \ \ \ \ 111 \vt \cr
\matrix{ 3 \cr 21 \cr 12 \cr 111} \!\!\!\!\!\! &
\left(\matrix{
1/3  & 1/6  & 1/3  & 1/6 \cr
-1/3 & 1/3  & -1/3 & 1/3 \cr
-1/3 & -1/6 & 1/6  & 1/3 \cr
1/3  & -1/3 & -1/6 & 1/6
}\right)
}\ ,
$$
\smallskip
$$
M(\Psi,R)_4 =
\matrix{
& 4\ \ \ 31 \ \ \ 22\ \ \ 211
\ \ \ 13 \ \ \ 121\ \ \ 112\ \ \ 1111 \vt \cr
\matrix{\ 4 \cr 31 \cr 22 \cr 211 \cr 13 \cr 121 \cr 112 \cr 1111}
\!\!\!\!\!\! &
\left( \
\matrix{
1 & 0    & 0    & 0    & -1   & 0    & 1    & -1 \cr
1 & 1    & 0    & 0    & -1   & -1   & 1    & 1 \cr
1 & -1   & 1    & -1   & -1   & 1    & -1   & 1 \cr
1 & 1    & 1    & 1    & -1   & -1   & -1   & -1 \cr
1 & 0    & -1   & 1    & 1    & 0    & -1   & 1 \cr
1 & 1    & -1   & -1   & 1    & 1    & -1   & -1 \cr
1 & -1   & 1    & -1   & 1    & -1   & 1    & -1 \cr
1 &  1   & 1    &  1   & 1    &  1   & 1    & 1 \cr
}
\ \right)
} \ ,
$$
\smallskip
$$
M(R,\Psi)_4 =
\matrix{
& 4\ \ \ \ \ \ \ 31 \ \ \ \ \ \ \ 22\ \ \ \ \ \ \ 211
\ \ \ \ \ \ \ 13 \ \ \ \ \ 121\ \ \ \ \ 112\ \ \ \ \ 1111 \vt \cr
\matrix{ 4 \cr 31 \cr 22 \cr 211 \cr 13 \cr 121 \cr 112 \cr 1111}
\!\!\!\!\!\! &
\left(\
\matrix{
1/4  \! & 1/12  \! & 1/8  &\! 1/24  & 1/4   &\! 1/12  & 1/8   & 1/24 \cr
-1/4 \! & 1/4   \! & -1/8 &\! 1/8   & -1/4  &\! 1/4   & -1/8  & 1/8 \cr
-1/4 \! & -1/12 \! & 1/8  &\! 5/24  & -1/4  &\! -1/12 & 1/8   & 5/24 \cr
1/4  \! & -1/4  \! & -1/8 &\! 1/8   & 1/4   &\! -1/4  & -1/8  & 1/8 \cr
-1/4 \! & -1/12 \! & -1/8 &\! -1/24 & 1/12  &\! 1/12  & 5/24  & 1/8 \cr
1/4  \! & -1/4  \! & 1/8  &\! -1/8  & -1/12 &\! 1/12  & -5/24 & 5/24 \cr
1/4  \! & 1/12  \! & -1/8 &\! -5/24 & -1/12 &\! -1/12 & 1/24  & 1/8 \cr
-1/4 \! & 1/4   \! & 1/8  &\! -1/8  & 1/12  &\! -1/12 & -1/24 & 1/24 \cr
}
\, \right)
} \, .
$$
\end{example}


\subsection{$\Phi$ and $R$}

The first row of $M(\Phi,R)_n$ is given by the following formula, which
is equivalent to Corollary 3.16 of \cite{Re} p. 42.

\begin{proposition} \label{PHINR}
The expansion of $\Phi_n$ in the basis $(R_I)$ is given by
$$
\Phi_n =
\ \sum_{|I|=n}\,
{ (-1)^{\ell(I)-1} \over {n-1\choose \ell(I)-1} } \ R_I \ .
$$
\end{proposition}

Let $I,J$ be two compositions of the same integer $n$ and let
$I = (I_1,\dots,I_s)$ be the ribbon decomposition of $I$ relatively to
$J$ as given by relation (\ref{DECOMPIJ}). Define $phr(I,J)$
by setting
$$
phr(I,J) = \ \prod_{i=1}^s \
{ (-1)^{\ell(I_i)-1} \over {|I_i|-1\choose \ell(I_i)-1} } \ .
$$
We can now give the expression of $\Phi^I$ in the basis of ribbon Schur
functions.

\begin{corollary}
For every composition $I$, one has
\begin{equation}
\Phi^I = \ \sum_{|J|=n} \ phr(J,I) \ R_J \ .
\end{equation}
\end{corollary}

\Proof This is a simple consequence of Propositions \ref{PHINR}
and \ref{MR}. \cqfd

On the other hand, the structure of the matrix $M(R,\Phi)_n$ is more
intricate.

\begin{proposition} \label{RPh}
1) Let $D(S_n,\Phi)$ be the diagonal matrix constructed with the elements of
the first row of $M(R,\Phi)_n$, i.e.  the diagonal matrix whose
entry of index $(I,I)$ is $1/sp(I)$ according to Proposition \ref{SPh}.
Then we can write in a unique way
$$
M(R,\Phi)_n = N(R,\Phi)_n \ D(S_n,\Phi)
$$
where $N(R,\Phi)_n$ is a matrix of order $2^{n-1}$ whose all
entries are integers.

\smallskip
2) The matrix $N(R,\Phi)_{n}$ can be block decomposed as follows
\smallskip
$$
N(R,\Phi)_{n} = \left( \
\matrix{
A_n^{(0)} &
\matrix{
A_{n}^{(00)} & \matrix{ A_n^{(000)} & \dots \vtt\cr -A_n^{(000)}
& \dots \vtt } \cr
-A_{n}^{(00)} & \matrix{ A_n^{(001)} & \dots \vtt\cr -A_n^{(001)}
& \dots \vtt }
} \vttt \cr
- A_n^{(0)} &
\matrix{
A_{n}^{(01)} & \matrix{ A_n^{(010)} & \dots \vtt \cr -A_n^{(010)}
& \dots \vtt }\cr
-A_{n}^{(01)} & \matrix{ A_n^{(011)} & \dots \vtt \cr -A_n^{(011)}
& \dots \vtt }
}}
\ \ \ \right)
$$
\medskip
where the blocks $(A_n^{(i_1,\dots,i_r)})_{r=1,\dots,n-1}$ are of order
$2^{n-1-r}$. Moreover $A_n^{(1)} = N(R,\Phi)_{n-1}$ and
$$
A_n^{(i_1,\dots,i_r)} = A_{n-1}^{(i_2,\dots,i_r)} \ ,
$$
for every $r \in \{ 1,\dots, n\! -\! 2 \}$. This shows in particular that the
matrix $N(R,\Phi)_n$ is completely determined by its last column.

\smallskip
3) Let $LC(R,\Phi)_n$ be the row vector of order $2^{n-1}$ which corresponds
to the reading of the last column of $N(R,\Phi)_n$ from top to bottom and
let $V_n$ be the vector of order $2^{n-2}$ defined in the statement of
Proposition \ref{RPsi}. Then, one has
$$
LC(R,\Phi)_n = V_n \, . \, \overline{V_n} \ .
$$
Thus $LC(R,\Phi)_n$, and hence all the matrix $M(R,\Phi)_n$, can be
recovered from $V_n$.
\end{proposition}

\begin{example}
{\rm The  matrices relating $R_I$ and $\Phi^I$
for $n = 2, 3, 4$ are given below.}
$$
M(\Phi,R)_2 =
\matrix{
& 2 \ \ \ 11 \vt \cr
\matrix{ 2 \cr 11 } \!\!\!\!\! &
\left(\matrix{
1 & -1 \cr
1 & 1 \cr
}\right) \vt
}
\ , \quad
M(R,\Phi)_2 =
\matrix{
& 2 \ \ \ \ \ \ 11 \vt \cr
\matrix{ 2 \cr 11 } \!\!\!\!\! &
\left(\matrix{
1/2 & 1/2 \cr
-1/2 & 1/2 \cr
}\right) \vt
}
\ ,
$$
\medskip
$$
M(\Phi,R)_3 = \!\!\!\!
\matrix{
& 3 \ \ \ 21 \ \ \ \ 12 \ \ \ \ 111 \vt \cr
\matrix{ 3 \cr 21 \cr 12 \cr 111} \!\!\!\!\!\! &
\left(\matrix{
1 \!\! & -1/2 \!\! & -1/2 \!\! & 1 \cr
1 \!\!  & 1   \!\! & -1   \!\! & -1 \cr
1 \!\! & -1   \!\! & 1    \!\! & -1 \cr
1 \!\! & 1    \!\! & 1    \!\! & 1
}\right)
}\, , \ \,
M(R,\Phi)_3 = \!\!\!\!
\matrix{
& 3\ \ \ \ \ \ 21\ \ \ \ \ \ 12\ \ \ \ \ \ 111 \vt \cr
\matrix{ 3 \cr 21 \cr 12 \cr 111} \!\!\!\!\!\! &
\left(\matrix{
1/3  \! & 1/4  \! & 1/4  \! & 1/6 \cr
-1/3 \! & 1/4  \! & -1/4 \! & 1/3 \cr
-1/3 \! & -1/4 \! & 1/4  \! & 1/3 \cr
1/3  \! & -1/4 \! & -1/4 \! & 1/6
}\right)
}\ ,
$$

\smallskip
$$
M(\Phi,R)_4 =
\matrix{
& 4\ \ \ \ \ \ 31 \ \ \ \ \ \ 22\ \ \ \ \ \ 211
\ \ \ \ \ \ 13 \ \ \ \ \ 121\ \ \ \ \ 112\ \ \ \ 1111 \vt \cr
\matrix{ 4 \cr 31 \cr 22 \cr 211 \cr 13 \cr 121 \cr 112 \cr 1111}
\!\!\!\!\!\! &
\left( \
\matrix{
1 & -1/3 & -1/3 & 1/3  & -1/3 & 1/3  & 1/3  & -1 \cr
1 & 1    & -1/2 & -1/2 & -1/2 & -1/2 & 1    & 1 \cr
1 & -1   & 1    & -1   & -1   & 1    & -1   & 1 \cr
1 & 1    & 1    & 1    & -1   & -1   & -1   & -1 \cr
1 & -1/2 & -1/2 & 1    & 1    & -1/2 & -1/2 & 1 \cr
1 & 1    & -1   & -1   & 1    & 1    & -1   & -1 \cr
1 & -1   & 1    & -1   & 1    & -1   & 1    & -1 \cr
1 &  1   & 1    &  1   & 1    &  1   & 1    & 1 \cr
}
\ \right)
} \ ,
$$

\smallskip
$$
M(R,\Phi)_4 =
\matrix{
& 4\ \ \ \ \ \ \ 31 \ \ \ \ \ \ \ 22\ \ \ \ \ \ \ 211
\ \ \ \ \ \ \ 13 \ \ \ \ \ 121\ \ \ \ \ 112\ \ \ \ \ 1111 \vt \cr
\matrix{ 4 \cr 31 \cr 22 \cr 211 \cr 13 \cr 121 \cr 112 \cr 1111}
\!\!\!\!\!\! &
\left(\
\matrix{
1/4  & 1/6  & 1/8  & 1/12  & 1/6  & 1/12  & 1/12  & 1/24 \cr
-1/4 & 1/6  & -1/8 & 1/6   & -1/6 & 1/6   & -1/12 & 1/8 \cr
-1/4 & -1/6 & 1/8  & 1/6   & -1/6 & -1/12 & 1/6   & 5/24 \cr
1/4  & -1/6 & -1/8 & 1/12  & 1/6  & -1/6  & -1/6  & 1/8 \cr
-1/4 & -1/6 & -1/8 & -1/12 & 1/6  & 1/6   & 1/6   & 1/8 \cr
1/4  & -1/6 & 1/8  & -1/6  & -1/6 & 1/12  & -1/6  & 5/24 \cr
1/4  & 1/6  & -1/8 & -1/6  & -1/6 & -1/6  & 1/12  & 1/8 \cr
-1/4 & 1/6  & 1/8  & -1/12 & 1/6  & -1/12 & -1/12 & 1/24 \cr
}
\ \right)
} \ .
$$
\end{example}


\subsection{$\Phi$ and $\Psi$}\label{PhiPsi}

The problem of expressing $\Phi_n$ as a linear combination of
the $\Psi^I$ is a classical question of Mathematical Physics, known
as the problem of the {\it continuous Baker-Campbell-Hausdorff exponents}
\cite{Mag}\cite{Chen7}\cite{Wil}\cite{BMP}\cite{MP}\cite{St}\cite{V}.
The usual formulation of this
problem is as follows. Let $H(t)$ be a differentiable function with values
in some operator algebra, and consider the solution $E(t;t_0)$ (the {\it
evolution operator}) of the Cauchy problem
$$
\left\{\matrix{
{\partial\over\partial t} E(t;t_0) = H(t)E(t;t_0) \cr
E(t_0;t_0)=1 }\right.
$$
The problem is to express the {\it continuous BCH exponent} $\Omega(t;t_0)$,
defined by $E(t;t_0) = \exp \Omega(t;t_0)$ in terms of $H(t)$.

Here we merely consider the (equivalent) variant
$\displaystyle {\partial E\over\partial t}=E(t)H(t)$ with $t_0=0$, and
we set $E(t)=\sigma(t)$, $H(t)=\psi(t)$, $\Omega(t)=\Phi(t)$.

The first answer to this question has been given by W. Magnus \cite{Mag}.
His formula provides an implicit relation between $\Phi(t)$ and $\psi(t)$,
allowing  the recursive computation of the coefficients $\Phi_n$.
{}From our point of view, it rather gives the explicit expression of
$\Psi_n$ in terms of the $\Phi^I$. Following \cite{Wil}, we shall derive
this formula from an identity of independent interest:

\begin{lemma} \label{DEXP}
The derivative of $\sigma(t)=\exp \Phi(t)$ is given by
\begin{equation}
\sigma'(t)={d\over dt}e^{\Phi(t)}
=\int_0^1 e^{(1-u)\Phi(t)} {d \Phi(t)\over dt} e^{u\Phi(t)} du \ .
\end{equation}
\end{lemma}

\Proof Expanding $\sigma(t)=e^{\Phi(t)}$ in powers of $\Phi(t)$, we have
$$
\sigma'(t) = {d\over dt}\sum_{n\ge 0} {1\over n!}\Phi(t)^n
=\sum_{r,s\ge 0}{\Phi(t)^r\Phi'(t)\Phi(t)^s\over (r+s+1)!}
$$
$$
=\sum_{r,s\ge 0} {r! s! \over (r+s+1)!}
{\Phi(t)^r \over r!} \Phi'(t) {\Phi(t)^s\over s!} \ ,
$$
and using the integral representation
$$
{r! s! \over (r+s+1)!}={\rm B}(r+1,s+1)=\int_0^1(1-u)^ru^s du
$$
we find
$$
\sigma'(t)=\sum_{r,s\ge 0}\int_0^1
{[(1-u)\Phi(t)]^r\over r!}\Phi'(t) {[u\Phi(t)]^s\over s!} du
$$
$$
= \int_0^1 e^{(1-u)\Phi(t)}\Phi'(t) e^{u\Phi(t)} du \ .
$$\cqfd

To obtain an expression for $\Psi_n$, we observe that Lemma \ref{DEXP}
can be rewritten
$$
\sigma'(t) =\sigma(t)\ \int_0^1 e^{-u\Phi(t)}\Phi'(t) e^{u\Phi(t)} du \ ,
$$
so that
\begin{equation}\label{PSIINT}
\psi(t) = \int_0^1 e^{-u\Phi(t)}\Phi'(t) e^{u\Phi(t)} du
\end{equation}
$$
= \int_0^1 \sum_{r\ge 0} {(-u)^r\over r!}\Phi(t)^r \Phi'(t)
\sum_{s\ge 0} {u^s\over s!}\Phi(t)^s\, du
$$
$$
= \sum_{r,s\ge 0} {(-1)^r\over (r+s+1)!}{r+s\choose r}
\Phi(t)^r\Phi'(t)\Phi(t)^s \ .
$$
Extracting the coefficient of $t^{n-1}$, we finally obtain:

\begin{proposition}
The expansion of $\Psi_n$ in the basis $(\Phi^K)$ is given by
$$
\Psi_n = \sum_{|K|=n}
\left[
\sum_{i=1}^{\ell (K)} (-1)^{i-1}{\ell (K)-1 \choose i-1} k_i
\right]
{\Phi^K \over \ell (K)! \pi (K)}
\ .
$$
\end{proposition}
\cqfd

Using the symbolic notation
$$
\{\Phi_{i_1}\cdots\Phi_{i_r}\, ,\, F\}
=\ad\Phi_{i_1} \ad\Phi_{i_2}\cdots \ad\Phi_{i_r} (F)
=[\Phi_{i_1},[\Phi_{i_2},[\ldots[\Phi_{i_r}\, ,\, F]\ldots]]]
$$
and the classical identity
$$
e^a b e^{-a} =\sum_{n\ge 0} {(\ad a)^n\over n!}\, b = \{e^a\, ,\, b\} \ ,
$$
we can rewrite (\ref{PSIINT}) as
$$
\psi(t) =\sum_{n\ge 0} {(-1)^n\over (n+1)!}\{\Phi(t)^n\, ,\,\Phi'(t)\}
=\left\{ {1-e^{-\Phi(t)} \over \Phi(t)} \, ,\, \Phi'(t) \right\}
$$
which by inversion gives the Magnus formula:
\begin{equation}\label{MAGNUS}
\Phi'(t)= \left\{ {\Phi(t)\over 1-e^{-\Phi(t)}  } \, ,\, \psi(t)
\right\}
= \sum_{n\ge 0} {B_n\over n!} (\ad \Phi(t))^n\, \psi(t)
\end{equation}
the $B_n$ being the Bernoulli  numbers.

As already mentioned, formula (\ref{MAGNUS}) permits the recursive computation
of the $\Phi_n$ in terms of iterated integrals. There exists, however, an
explicit expression of this type, which is due to Bialynicki-Birula, Mielnik
and
Pleba\'nski \cite{BMP} (see also \cite{MP})
and rediscovered by several authors (see \eg \cite{St}\cite{V}).
We shall here only state a version
of this result, postponing the discussion to section \ref{BCHCONT}, where
the formalism of the internal product will be needed.

Recall that an index $i\in\{1,2,\ldots,n-1\}$ is said to be a {\it descent}
of a permutation $\sigma\in\S_n$ if $\sigma(i)>\sigma(i+1)$. We denote
by $d(\sigma)$ the number of descents of the permutation $\sigma$.

\begin{theorem} {\rm (Continuous BCH formula)} The expansion of
$\Phi(t)$ in the basis $(\Psi^I)$ is given by the series
\begin{equation}\label{BCHCP}
\Phi(t)= \sum_{r\ge 1}\int_0^t dt_1\cdots \int_0^{t_{r-1}}dt_r
\sum_{\sigma\in \S_r}
{(-1)^{d(\sigma)} \over r} {r-1\choose d(\sigma)}^{-1}
\psi(t_{\sigma(r)})\cdots \psi(t_{\sigma(1)}) \ .
\end{equation}
Thus, the coefficient of $\Psi^I=\Psi_{i_1}\cdots\Psi_{i_r}$ in the
expansion of $\Phi_n$ is equal to
$$
n\int_0^1 dt_1\cdots \int_0^{t_{r-1}} dt_r
\sum_{\sigma\in\S_r} {(-1)^{d(\sigma)}\over r}
{r-1\choose d(\sigma)}^{-1}
t_{\sigma(r)}^{i_1-1}\cdots t_{\sigma(1)}^{i_r-1} \ .
$$
\end{theorem}

\begin{example}
{\rm Here are the transition matrices corresponding to the two bases
$\Psi$ and $\Phi$, up to the order $n = 4$}
$$
M(\Psi,\Phi)_2 = M(\Phi,\Psi)_2 =
\matrix{
& 2 \ \ \ 11 \vt \cr
\matrix{ 2 \cr 11 } \!\!\!\!\! &
\left(\matrix{
1 & 0 \cr
0 & 1 \cr
}\right) \vt
}
\ ,
$$
\bigskip
$$
M(\Psi,\Phi)_3 = \!\!\!
\matrix{
& 3 \ \ \ 21 \ \ \ 12 \ \ \ 111 \vt \cr
\matrix{ 3 \cr 21 \cr 12 \cr 111} \!\!\!\!\!\! &
\left(\matrix{
1 & 1/4  & -1/4 & 0 \cr
0 & 1  & 0 & 0 \cr
0 & 0  & 1  & 0 \cr
0 & 0  & 0  & 1
}\right)
}\, , \ \,
M(\Phi,\Psi)_3 = \!\!\!
\matrix{
& 3 \ \ \ 21 \ \ \ 12 \ \ \ 111 \vt \cr
\matrix{ 3 \cr 21 \cr 12 \cr 111} \!\!\!\!\!\! &
\left(\matrix{
1 & -1/4    & 1/4 & 0 \cr
0 & 1  & 0  & 0 \cr
0 & 0  & 1  & 0 \cr
0 & 0  & 0  & 1
}\right)
}\ ,
$$
\bigskip
$$
M(\Psi,\Phi)_4 =
\matrix{
& 4\ \ \ \ 31\ \ \ \ 22\ \ \ \ 211
\ \ \ 13\ \ \ \ 121\ \ \ \ 112\ \ \ \ 1111 \vt \cr
\matrix{\ 4 \cr 31 \cr 22 \cr 211 \cr 13 \cr 121 \cr 112 \cr 1111}
\!\!\!\!\!\! &
\left( \
\matrix{
1 & 1/3  & 0    & 1/12 & -1/3 & -1/6 & 1/12 & 0 \cr
0 & 1    & 0    & 1/4  & 0    & -1/4 & 0    & 0 \cr
0 & 0    & 1    & 0    & 0    & 0    & 0    & 0 \cr
0 & 0    & 0    & 1    & 0    & 0    & 0    & 0  \cr
0 & 0    & 0    & 0    & 1    & 1/4  & -1/4 & 0 \cr
0 & 0    & 0    & 0    & 0    & 1    & 0    & 0  \cr
0 & 0    & 0    &  0   & 0    & 0    & 1    & 0  \cr
0 & 0    & 0    &  0   & 0    &  0   & 0    & 1 \cr
}
\ \right)
} \ ,
$$
\bigskip
$$
M(\Phi,\Psi)_4 =
\matrix{
& 4 \ \ \ \ 31 \ \ \ \ 22 \ \ \ 211
\ \ \ \ 13\ \ \ \ 121\ \ \ \ 112 \ \ \ \ 1111 \vt \cr
\matrix{ 4 \cr 31 \cr 22 \cr 211 \cr 13 \cr 121 \cr 112 \cr 1111}
\!\!\!\!\!\! &
\left(\
\matrix{
1 & -1/3 & 0    & 0    & 1/3  & 0    & 0    & 0 \cr
0 & 1    & 0    & -1/4 & 0    & 1/4  & 0    & 0 \cr
0 & 0    & 1    & 0    & 0    & 0    & 0    & 0 \cr
0 & 0    & 0    & 1    & 0    & 0    & 0    & 0  \cr
0 & 0    & 0    & 0    & 1    & -1/4 & 1/4 & 0 \cr
0 & 0    & 0    & 0    & 0    & 1    & 0    & 0  \cr
0 & 0    & 0    &  0   & 0    & 0    & 1    & 0  \cr
0 & 0    & 0    &  0   & 0    &  0   & 0    & 1 \cr
}
\, \right)
} \, .
$$
\end{example}

\newpage
\section{Connections with Solomon's descent algebra}\label{DESCENT}

Let us denote by $\D (\sigma)$ the descent set of a permutation
$\sigma$, and for
each subset $A\subseteq \{1,\ldots,n-1\}$, consider the element
$$
D_{=A} = \sum_{\D (\sigma)=A} \ \sigma \ \ \, \in K [{\goth S}_n] \ .
$$
It has been shown by Solomon (\cf \cite{So}) that the subspace of
$K [{\goth S}_n]$ generated by the elements $D_{=A}$ is in fact a subalgebra
of $K [{\goth S}_n]$. More precisely, there exist nonnegative
integers $d_{AB}^C$ such that
$$
D_{=A}\, D_{=B} = \ \sum_C\ d_{AB}^C \, D_{=C} \ ,
$$
for  $A,B \subset \{1,\ldots,n-1\}$. This subalgebra is called the
{\it descent algebra} of ${\goth S}_n$ and is denoted by $\Sigma_n$. A similar
algebra can be defined for any Coxeter group (\cf \cite{So}).

\smallskip
The dimension of $\Sigma_n$ as a vector space is obviously $2^{n-1}$, that is,
the same as the dimension of the space  $\Sym_n$ of noncommutative formal
symmetric functions of weight $n$. This is not a mere coincidence : there is
a canonical isomorphism between these two spaces, and we shall see below that
the whole algebra $\Sym$ can be identified with the direct sum
$\Sigma = \bigoplus_{n\ge 0}\Sigma_n$ of all descent algebras, endowed
with some natural operations.


\subsection{Internal product}\label{INTERNAL}

Subsets of $[n-1] = \{1,\ldots,n-1\}$ can be represented by compositions
of $n$ in the following way. To a subset $A=\{a_1<a_2<\ldots<a_k\}$
of $[n-1]$, one associates the composition
$$
c(A) = I = (i_1,\ldots,i_{k+1}) =(a_1,a_2-a_1,\ldots,a_k-a_{k-1},n-a_k) \ .
$$
A permutation compatible with a ribbon diagram of associated composition
$I$ ({\it i.e.} yielding a standard skew tableau when written in the diagram)
has then for descent set $c^{-1}(I)$. It is thus natural to define a linear
isomorphism $\alpha :\ \Sigma_n \longrightarrow \Sym_n$ by setting
$$
\alpha(D_{=A}) = R_{c(A)} \ .
$$
This allows us to define an algebra structure on each homogeneous
component $\Sym_n$ of $\Sym$ by transporting the product of $\Sigma_n$.
More precisely, we shall use the {\it opposite} structure, for
reasons that will become apparent later. The so-defined product will be
denoted by $*$. Hence, for  $F,G\in \Sym_n$, we have
$$
F*G = \alpha(\alpha^{-1}(G)\, \alpha^{-1}(F)) \ .
$$
We then extend this  product to $\Sym$ by setting $F*G=0$ if $F$ and
$G$ are homogeneous of different weights. The operation $*$ will be
called the {\it internal product} of $\Sym$, by analogy with the
commutative operation of which it is the natural noncommutative
analog. Indeed, let us consider another natural basis of $\Sigma_n$,
constituted by the elements
$$
D_{\subseteq A} = \, \sum_{B\subseteq A}\ D_{=B} \ .
$$
Then, according to formula (\ref{R2S}),
$$
\alpha (D_{\subseteq A}) = S^I \ ,
$$
where $I=c(A)$ (see also \cite{Re}). Remark that in particular
$$\alpha(id) = S_n
$$
so that for  $F\in\Sym_n$,
$$
F*S_n = S_n*F =F \ ,
$$
and that
$$
\alpha (\omega_n) = \Lambda_n \ ,
$$
where $\omega_n$ denotes the maximal permutation
$(n \, n-1 \, \cdots\, 1) = D_{=[1,n-1]}$ of ${\goth S}_n$. Finally the
multiplication formula
for $D_{\subseteq A}\, D_{\subseteq B}$ (\cf \cite{So}) can be be rewritten
as follows.

\begin{proposition}
For any two compositions $I = (i_1,\dots,i_p)$ and $J = (j_1,\dots,j_q)$,
one has
\begin{equation}\label{ipcomp}
S^I * S^J= \sum_{M\in \mat (I,J)} \, S^M
\end{equation}
where $\mat (I,J)$ denotes the set of matrices of nonnegative integers
$M = (m_{ij})$ of order $p\times q$ such that $\sum_s \, m_{rs}=i_r$ and
$\sum_r \, m_{rs}=j_s$ for $r \in [1,p]$ and $s \in [1,q]$, and
where
$$
S^M = S_{m_{11}}\, S_{m_{12}}\, \cdots\, S_{m_{1 p}}\, \cdots \,
S_{m_{q1}}\cdots S_{m_{qp}} \ .
$$
\end{proposition}

It is  well known that the same formula holds
for {\it commutative} symmetric functions. In this case, a product of
complete functions $S^I$ is the Frobenius characteristic of a permutation
representation (see \cite{JK}). Thus the passage to commutative symmetric
functions transforms the noncommutative $*$-product into the ordinary
internal product (this is not surprising, since the descent algebra
was originally introduced as a noncommutative analog of
the character ring in the group algebra).

\smallskip
It is thus natural to identify $\Sym$ with the direct sum
$\Sigma = \bigoplus_{n\ge 0}\Sigma_n$ of all descent algebras. The ordinary
product of $\Sym$ then corresponds to a natural product on $\Sigma$, which we
will call {\it outer product}. In fact $\Sigma$ can be interpreted as a
subalgebra of the convolution algebra of a free associative algebra
(considered as a Hopf algebra for the standard comultiplication making the
letters primitive). More precisely, it is the subalgebra generated by the
projectors $q_n$
onto the homogeneous components $K_n\< A\>$
of the free algebra  $K\< A\>$ (\cf \cite{Re}).
The convolution product, denoted $*$ in \cite{Re}, corresponds then
to the ordinary product of noncommutative symmetric functions, and the
composition of endomorphisms to the internal product. This construction
has been recently extended to the case of any graded bialgebra
by Patras (\cf \cite{Pa}).

\smallskip
The noncommutative internal product satisfy some compatibility relations
with the ordinary product, similar to those encountered in the commutative
case. For example one has the following identity, whose commutative
version is a fundamental tool for  calculations with tensor products of
symmetric group representations (see \cite{Ro}, \cite{Li56} or \cite{Th}).

\begin{proposition}\label{mackey}
Let $F_1,F_2,\ldots,F_r,G \in \Sym$. Then,
$$
(F_1F_2\cdots F_r)*G
= \mu_r\left[ (F_1\otimes\cdots\otimes F_r)* \Delta^r G \right]
$$
where in the right-hand side, $\mu_r$ denotes the
$r$-fold ordinary multiplication and $*$ stands for the operation induced
on $\Sym^{\otimes n}$ by $*$.
\end{proposition}

\Proof The formula being linear in each of its arguments, it is
sufficient to establish it in the case where the $F_i$ and $G$ are products
of complete functions. When the $F_k$ are of the form
$F_k=S_{i_k}$ and $G=S^J=S_{j_1}S_{j_2}\cdots S_{j_s}$, the formula
\begin{equation}\label{ipcomp1}
(S_{i_1}S_{i_2}\cdots S_{i_r})*S^J
=
\mu_r\left[(S_{i_1}\otimes\cdots\otimes S_{i_r})*\Delta^r S^J \right]
\end{equation}
is  equivalent to the multiplication formula (\ref{ipcomp}).
Next, let $I^{(k)}=(i_1^{(k)},i_2^{(k)},\ldots,i_{n_k}^{(k)})$ for every
$k=1,\ldots,r$, and consider the transformations
$$
\mu_r
\left[(S^{I^{(1)}}\otimes\cdots\otimes S^{I^{(r)}})*\Delta^r G
\right]
$$
$$
=
\mu_r\left[ \ \sum_{(G)}\
(S^{I^{(1)}}*G_{(1)})\otimes(S^{I^{(2)}}*G_{(2)})\otimes
\cdots\otimes (S^{I^{(r)}}*G_{(r)}) \right]
$$
(using Sweedler's notation for $\Delta^rG$)
$$
=\mu_r\left[ \, \sum_{(G)}\,
\mu_{n_1} \!\!
\left(
( S_{i_1^{(1)}}\otimes\cdots\otimes S_{i_{n_1}^{(1)}})*\Delta^{n_1} G_{(1)}
\right)
\! \otimes\cdots\otimes
\mu_{n_r}\!\!
\left(
(S_{i_1^{(r)}}\otimes\cdots\otimes S_{i_{n_r}^{(r)}})*\Delta^{n_r} G_{(r)}
\right)
\right]
$$
(by application of formula (\ref{ipcomp1}))
$$
=
\mu_r\circ(\mu_{n_1}\otimes\cdots\otimes\mu_{n_r})
\left[
\left(
S_{i_1^{(1)}}\otimes\cdots\otimes S_{i_{n_1}^{(1)}}\otimes S_{i_1^{(2)}}
\otimes\cdots\otimes S_{i_{n_r}^{(r)}}
\right) *
(\Delta^{n_1}\otimes\cdots\otimes\Delta^{n_r})\circ\Delta^r G
\right]
$$
$$
=
\mu_n\left[\left(
S_{i_1^{(1)}}\otimes\cdots\otimes S_{i_{n_1}^{(1)}}
\otimes \cdots \otimes
S_{i_1^{(r)}} \otimes\cdots\otimes S_{i_{n_r}^{(r)}}
\right) * \Delta^N G \right]
$$
(by associativity and coassociativity, with $N=n_1+\cdots +n_r$)
$$
= \left( S^{I^{(1)}}S^{I^{(2)}}\cdots S^{I^{(r)}}\right)*G\ .
$$\cqfd

\begin{example}
{\rm To compute $S^{212}*S^{23}$, one has to enumerate the matrices with
row sums vector $I=(2,1,2)$ and column sums vector $J=(2,3)$, which yields
$$
S^{212}*S^{23}=2\, S^{212} + S^{2111}+S^{1112}+S^{11111} \ .
$$
Applying Proposition \ref{mackey} and taking into account the fact that
$S_k*F=F$ for $F\in\Sym_k$ and $S_k*F=0$ otherwise, we obtain as well
$$
\mu\left[(S^{21}\otimes S^2)*\Delta S^{23} \right]
=
\mu\left[(S^{21}\otimes S^2)*
\left\{ {(S^2\otimes 1+S^1\otimes S^1 + 1\otimes S^2)\, \times \atop
         (S^3\otimes 1+S^2\otimes S^1+S^1\otimes S^2 + 1\otimes S^3)}
\right\}
\right]
$$
$$
=\mu\left[(S^{21}\otimes S^2)*(S^{21}\otimes S^2 +S^{12}\otimes S^{11}
 +S^3\otimes S^2) \right]
$$
$$=2\, S^{212}+S^{1112}+S^{11111}+S^{2111} = S^{212}*S^{23} \ .$$
}
\end{example}

\begin{example}
{\rm One can recognize the involution $\eta : F\mapsto F*\lambda(1)$
by computing the products $S^I*\lambda(1)$. If $I=(i_1,\ldots,i_r)$,
$$
S^I*\lambda(1)
=
\mu_r\left[ (S_{i_1}\otimes\cdots\otimes S_{i_r}) *
(\lambda(1)\otimes\cdots\otimes\lambda(1) ) \right]
$$
$$
= \Lambda_{i_1}\, \Lambda_{i_2}\, \cdots\, \Lambda_{i_r} = \Lambda^I \ .
$$
Hence, $\eta$ is the involutive automorphism whichs sends $S_k$ to $\Lambda_k$.
One can check that its action on ribbons is given by
\begin{equation}
\eta(R_I) = R_I*\lambda(1) = R_{\overline{I^\sim}} \ .
\end{equation}
}
\end{example}

The involution $F\mapsto \lambda(1) * F$ is more easily identified
in terms of permutations. Since the descent set of $\sigma\omega$
is the complement in $\{1,\ldots,n-1\}$ of the descent set of $\sigma$,
one sees that it is the antiautomorphism
$$
\lambda(1)*R_I = R_{I^\sim} = \omega (R_I) \ .
$$

\smallskip
Other properties of the internal product can be
lifted to the noncommutative case. For example,
in \cite{At}, Atkinson defines maps
$\varepsilon_I:\ \Sym_n \longrightarrow
\Sym_{i_1}\otimes\Sym_{i_2}\otimes\cdots\otimes\Sym_{i_r}$ by setting
\begin{equation}
\varepsilon_I(F) = (S_{i_1}\otimes S_{i_2}\otimes \cdots\otimes S_{i_r})
* \Delta^r(F)
\end{equation}
and shows that they are algebra homomorphisms for the internal product.
He then uses them to construct a simplified version of the representation
theory of the descent algebra (the first theory was due to Garsia and
Reutenauer (\cf \cite{GaR})). The fact that these maps are homomorphisms
is equivalent to the following important property whose proof may be
adapted from Atkinson's paper.

\begin{proposition}\label{ATK}
The iterated coproduct $\Delta^r$ is a homomorphism for the internal
product from $\Sym$ into $\Sym^{\otimes r}$. In other words,  for
$F,G \in \Sym$
$$
\Delta^r(F*G) = \Delta^rF * \Delta^r G \ .
$$
\end{proposition}

In the commutative case, $\sym_n$ endowed with the internal product is
interpreted as the representation ring $R(\S_n)$ of the symmetric group, and
$\sym_{i_1}\otimes\sym_{i_2}\otimes\cdots\otimes\sym_{i_r}$ as the
representation ring of a Young subgroup
$\S_I = \S_{i_1}\times\S_{i_2}\times\cdots\times\S_{i_r}$. With
this interpretation, the maps $\varepsilon_I$ correspond
to the restriction of representations from $\S_n$ to $\S_I$.

\smallskip
The following consequence is worth noting.
Remark that it follows from the structure theory of Hopf algebras and
from Proposition \ref{LIE} that $\Sym$ is the universal enveloping
algebra of the free Lie algebra $L(\Psi)$ constituted
by its primitive elements.

\begin{corollary}
The internal product preserves the primitive
Lie algebra $L(\Psi)$ of $\Sym$.
\end{corollary}


\subsection{Lie idempotents}\label{LIEID}

A Lie idempotent is an idempotent of the group algebra $K[\S_n]$ which
acts as a projector from the free algebra
$K\< A\>$ onto the free Lie algebra $L(A)$
(for the right action of the symmetric group on words).
It turns out that most of the Lie idempotents which have been encountered
up to now are elements of the descent algebra (\cf \cite{BBG} or \cite{Re}).
In this section, we show that these elements appear quite naturally in the
theory of noncommutative symmetric functions.

\smallskip
In the commutative case, the products of power sums $\psi^I$ are
idempotent (up to a scalar factor) for the internal product.
The noncommutative case is more complicated, but
the simplest noncommutative internal products of power sums
already lead to two interesting Lie idempotents.

\begin{proposition} \label{Dynkin}
For all $n \geq 1$, one has $\Psi_n *\Psi_n = n\, \Psi_n\ $.
\end{proposition}

\Proof Recall that the generating series for the $\Psi_n$ is given by
$$
\psi(t) = \ \sum_{n\ge 1}\ t^{n-1}\, \Psi_n
=
\lambda(-t)\, \sigma'(t)
$$
and that $\Delta\psi(t) =\psi(t)\otimes 1+1\otimes\psi(t)$. Then, using the
fact that $\Psi_i*\Psi_j=0$ for $i\not = j$ and applying Proposition
\ref{mackey}, we can write
$$
\sum_{n\ge 1} \ (xy)^{n-1}(\Psi_n*\Psi_n) = \psi(x)*\psi(y)
=
\mu\left[
(\lambda(-x)\otimes\sigma'(x))*(\psi(y)\otimes 1+ 1\otimes\psi(y))
\right]
$$
$$
=\mu\left[(\lambda(-x)*1)\otimes(\sigma'(x)*\psi(y))\right]
$$
(since $\sigma'(x)$ has no term of weight $0$)
$$
=
\left(\ \sum_{n\ge 1}\ nx^{n-1}\, S_n \right)
*
\left(\ \sum_{n\ge 1} \ y^{n-1}\, \Psi_n \right)
=
\ \sum_{n\ge 1}\ (xy)^{n-1}\, n\, \Psi_n \ ,
$$
the last equality following from the fact that $S_n * F = F$ for
$F \in \Sym_n$. \cqfd

This proposition is in fact equivalent to Dynkin's characterization of
Lie polynomials (see {\it e.g.} \cite{Re}). Indeed, recall that the
{\it standard left bracketing} of a word $w=x_1x_2\cdots x_n$ in the
noncommuting indeterminates $x_i$ is the Lie polynomial
$$
L_n(w) = [\cdots [[[x_1,x_2],x_3],x_4],\ldots,x_n ] \ .
$$
In terms of the right action of the symmetric group $\S_n$ on the homogeneous
component $K_n\langle A\rangle$ of degree $n$ of the free associative
algebra $K\langle A\rangle$, defined on words by
$$
x_1\, x_2\, \cdots\,  x_n \cdot \sigma
=
x_{\sigma(1)}\, x_{\sigma(2)}\, \cdots \, x_{\sigma(n)} \ ,
$$
one can write
$$
L_n(w) =
\ \sum_{\sigma\in\S_n} \ a_\sigma\, (w\cdot\sigma) = w\cdot\theta_n
$$
where $\theta_n$ is an element of $\Z [\S_n]$. It is not difficult to show
that $\theta_n$ is in fact in $\Sigma_n$ and that one has  (\cf \cite{Ga})
$$
\theta_n = \ \sum_{k=0}^{n-1}\ (-1)^k\, D_{=\{1,2,\ldots,k\}} \ .
$$
To see this, one just has to write the permutations appearing in the first
$\theta_i$ as ribbon tableaux and then to use induction. For example,
$$
\theta_3 =
[[1,2],3]
=
123 \ -\ \matrix{2&\cr 1 & \!\!\! 3\cr} \ - \
\matrix{3&\cr 1 & \!\!\! 2\cr} \ +\ \matrix{3\cr 2\cr 1\cr}
$$
and it is clear that when expanding $\theta_4=[\theta_3,4]$ one will only
get those (signed) tableaux obtained from the previous ones by adding $4$
at the end of the last row, minus those obtained by adding $4$ on the top of
the first column. Thus, in terms of noncommutative symmetric functions, we
have from Corollary \ref{psihook}
$$
\alpha(\theta_n) = \Psi_n \ ,
$$
so that Proposition \ref{Dynkin} becomes Dynkin's theorem :
$\theta_n^2 =n\, \theta_n$, or more explicitely, a noncommutative polynomial
$P\in K\langle X\rangle$ is a Lie polynomial iff $L_n(P)=nP$.

\smallskip
The same kind of argument works as well for the power sums of the
second kind $\Phi_n$.

\begin{proposition}
For every $n \geq 1$, one has $\Phi_n*\Phi_n = n\, \Phi_n\ $.
\end{proposition}

\Proof Using the generating series
$$
\Phi(t) = \ \sum_{n\ge 1}\ \Phi_n\, {t^n\over n}
=\log \sigma(t)
=\ \sum_{k\ge 1}\ {(-1)^{k-1}\over k} \ (tS_1+t^2S_2+\cdots)^k \ ,
$$
we have
$$
\Phi(x)*\Phi(y) =
\ \sum_{n\ge 1}\ {(-1)^{n-1}\over n}\ (xS_1+x^2S_2+\cdots )^n *\Phi(y) \ .
$$
But, using Proposition \ref{mackey}, one has for  $n > 1$
$$
(xS_1+x^2S_2+\cdots )^n*\Phi(y)
=
\mu_n\left[
\left(\sum_{i_1\ge 1}x^{i_1}S_{i_1}\right)
\otimes\cdots\otimes
\left(\sum_{i_n\ge 1}x^{i_n}S_{i_n}\right) *
\Delta^n\Phi(y)
\right]
= 0 \ ,
$$
since $\Phi(y)$ is primitive and $\sum_{i\ge 1}x^iS_i$ has no term of weight
zero. Thus, using again the fact that $S_n$ is a unit for the internal
product on $\Sym_n$, we get
$$
\Phi(x)*\Phi(y)
=
\left(\ \sum_{i\ge 1}\ x^i\, S_i \, \right) *
\left(\ \sum_{i\ge 1}\ y^i\, {\Phi_i\over i}\, \right)
=
\ \sum_{i\ge 1}\ (xy)^i\, {\Phi_i\over i} =\Phi(xy) \ .
$$\cqfd

The element $e_n^{[1]} = \alpha^{-1}(\Phi_n/n)$ is thus an idempotent
of the descent algebra $\Sigma_n$. In terms of permutations, the
formula $\phi(t)=\log\left(1 +(tS_1+t^2S_2+\cdots)\right)$ shows that
$$
e_n^{[1]}= \sum_{A\subseteq \{1,\ldots,n-1\} } \,
 {(-1)^{|A|}\over |A|+1} \ D_{\subseteq A} \ .
$$
(compare with \cite{Re} p. 66-67). This idempotent is also a projector onto
the free Lie algebra. One way to see this is to compute the products
$\Phi_n*\Psi_n$ and $\Psi_n*\Phi_n$ and to use Dynkin's characterization
of Lie polynomials.

\begin{proposition}
For every $n \geq 1$, one has (i)  $\Psi_n *\Phi_n = n\, \Phi_n$ and (ii)
$\Phi_n *\Psi_n=n\, \Psi_n\ $.
\end{proposition}

\Proof {\it (i)} Using Proposition \ref{mackey} and the fact
that $\Phi(y)$ is primitive for $\Delta$, we have
$$
\psi(x)*\phi(y) = (\lambda(-x)\, \sigma'(x))*\Phi(y)
=
\mu\left[(\lambda(-x)\otimes\sigma'(x))*
(\Phi(y)\otimes 1+1\otimes\Phi(y))\right]
$$
$$
=
\sigma'(x)*\Phi(y) = y\sum_{n\ge 1} (xy)^{n-1}\Phi_n \ .
$$

\smallskip
{\it (ii)} As above, one can write
$$
\Phi(x)*\psi(y)
=
\ \sum_{n\ge 1}\ {(-1)^{n-1}\over n}\,
\left(\ \sum_{i\ge 1}\ x^i\, S_i \right)^n \!\! * \psi(y)
=
\left(\ \sum_{i\ge 1}\ x^i\, S_i \right)*\psi(y)
=
x\, \psi(xy) \ .
$$\cqfd

\medskip
As already mentioned in Note \ref{REGRUB}, ribbon Schur functions
appear in the commutative theory as the characteristics associated
to a certain decomposition of the regular representation of $\S_n$.
To give a precise statement, we need the notion (due to MacMahon,
{\it cf.} \cite{MM}) of {\it major index} of a permutation. Let
$\sigma\in\S_n$ have descent set
$$
\D (\sigma)
=
\{d_1,\ldots,d_r\} \subseteq \{1,\ldots,n-1\} \ .
$$
Then, by definition, $\maj (\sigma) = d_1+\cdots +d_r$. We also
define the major index of a composition $I$ to be the major
index of any permutation with descent composition $I$. That is,
if $I=(i_1,i_2,\ldots,i_m)$, then
$$
\maj (I) = (m-1)i_1+(m-2)i_2+\cdots +i_{m-1} \ .
$$
Now, if ${\cal H}_n \subset \C[x_1,\ldots,x_n]$ denotes the vector
space of $\S_n$-harmonic polynomials, and ${\cal H}_n^k$ its homogeneous
component of degree $k$, it is a classical result that its graded
characteristic as a $\S_n$-module is given by
\begin{equation}
{\cal F}_q({\cal H}_n)
:=
\ \sum_{k\ge 0} \ q^k\, {\cal F}({\cal H}_n^k)
= (q)_n\,  S_n \left( {X\over 1-q} \right) \ ,
\end{equation}
where as usual $(q)_n=(1-q)(1-q^2)\cdots(1-q^n)$, the symmetric
functions of $X/(1-q)$ being defined by the generating series
\begin{equation}
\sigma\left( {X\over 1-q}\, ,\, t \right)
= \ \prod_{k\ge 0}\ \sigma(X, tq^k) \ .
\end{equation}
On the other hand, it is also known that
\begin{equation}
(q)_n\,  S_n \left( {X\over 1-q} \right)
=
\ \sum_{|C|=n} \, q^{\maj (C)}\, R_C \ ,
\end{equation}
where the sum runs over all the compositions $C$ of $n$, so that
\begin{equation}
{\cal F}({\cal H}_n^k) = \sum_{\maj (C)=k} R_C \ .
\end{equation}
It is therefore of interest to investigate the properties
of the noncommutative symmetric functions $K_n(q)$ defined by
\begin{equation}
K_n(q) = \, \sum_{|C|=n} \, q^{\maj (C)} \, R_C \ .
\end{equation}
These functions can be seen as noncommutative analogs of the Hall-Littlewood
functions indexed by column shapes
$Q_{(11\ldots 1)} (X/(1-q)\, ;\, q)$ ({\it cf.} \cite{McD}). One can
describe their generating function
$$
\k(q) =\ \sum_{n\ge 0} \ {K_n(q)\over (q)_n}
$$
in the following way.

\begin{proposition}\label{FACTK}
One has
$$
\k(q) =
\overleftarrow{\ \prod_{k\ge 0}}\ \sigma(q^k)
=
\cdots\, \sigma(q^3)\, \sigma(q^2)\, \sigma(q)\, \sigma(1) \ .
$$
\end{proposition}

\Proof Expanding the product, we have
$$
\overleftarrow{\ \prod_{k\ge 0}}\ \sigma(q^k)
=
\cdots(\cdots + q^{n_1i_1}S_{i_1} + \cdots)
\cdots(\cdots + q^{n_2i_2}S_{i_2} + \cdots)
\cdots(\cdots + q^{n_ri_r}S_{i_r} + \cdots)\cdots
$$
$$
=
\ \sum_I \ \big( \sum_{n_1>n_2>\ldots >n_r \ge 0} \
q^{n_1i_1+n_2i_2+\cdots +n_ri_r} \ \big) \ S^I
$$
$$
=
\ \sum_I \ q^{\maj(I)} \
\big( \sum_{m_1\ge m_2\ge\ldots\ge m_r\ge 0}
\ (q^{i_1})^{m_1}\, (q^{i_2})^{m_2}\, \cdots \, (q^{i_r})^{m_r} \ \big)
\ S^I
$$
$$
=
\ \sum_I \ {q^{\maj (I)} \over
(1-q^{i_1})(1-q^{i_1+i_2})\cdots (1-q^{i_1+i_2+\cdots +i_r})} \ S^I \ .
$$
Let $F_n(q)$ be the term of weight $n$ in this series. We want to
show that $(q)_nF_n(q)=K_n(q)$. Working with subsets of
$[n-1]=\{1,\ldots,n-1\}$ rather than with compositions of $n$, we can write
$$
(q)_n \, F_n(q)
=
\ \sum_{A\subseteq [n-1]} \, S^{c(A)} \, f(A)
$$
where $\displaystyle f(A)=\prod_{t\in A}q^t\prod_{s\not\in A}(1-q^s)$,
so that
$$
(q)_n\, F_n(q)
=
\ \sum_{A\subseteq [n-1]}\, f(A)\ \big(\ \sum_{B\subseteq A}\ R_{c(B)} \ \big)
=
\ \sum_{B\subseteq [n-1]} \ R_{c(B)} \ \big(\ \sum_{A\supseteq B}\ f(A) \ \big)
$$
But, denoting by $\overline A$ the complementary subset of $A$ in $[n-1]$ and
by $\Sigma(A)=\maj (c(A))$ the sum of all elements of $A$, we see that
$$
f(A)
=
q^{\Sigma(A)}\
\big( \ \sum_{\overline C \subseteq \overline A}\ (-1)^{|\overline C|}\,
q^{\Sigma(\overline C)} \ \big)
=
\ \sum_{\overline C \subseteq \overline A}
\ (-1)^{|\overline C|}\, q^{\Sigma (A\cup\overline C)}
$$
$$
= \ \sum_{B\supseteq A}\ (-1)^{|B|-|A|}\, q^{\Sigma(B)} \ .
$$
It follows now by M\"obius inversion in the lattice of subsets of $[n-1]$
that
$$
\ \sum_{A\supseteq B}\ f(A) = q^{\Sigma(B)} = q^{\maj (c(B))} \ ,
$$
as required. \cqfd

This factorization of $\k(q)$ shows in particular that
$\k(q)$ is grouplike for $\Delta$, which is well suited for computing internal
products of the form $F*\k(q)$ by means of Proposition \ref{mackey}.
For example, with $J=(j_1,\ldots,j_m)$, we have
$$
S^J*\k(q) = \mu_m\left[
(S_{j_1}\otimes\cdots\otimes S_{j_m}) *
(\k(q)\otimes\cdots\otimes \k(q)) \right]
$$
$$
= { K_{j_1}(q)\cdots K_{j_m}(q) \over
   (q)_{j_1}\cdots (q)_{j_m} } \ ,
$$
and we obtain :

\begin{proposition}\label{S*K}
Let $J=(j_1,\ldots,j_m)$ be a composition of $n$. Then one has
\begin{equation}
S^J*K_n(q) = {n\atopwithdelims[] j_1,j_2,\ldots, j_m}_q \,
K_{j_1}(q)\, K_{j_2}(q)\, \cdots\, K_{j_m}(q) \ .
\end{equation}
\end{proposition}

Denote by $\kappa_n(q)$ the element of $\C[q][\S_n]$ such
that $\alpha(\kappa_n(q)) ={1\over n} K_n(q)\in \C[q]\otimes\Sym$.
Let $\zeta\in\C$ be a primitive $n$th root of unity.
It has been shown by Klyachko ({\it cf.} \cite{Kl}) that
$\kappa_n(\zeta)$ is a Lie idempotent. To derive this
result, we begin by considering the specialization at $q=\zeta$ of
the multinomial coefficient in Proposition \ref{S*K}, which shows
that $S^I*K_n(\zeta)=0$ as soon as $\ell(I)>1$. Hence, we get
$$
{\Phi_n\over n}*K_n(\zeta) = \log \sigma(1)*K_n(\zeta)
$$
$$
=
\ \sum_{r\ge 1} \ {(-1)^{r-1}\over r} \ \big(\
\sum_{\ell(I)=r} \, S^I * K_n(\zeta) \ \big)
=
\ \sum_{\ell(I)=1}\, S^I *K_n(\zeta) = K_n(\zeta) \ ,
$$
from which it follows that the range of the associated endomophism
$\kappa_n(\zeta)$ of $\C_n\< A\>$ is contained in the free Lie algebra.
To show that $\kappa_n(\zeta)$ is indeed an idempotent, we must
prove that $\kappa_n(\zeta)*\Phi_n =\Phi_n$. As observed in \cite{BBG},
it is possible to get a little more.

\begin{proposition}
For every $n \geq 1$, one has $K_n(q) *\Phi_n = (q)_{n-1}\, \Phi_n\ $.
\end{proposition}

\Proof Let $\k_N(q)=\sigma(q^{N-1})\, \sigma(q^{N-2})\, \cdots\, \sigma(1)$.
According to Proposition \ref{mackey}, we have
$$
\k_N(q)* \Phi_n = \mu_N\left[
\left( \sigma(q^{N-1})\otimes \sigma(q^{N-2})\otimes \cdots\otimes \sigma(1)
\right) *
\Delta^N \Phi_n \right]
$$
$$
= \left( q^{n(N-1)} + q^{n(N-2)} +\cdots +q^n +1 \right) \Phi_n
$$
by primitivity of $\Phi_n$, and taking now the limit for
$N\rightarrow\infty$, we find $\k(q) *\Phi_n=(1-q^n)^{-1}\Phi_n$, whence
$K_n(q)*\Phi_n=(q)_{n-1}\Phi_n$. \cqfd

Taking into account the fact that $(\zeta)_{n-1}=n$, we obtain
Klyachko's result.

\begin{corollary}
Let $\zeta$ be a primitive $n$th root of unity. Then, the element
$\kappa_n(\zeta)$ of $\C[\S_n]$ defined by
$$
\kappa_n(\zeta) =\ {1\over n}\sum_{\sigma\in\S_n}\ \zeta^{\maj (\sigma)}\,
\sigma
$$
is a Lie idempotent.
\end{corollary}

\begin{note}\label{PHIPSI}
{\rm The previous results show that it is natural to define noncommutative
symmetric functions of the alphabet
${1 \over 1-q}\,A$ by setting
$$
S_n({1 \over 1-q}\, A) = {K_n(q) \over (q)_n}\ .
$$
It can be shown that
$$
\lim_{q\rightarrow 1} \ \, (1-q^n) \, \Psi_n({1 \over 1-q}\, A) = \Phi_n(A)
\ ,
$$
which shows that the two families of noncommutative power
sums are closely related.
}
\end{note}

Recall that the Witt numbers $\ell_n(k)$ are defined by
$$
\ell_n(k) = {1\over n}\ \sum_{d|n}\ \mu(d)\, k^{n/d} \ ,
$$
$\mu$ being the usual M\"obius function. We can then give the following
simple characterization of Lie idempotents of the descent algebra in terms
of noncommutative symmetric functions.

\begin{theorem}\label{CARLIEID}
A symmetric function $J_n\in\Sym_n$ is the image under $\alpha$ of
a Lie idempotent of $\Sigma_n$ iff it has the form
$$
J_n = {1\over n} \Psi_n \, + \, F_n
$$
where $F_n\in L^2(\Psi)= [L(\Psi)\, ,\, L(\Psi)]$. Thus, the Lie
idempotents form an affine subspace in $\Sigma_n$. The dimension
of this affine subspace is $d_1 = 1$ for $n = 1$ and
$d_n = \ell_n(2)-1$ for $n\ge 2$.
\end{theorem}

\Proof This will once more follow from Proposition \ref{mackey}.
First, we see that $\Psi^I * \Psi_n = 0$ when $\ell(I)\ge 2$.
Hence $F_n * \Psi_n = 0$ for every $F_n\in L^2(\Psi)\cap\Sym_n$.
Since $F_n$ is primitive, we also have
$$
\Psi_n*F_n
=
\mu\left[ \left(
\lambda(-1)\otimes\sigma'(1) \right) * (F_n\otimes 1 + 1\otimes F_n)
\right]
$$
$$
\qquad = \mu[ 0 + 1\otimes(\sigma'(1)*F_n)] = n\, F_n
$$
since $\sigma'(1)$ has no term of weight zero.
So, for $F_n\in L^2(\Psi)\cap\Sym_n$ and $J_n={1\over n}\Psi_n + F_n$,
one has $\Psi_n*J_n=n\, J_n$ and $J_n*\Psi_n=\Psi_n$, which shows that
$J_n$ is the image of a Lie idempotent.

\smallskip
Conversely, suppose that $\Psi_n*J_n=n\, J_n$ and $J_n*\Psi_n=\Psi_n$.
Defining $F_n$ by the relation $J_n={1\over n}\Psi_n + F_n$, we see that
$F_n$ must satisfy $F_n*\Psi_n=0$, so that $F_n$ must be of degree $\ge 2$
in the generators $\Psi_k$, and that $\Psi_n* F_n=n\, F_n$, which
implies that $F_n$ is primitive. Indeed, Proposition \ref{ATK} shows that
$$
\Delta(n\, F_n) =  \Delta\Psi_n * \Delta F_n
=
(\Psi_n\otimes 1+1\otimes\Psi_n)*
\big( \ \sum_{(F)}\ F_{(1)}\otimes F_{(2)} \ \big)
$$
$$
= (\Psi_n*F_n)\otimes 1 + 1\otimes(\Psi_n*F_n)
$$
since any element $G\in\Sym$ satisfies $\Delta G=G\otimes 1+
1\otimes G + \sum_i P_i\otimes Q_i$, where $P_i$ and $Q_i$ are
of weight $\ge 1$.

\smallskip
The formula for the dimension follows from the Poincar\'e-Birkhoff-Witt
theorem applied to the Lie algebra $L(\Psi)$ endowed with the
weight graduation. If $h_n:= {\rm dim\, }L_n(\Psi)$, where
$L_n(\Psi)$ is the subspace of elements of weight $n$, then
$$
{1-t\over 1-2t} = \prod_{n\ge 1}\left( {1\over 1-t^n}\right)^{h_n} \ .
$$
Taking logarithms and applying M\"obius inversion,
one finds $h_n=\ell_n(2)$.  \cqfd

\def\ad{{\rm ad\,}}
\begin{note}
{\rm The fact that the dimension of the homogeneous component $L_n(\Psi)$
of {\it weight} $n$ of the Lie algebra $L(\Psi)$ is $l_n(2)$ for $n \geq 2$
can be combinatorially interpreted. For $n \geq 2$, there is an
explicit bijection between $L_n(\Psi)$ and the homogeneous component
$L_n(a,b)$ of {\it degree} $n$ of the free Lie algebra $L(a,b)$ on a two
letters alphabet. This bijection follows from Lazard's elimination theorem
(see \cite{Bour}) which says in particular that the $K$-module $L(a,b)$ is
the direct sum of two free Lie algebras
$$
L(a,b) = K \, a \oplus L(\{ (\ad  a)^n \cdot b \ , \ n \geq 0\}) \ .
$$
The desired bijection is obtained by considering the Lie morphism from
$L(\Psi)$ into $L(a,b)$ mapping $\Psi_n$ onto $(\ad a)^n \cdot b$ for
$n \geq 1$.}
\end{note}

Let us define a {\it quasi-idempotent} of a $K$-algebra ${\cal A}$ as an
element $\pi$ of ${\cal A}$ such that $\pi \cdot \pi = k \, \pi$ for some
constant $k \in K$. Using this terminology, we can restate Theorem
\ref{CARLIEID}
in the following way.

\begin{corollary}
Let $\pi_n$ be an homogeneous element of $\Sym_n$. The following
assertions are equivalent:

\medskip
\hspace{10mm}
1) $\, \pi_n$ is the image under $\alpha$ of a Lie quasi-idempotent
of $\Sigma_n$.

\smallskip
\hspace{10mm}
2) $\, \pi_n$ belongs to the Lie algebra $L(\Psi)$.

\smallskip
\hspace{10mm}
3) $\, \pi_n$ is a primitive element for $\Delta$.
\end{corollary}


\subsection{Eulerian idempotents}\label{EULERID}

The generating series corresponding to  other interesting families of
idempotents also have a simple description in  terms of noncommutative
symmetric functions. Consider for
example the coefficient of $x^r$ in the expansion of $\sigma(1)^x$,
that is, the series
\begin{equation}
E^{[r]} :=
{1\over r!}
\left(\ \sum_{k\ge 1}\ {\Phi_k\over k} \right)^r
=
\ \sum_{n\ge r} \ E_n^{[r]}
\end{equation}
where, using again the notation $\pi(I)=i_1\, i_2\, \cdots\, i_r$,
$$
E_n^{[r]} = {1\over r!}\ \sum_{|I|=n,\ell(I)=r}\ {\Phi^I\over \pi(I)} \ .
$$
It can be shown that the elements $e_n^{[r]}=\alpha^{-1}(E_n^{[r]})$
are idempotents of $\Sigma_n$, called {\it Eulerian idempotents}. They have
been introduced independently by Mielnik and Pleba\'nski \cite{MP} as
ordering operators acting of series of iterated integrals and
by Reutenauer (\cf \cite{Re86}) as projectors associated with
the canonical decomposition of the free associative algebra interpreted as
the universal enveloping algebra of a free Lie algebra. They also appear in
the works of Feigin and Tsygan \cite{FT} and of
Gerstenhaber and Schack (\cf \cite{GeS87}), where they are used
to provide  Hodge-type decompositions for various cohomology theories.
A geometric construction of these idempotents has been given
by Patras \cite{Pa1}. It seems that the  Eulerian idempotent
$e_n^{[1]}$ first appeared  in
\cite{Ba} and \cite{So1}.

\smallskip
Here is a simple argument, due to Loday (\cf \cite{Lod89} or \cite{Lod92}),
showing that the $e_n^{[r]}$ are indeed idempotents. Starting from the
definition of $E^{[1]}$, {\it i.e.} $\sigma(1)=\exp E^{[1]}$, we have
for any integer $p$
\begin{equation}
\sigma(1)^p= \exp (p\, E^{[1]})
=
\ \sum_{k\ge 0}\ p^k\, {(E^{[1]})^k\over k!}
=
\ \sum_{k\ge 0}\ p^k\, E^{[k]}
\end{equation}
(setting $E^{[0]}=1$). Now, denoting by $S_n^{[p]}$ the term of
weight $n$ in $\sigma(1)^p$, that is
$$
S_n^{[p]}= \sum_{|I|=n, \ell(I)\le p} S^I \ ,
$$
we have
\begin{equation}\label{S2E}
S_n^{[p]} = p E_n^{[1]}+p^2E_n^{[2]}+\cdots+p^n E_n^{[n]}
\end{equation}
so that the transition matrix from the $S_n^{[i]}$ to the $E_n^{[j]}$
is the Vandermonde matrix over $1,2,\ldots,n$, and hence is invertible.
Using the easily established fact that
\begin{equation}\label{compL}
S_n^{[p]}*S_n^{[q]}=S_n^{[pq]}
\end{equation}
(see below), one deduces the existence of a decomposition
$E_n^{[i]}*E_n^{[j]}=\sum_m a_{ijm}\, E_n^{[m]}$. Substituting this
relation in (\ref{compL}) by means of (\ref{S2E}), one obtains that
the coefficients $a_{ijm}$ must satisfy the equation
$$
\sum_{1\le i,j\le n} \, p^i\, q^j \, a_{ijm} = (pq)^m
$$
whose only solution is $a_{ijm}=0$ except for $i=j=m$ in which case
$a_{iii}=1$. Note also that equation (\ref{S2E}) with $p=1$ shows that the
sum of the $e^{[k]}_n$ is the identity of $\Sigma_n$, so that the
$e_n^{[k]}$ form a complete family of othogonal idempotents.

\smallskip
Equation (\ref{compL}) is an instance of a quite general phenomenon. Indeed,
in any Hopf algebra ${\cal A}$ with multiplication $\mu$ and comultiplication
$\Delta$, one can define {\it Adams operations} $\psi^p$ by setting
$\psi^p(x) =\mu_p\circ\Delta^p(x)$ (see \cite{FT}, \cite{GeS91}, \cite{Pa2}
 or \cite{Lod92b}). When
we take for ${\cal A}$ the algebra $\Sym$ of noncommutative symmetric
functions, these Adams operations are given by $\psi^p(F)=[\sigma(1)^p]*F$.
Indeed, using Proposition \ref{mackey}, we have
$$
[\sigma(1)^p]*F
=
\mu_p[(\sigma(1)\otimes\cdots\otimes\sigma(1))*\Delta^pF]
=
\mu_p\circ\Delta^p(F)
=
\psi^p(F)
$$
since $\sigma(1)$ is the unit element for $*$. This shows
that $\sigma(1)^p*\sigma(1)^q =\sigma(1)^{pq}$, which
is equation (\ref{compL}).
In this particular case, the
Adams operations can be interpolated into a one-parameter group : for any
scalar $z\not =0$, one can define $\psi^z(F)=\sigma(1)^z*F$.

\smallskip
{}From (\ref{compL}) we see that the $S_n^{[k]}$ generate a commutative
$*$-subalgebra of $\Sym_n$ which is of dimension $n$ according to (\ref{S2E}).
We shall call it the {\it Eulerian subalgebra} and denote it by $\E_n$.
The
corresponding subalgebra of $\Sigma_n$ will also be called Eulerian, and will
be denoted by ${\cal E}_n$.

\begin{example} {\rm
{}From (\ref{S2E}) we have $S_n^{[p]}*E_n^{[i]}=p^i E_n^{[i]}$, so that the
minimal polynomial of $S_n^{[p]}$ is
$f_n^{[p]}(x):=\prod_{1\le i\le n}(x-p^i)$. In \cite{GeS87}, Gerstenhaber
and Schack proved directly that the minimal polynomial of
$s_n:=\alpha^{-1}(S_n^{[2]}-S_n^{[1]})$ was $f_n^{[2]}(x-2)$,
and then obtained the Eulerian idempotents by Lagrange interpolation, that
is  by setting
$$
e_n^{[j]} = \ \prod_{i\not =j}\ (\lambda_i-\lambda_j)^{-1}\, (s_n-\lambda_i)
\ ,
$$
where $\lambda_i = 2^i\!-\! 2$. More precisely their idempotents are the
images of those presented here by the automorphism of $\Q [\S_n]$ defined
on permutations by $\sigma \mapsto \sgn (\sigma) \sigma$.
}
\end{example}


\subsection{Eulerian symmetric functions}\label{EULER}

The commutative ribbon Schur functions are known to be
related to the combinatorics of Eulerian and Euler numbers. These
numbers count sets of permutations with constrained descents.
To such a set, one can associate the sum of its elements in
the descent algebra, and interpret it as a noncommutative
symmetric function. In fact, most of the formulas satisfied by these
numbers and their commutative symmetric analogs
are specializations of identities at the level of
noncommutative symmetric functions.


\subsubsection{Noncommutative Eulerian polynomials}\label{EULPOL}

Let us recall that the Eulerian number $A(n,k)$ is defined as the
number of permutations in ${\goth S}_n$ with exactly $k-1$ descents.
Thus, $A(n,k)$ is equal to the number of standard ribbon-shaped
Young tableaux, whose shape is encoded by a composition with
exactly $k$ parts, so that
$A(n,k)/n!$ is the image by the specialization
$S_i\longmapsto 1/i!$ of the symmetric function (introduced
by Foulkes in \cite{F2}) :
\begin{equation}
{\bf A}(n,k) = \sum_{{\scriptstyle |I|=n}\atop{\scriptstyle \ell(I)=k}} R_I\ .
\end{equation}
These symmetric functions remain meaningful (and can in fact
be better understood) in the noncommutative setting.

\begin{definition} Let $t$ be an indeterminate commuting with
the $S_i$. The noncommutative Eulerian polynomials are defined by
\begin{equation}
{\cal A}_n(t) =
\ \sum_{k=1}^n \ t^k\, \Big(
\sum_{{\scriptstyle |I|=n}\atop{\scriptstyle \ell(I)=k}} R_I \,
\Big)
=
\ \sum_{k=1}^n \ {\bf A}(n,k)\, t^k \ .
\end{equation}
\end{definition}

One has for these polynomials the following identity, whose commutative
version is due to  D\'esarm\'enien (see \cite{Desar}).

\begin{proposition} The generating series of the ${\cal A}_n(t)$ is
\begin{equation}
{\cal A}(t) := \ \sum_{n\ge 0} \, {\cal A}_n(t)
=
(1-t) \, \left( 1 - t\, \sigma(1-t) \right)^{-1} \ .
\end{equation}
\end{proposition}

\Proof Using Proposition \ref{MATRANSR}, it is not difficult to see that
$$
\sum_{|I|=n} \ t^{\ell(I)} \, R_I
=
t\
\left(
\matrix{
1 &\!\! 1 \cr
}\right)^{\otimes (n-1)}
\,
\left(
\matrix{
1  & 0 \cr
-t & t
}
\right)^{\otimes (n-1)}
\,
\left(
\matrix{
S_n \cr \vdots \cr S^{1^n} \cr
}
\right)
=
t \
\left(
\matrix{
1\!-\!t &\! t \cr
}\right)^{\otimes (n-1)}
\,
\left(
\matrix{
S_n \cr \vdots \cr S^{1^n} \cr
}
\right)
\ ,
$$
which means that
\begin{equation} \label{SUMRUBn}
\sum_{|I|=n} \ t^{\ell(I)} \, R_I
=
\sum_{|I|=n} \ (1-t)^{n-\ell(I)} \, t^{\ell(I)} \, S^I \ .
\end{equation}
Hence,
$$
{\cal A}(t)
=
\ \sum_I \ t^{\ell(I)} \, R_I
=
\ \sum_{I} \ (1-t)^{|I|-\ell(I)} \, t^{\ell(I)} \, S^I \ .
$$
and it follows that
$$
{\cal A}(t)
=
\ \sum_{i\geq 0} \
\left( \displaystyle {t \over 1-t} \right)^i
\left(
\ \sum_{j\geq 1} \ (1-t)^j \, S_j
\right)^i
=
\ \sum_{i\geq 0} \
\left( \displaystyle {t \over 1-t} \right)^i
\,
(\, \sigma(1-t) - 1 \,)^i
$$
$$
=
(\, 1 - \displaystyle{t \over 1-t} \, ( \, \sigma(1-t) - 1\, ) \, )^{-1} \ .
$$\cqfd

\medskip
Let us introduce the notation ${\cal A}_n^*(t) = (1-t)^{-n}\, {\cal A}_n(t)$.
Using (\ref{SUMRUBn}), we see that
$$
{\cal A}^*(t)
:=
\ \sum_{n\ge 0} \, {\cal A}_n^*(t)
=
\sum_{I} \
\left( \displaystyle {t \over 1-t} \right)^{\ell(I)} \, S^I \ .
$$
This last formula can also be written in the form
\begin{equation} \label{GEN*}
{\cal A}^*(t)
=
\ \sum_{k\ge 0} \ \left(
{t\over 1-t}\right)^k \left( S_1+S_2+S_3+\cdots\, \right)^k
\end{equation}
or
\begin{equation}\label{GEN_A}
{1\over 1-t\, \sigma(1)}
=
\ \sum_{n\ge 0}\ {{\cal A}_n(t)\over (1-t)^{n+1}} \ .
\end{equation}

We can also prove equation (\ref{GEN*}) directly as follows. Writing
$t/(1-t)=x$, the right-hand side is
$$
\sum_I \ S^I\, x^{\ell(I)} = \ \sum_I \, x^{\ell(I)} \
\big(\ \sum_{J\preceq I}\ R_J \ \big)
$$
$$
=\ \sum_J \ R_J \ \big( \ \sum_{J\preceq I}\ x^{\ell(I)}\ \big)
=\ \sum_J \ R_J \, x^{\ell(J)}\, (1+x)^{|J|-\ell(J)}
$$
(using the bijection between compositions $I$  of $n=|J|$
and subsets of $\{1,2,\ldots,n-1\}$ and the fact that
reverse refinement corresponds to inclusion)
$$
=\ \sum_J\ R_J\ {t^{\ell(J)}\over(1-t)^{\ell(J)}}\cdot
\Bigl( {1\over 1-t} \Bigr)^{n-\ell(J)}
=\ \sum_{n\ge 0}\ {\cal A}_n^*(t) \ .
$$

It is not difficult to show, using these identities, that most
classical formulas on Eulerian numbers (see {\it e.g.} \cite{FS})
admit a noncommutative symmetric analog. Several of these formulas
are already in the literature, stated in terms
of permutations. The following ones have a rather simple interpretation
in the noncommutative setting : they just give the transition matrices
between several natural bases of the Eulerian subalgebras.

\smallskip
First, expanding the factors $(1-t)^{-(n+1)}$ in the right-hand side
of (\ref{GEN_A}) by the binomial
theorem, and taking the coefficient of $t^k$ in the term of
weight $n$ in both sides, we obtain
\begin{equation}
S_n^{[k]}
=
\ \sum_{i=0}^k \ {n+i \choose i} \, {\bf A}(n,k-i) \ ,
\end{equation}
an identity due to Loday, which shows in particular that the noncommutative
Eulerian functions also span the Eulerian subalgebra. Similarly, one has
$$
{ {\cal A}_n(t) \over (1-t)^{n+1} }
=
\ \sum_{k\ge 0}\ t^k\, S_n^{[k]} \ ,
$$
so that
\begin{equation}
{\bf A}(n,p)
=
\ \sum_{i=0}^p\ (-1)^i\, {n+i\choose i}\, S_n^{[p-i]} \ .
\end{equation}
Another natural basis of the Eulerian subalgebra $\E_n$ is constituted
by the elements
\begin{equation}
M_n^{[k]}
:=
\sum_{|I|=n, \ell(I)=k}\, S^I \ .
\end{equation}
Indeed, putting $x=t/(1-t)$, we have
$$
\sum_{k\ge 0}\ x^k \, (S_1+S_2+\cdots)^k
=
\ \sum_{n\ge 0}\ {\cal A}_n^*\left( {x\over 1+x}\right)
=
\ \sum_{n\ge 0}\ (1+x)^n\, {\cal A}_n \left( {x\over 1+x}\right) \ ,
$$
so that
\begin{equation}
\sum_{k=1}^n\ x^k\, M_n^{[k]}
=
\ \sum_{j=1}^n\ x^j\, (1+x)^{n-j}\, {\bf A}(n,j) \ .
\end{equation}
Another kind of generating function can be obtained by writing
$$
\sigma(1)^x =
[1+(S_1+S_2+\cdots )]^x
=
\ \sum_{k\ge 0}\ {x \choose k}\, M^{[k]} \ .
$$
Comparing with $\sigma(1)^x = \exp\, ( x\, E^{[1]})$, it follows that
\begin{equation}
\sum_{k=1}^n\, x^k\, E_n^{[k]}
=
\ \sum_{k=1}^n \ {x\choose k}\, M_n^{[k]} \ .
\end{equation}
One obtains similarly the following expansion of the $E_n^{[k]}$
on the basis ${\bf A}(n,i)$, which is a noncommutative analog
of Worpitzky's identity (see \cite{Ga} or \cite{Lod89}) :
\begin{equation}
\sum_{k=1}^n \ x^k\, E_n^{[k]}
=
\ \sum_{i=1}^n \ {x-i+1\choose n}\, {\bf A}(n,i) \ .
\end{equation}


\subsubsection{Noncommutative trigonometric functions}\label{TRIG}

The Euler numbers (not to be confused with the
Eulerian numbers of the preceding section), defined by
$$
\sum_{n\ge 0}\ E_n\, {x^n\over n!}=\tan x + \sec x
$$
can also be given a combinatorial interpretation (see \cite{Andre}) which
is again best understood in terms of symmetric functions (see \cite{F1},
\cite{F2}
or \cite{Desar}). As shown by D. Andr\'e, $E_n$ is equal to the number of
{\it alternating} permutations of ${\goth S}_n$ ($\sigma$ is alternating
if $\sigma_1 < \sigma_2 > \sigma_3 < \ldots\ $). An alternating permutation
is thus the same as a skew standard tableau of ``staircase ribbon'' shape,
{\it i.e.} a ribbon shape indexed by the composition
$C_{2k}=(2^k)=(2,2,\ldots,2)$ if $n=2k$ or $C_{2k+1}=(2^k1)=(2,2,\ldots,2,1)$
if $n=2k+1$. So the staircase ribbons
provide symmetric analogs of the $E_n/n!\, $. As in the preceding section,
it is possible to replace the commutative ribbons by the noncommutative
ones. In fact, most of the important properties of the tangent
numbers $E_{2k+1}$ are specializations of general identities valid
for the coefficients of the quotient of an odd power series by an
even one. Such identities can also be given for noncommutative
power series, but one has to distinguish between right and left
quotients.
The existence of a noncommutative
interpretation of trigonometric functions has been first observed by
Longtin (\cf \cite{Lo}, see also \cite{MaR}).

\begin{definition} The noncommutative trigonometric functions associated
to the noncommutative generic series $\sigma(t)=\, \sum_{n\ge 0}\, S_n \, t^n$
are
$$
SIN = \ \sum_{i\ge 0}\ (-1)^i\, S_{2i+1}\ , \quad
COS = \ \sum_{i\ge 0}\ (-1)^i\, S_{2i}
$$
$$
SEC = (COS)^{-1}\ ,\quad TAN_r = SIN\cdot (COS)^{-1}\ ,\quad
TAN_l = (COS)^{-1} \cdot SIN \ .
$$
\end{definition}

\begin{definition} The noncommutative Euler symmetric functions are
defined by
$$
T_{2n} = R_{(2^n)}\ ,\quad T_{2n+1}^{(r)} = R_{(1\,2^n)}\ ,
\quad T_{2n+1}^{(l)} =R_{(2^n\, 1)}\ .
$$
\end{definition}

These symmetric functions give, as in the commutative case, the
expansion of the secant and of the tangent :

\begin{proposition} \label{SECTT}
One has the following identities :
\begin{equation}
SEC = 1 + \ \sum_{n\ge 1} \ T_{2n} \ ,
\end{equation}
\begin{equation}
TAN_l =\ \sum_{n\ge 0}\ T_{2n+1}^{(l)} \ ,\quad
TAN_r =\ \sum_{n\ge 0}\ T_{2n+1}^{(r)} \ .
\end{equation}
\end{proposition}

\Proof The three identities being proved in the same way, we only give
the proof for the $SEC$ function. In this case,
it is sufficient to prove that
\begin{equation} \label{PRODEUL1}
\left( \ \sum_{i\geq 0} \ (-1)^i \, R_{2i} \, \right) \,
\left( \, 1 + \ \sum_{j\geq 1} \ R_{2^j} \, \right)
=
\ \sum_{i\geq 0} \ (-1)^i \, R_{2i} \
+
\ \sum_{i\geq 0,j\geq 1} \ (-1)^i \, R_{2i} \, R_{2^j}
=
1 \ .
\end{equation}
But, according to Proposition \ref{MR},
$$
R_{2i} \, R_{2^j} =
\left\{ \
\matrix{
R_{2i,2^j} + R_{2i+2,2^{j-1}} & \ \ \hbox{when} \ j \geq 2 \vtr{3} \cr
R_{2i,2} + R_{2i+2} & \ \ \hbox{when} \ j = 1 \cr
}
\ \right.
$$
for $i \geq 1$. Hence the product in the left-hand side of relation
(\ref{PRODEUL1}) is equal to
$$
\sum_{i\geq 0} \ (-1)^i \, R_{2i} \
+
\ \sum_{j \geq 1} \, R_{2^j}
+
\ \sum_{i\geq 1} \ (-1)^i\, (\, R_{2i,2} + R_{2i+2}\, )
+
\ \sum_{i\geq 1,j\geq 2} \ (-1)^i \, (\, R_{2i,2^j} + R_{2i+2,2^{j-1}} \, )
\ ,
$$
which can be rewritten as follows
$$
1 - R_2
+
\ \sum_{j \geq 1} \, R_{2^j}
+
\ \sum_{i\geq 1} \ (-1)^i\, R_{2i,2}
+
\ \sum_{i\geq 1,j\geq 2} \ (-1)^i \, R_{2i,2^j}
-
\ \sum_{i\geq 2, j\geq 1} \ (-1)^i R_{2i,2^j}
$$
$$
=
1 - R_2
+
\ \sum_{j \geq 1} \, R_{2^j}
+
\ \sum_{i\geq 1,j\geq 1} \ (-1)^i \, R_{2i,2^j}
-
\ \sum_{i\geq 2, j\geq 1} \ (-1)^i R_{2i,2^j}
$$
$$
=
1 - R_2
+
\ \sum_{j \geq 1} \, R_{2^j}
-
\ \sum_{j\geq 1} \ R_{2,2^j}
=
1 \ .
$$\cqfd

It can also be convenient to consider the symmetric hyperbolic functions
$$
COSH =\ \sum_{i\ge 0} \ S_{2i} \, , \quad SINH =\ \sum_{i\ge 0} \ S_{2i+1} \, ,
\quad SECH = (COSH)^{-1} \, , \ \
$$
$$
TANH_l = SINH \cdot SECH  \quad \hbox{and} \quad  TANH_r = SECH\cdot SINH \ .
$$
Then, arguing as in the proof of Proposition \ref{SECTT}, one can check by
means of Proposition \ref{MR} that one has the following expansions :
$$
SECH = 1 + \ \sum_{n\ge 1}\ (-1)^n\, T_{2n}\, ,\ \
$$
$$
TANH_l =\ \sum_{n\ge 0}\ (-1)^n\, T_{2n+1}^{(l)} \quad {\rm and} \quad
TANH_r =\ \sum_{n\ge 0}\ (-1)^n\, T_{2n+1}^{(r)} \ .
$$
As an illustration of the use of these functions let us consider the
sums of hooks
\begin{equation}
H_n =\ \sum_{k=0}^{n-1}\ R_{1^k,n-k} \ .
\end{equation}
In the commutative case these functions appear for example as the moments
of the symmetric Bessel polynomials (see \cite{Le}, \cite{LT}). Here, we
compute the inverse of their generating series :

\begin{proposition} Let ${\bf H}=\, \sum_{n\ge 0}\, H_n$ with $H_0 = 1$.
Then, one has
$$
{\bf H}^{-1} = 1 -\, \sum_{n\ge 0}\ (-1)^n\, T_{2n+1}^{(r)} = 1 - TANH_r \ .
$$
\end{proposition}

\Proof Since $1-TANH_r = COSH^{-1} \, (\, COSH-SINH\, )$, the identity to be
proved is  equivalent to
$$
(\ \, \sum_{i\geq 0} \ (-1)^i \, R_{i} \, ) \
(\, 1+ \, \sum_{k\geq 0,l\geq 1} \ R_{1^k,l} \, )
=
\ \sum_{j\geq 0} \ R_{2i} \ ,
$$
which can be rewritten as
\begin{equation} \label{PRODEUL2}
\sum_{i\geq 0} \ (-1)^i \, R_{i}
+
\ \sum_{k,i\geq 0,l\geq 1} \ (-1)^i \, R_i \, R_{1^k,l}
=
\ \sum_{j\geq 0} \ R_{2i} \ .
\end{equation}
But Proposition \ref{MR} shows that
$$
R_i \, R_{1^k,l}
=
\left\{
\matrix{
R_{i,1^k,l} + R_{i+1,1 ^{k-1},l} & \ \ \hbox{when} \ k\geq 1 \vtr{3} \cr
R_{i,l} + R_{i+l} & \ \ \hbox{when} \ k = 0
}
\ \right. \ .
$$
Hence the left-hand side of equation (\ref{PRODEUL2}) can be
rewritten in the following way
$$
\sum_{i\geq 0} \ (-1)^i \, R_{i}
+
\, \sum_{k\geq 0,l\geq 1} \ R_{1^k,l}
+
\, \sum_{i,l\geq 1} \ (-1)^i \, (\, R_{i,l} + R_{i+l}\, )
+
\, \sum_{i,k,l\geq 1} \ (-1)^i \, (\, R_{i,1^k,l} + R_{i+1,1^{k-1},l}\, )
\ ,
$$
which is itself equal to
$$
\sum_{i\geq 0} \ (-1)^i \, R_{i}
+
\, \sum_{k\geq 0,l\geq 1} \ R_{1^k,l}
+
\, \sum_{i,l\geq 1} \ (-1)^i \, R_{i,l}
+
\, \sum_{i,l\geq 1} \ (-1)^i \, R_{i+l}
$$
$$
+
\, \sum_{i,k,l\geq 1} \ (-1)^i \, R_{i,1^k,l}
+
\, \sum_{i\geq 2,k\geq 0,l\geq 1} \ (-1)^{i-1} \, R_{i,1^{k},l}
\ .
$$
Using now the fact that one has
$$
\sum_{i,l\geq 1} \ (-1)^i \, R_{i+l}
=
- \ \sum_{k\geq 1} \ R_{2k}
$$
and
$$
\sum_{i,k,l\geq 1} \ (-1)^i \, R_{i,1^k,l}
=
\, \sum_{i\geq 2,k,l\geq 1} \, (-1)^i \, R_{i,1^k,l}
- \, \sum_{k\geq 2,l\geq 1} \, R_{1^k,l} \ ,
$$
it is easy to see that this last expression is equal to
$$
1 - \ \sum_{i\geq 0} \ R_{2i+1}
+
\ \sum_{l\geq 1} \ R_{l}
+
\ \sum_{l\geq 1} \ R_{1,l}
+
\, \sum_{i,l\geq 1} \ (-1)^i \, R_{i,l}
+
\, \sum_{i\geq 2,l\geq 1} \ (-1)^{i-1} \, R_{i,l}
=
\ \sum_{i\geq 0} \ R_{2i} \ .
$$\cqfd
%


\subsection{Continuous Baker-Campbell-Hausdorff formulas}\label{BCHCONT}

In this section, we shall obtain the main results of \cite{BMP} and \cite{MP}
by means of computations with the internal product and noncommutative Eulerian
polynomials. To explain the method, we first treat the problem of the
continuous BCH exponents (\cf Section \ref{PhiPsi}).

For us, the problem is to express $\Phi(t)$ in terms of the $\Psi^I$. The
starting point is to write
\begin{equation}\label{REPRO}
\Phi(t) = \Phi(1) * \sigma(t)
\end{equation}
using $\sigma(t)$ as a reproducing kernel, and expressing it in the form
$$
\sigma(t) = 1+\sum_{r\ge 1}\int_0^t dt_1\cdots\int_0^{t_{r-1}}dt_r \
\psi(t_r)\cdots\psi(t_1) \ .
$$
The problem is thus reduced to the computation of
$\Phi(1)*\psi(t_r)\cdots\psi(t_1)$,
which is given by the following lemma:

\begin{lemma} \label{LEMBCH}
Let $F_1,\ldots,F_r$ be primitive for $\Delta$. Then,
$$
\Phi(1)*(F_1\cdots F_r) =
\sum_{\sigma\in\S_r} {(-1)^r\over r} {r-1\choose d(\sigma)}^{-1}
F_{\sigma(1)}\cdots F_{\sigma(r)} \ .
$$
That is, if $\phi_r := e_r^{[1]}=\alpha^{-1}(\Phi_r/r)$,
$$
\Phi(1)*(F_1\cdots F_r) = (F_1\cdots F_r)\cdot\phi_r \ ,
$$
the right action of a permutation $\sigma\in\S_r$ on $F_1\cdots F_r$
being as usual $(F_1\cdots F_r)\cdot \sigma= F_{\sigma(1)}\cdots
F_{\sigma(r)}$.
\end{lemma}

\Proof Since $\Phi(1) =\log (1+(\sigma(1)-1))$,
$$
\Phi(1)*(F_1\cdots F_r) =\sum_{k\ge 1} {(-1)^k\over
k}(\sigma(1)-1)^k*(F_1\cdots F_r)
\ ,
$$
and by Proposition \ref{mackey}
$$
(\sigma(1)-1)^k*(F_1\cdots F_r) =
\mu_k\left[
(\sigma(1)-1)^{\otimes k} *
\sum_{A_1,\ldots,A_k}
F_{a_1^1}\cdots F_{a_{m_1}^1} \otimes
\cdots\otimes F_{a_1^k}\cdots F_{a_{m_k}^k}\right]
$$
where the sum runs over all decompositions $\{1,\ldots,r\}=A_1\cup\cdots\cup
A_k$
into disjoint subsets $A_i=\{a_1^i<\ldots <a_{m_i}^i\}$. This expression
is equal to
$$
\sum_{d(\sigma)\le k-1} F_{\sigma(1)}\cdots F_{\sigma(r)} \ ,
$$
so that
$$
\Phi(1)* (F_1\cdots F_r) =
(F_1\cdots F_r)\cdot \sum_{k\ge 1}{(-1)^{k-1}\over k}\sum_{d(\sigma)\le
k-1}\sigma
= (F_1\cdots F_r)\cdot\phi_r \ .
$$\cqfd

Using the step function
$$
\theta(t) =
\left\{\matrix{ 1 & t>0 \cr 0 & t<0\cr }\right.
$$
and the notations
$$
\theta_{i,j} =\theta(t_i-t_j), \quad \Theta_n=\theta_{1,2}+\theta_{2,3}+\cdots+
\theta_{n-1,n} \ ,
$$
formula (\ref{BCHCP}) can be rewritten in the form of \cite{MP}

\begin{equation}
\Phi(t)= \sum_{r\ge 1}\int_0^t\cdots\int_0^t dt_1\cdots dt_r
{(-1)^{\Theta_r}\over r} {r-1\choose \Theta_r}^{-1}
\psi(t_r)\cdots \psi(t_1)
\end{equation}
which can be converted into a Lie series by means of Dynkin's theorem, since
we know that $\Phi(t)$ is in the Lie algebra generated by the $\Psi_i$, that
is,

\begin{equation}
\Phi(t)= \sum_{r\ge 1}\int_0^t\cdots\int_0^t dt_1\cdots dt_r
{(-1)^{\Theta_r}\over r^2} {r-1\choose \Theta_r}^{-1}
\{\psi(t_r)\cdots \psi(t_1)\} \ ,
\end{equation}
where $\{\psi(t_r)\cdots \psi(t_1)\}=\ad\psi(t_r)\cdots\ad\psi(t_2)
(\psi(t_1))$.

Looking at the proof of Lemma \ref{LEMBCH}, one observes that the argument
only depends of the fact that $\Phi(1)$ is of the form $g(\sigma(1)-1)$,
where $g(t)=\sum_{n\ge 0}g_nt^n$ is a power series in one variable.
More precisely, one has the following property, which allows for a similar
computation of any series of the form $g(\sigma(t)-1)$.

\begin{lemma}\label{LEMG}
Let $g(t)=\sum_{n\ge 0} g_nt^n \in K[[t]]$ be a formal power series in
one variable, $G(t):= g(\sigma(t)-1) =\sum_{n\ge 0} G_n t^n$, $G_n\in \Sym_n$,
and $\gamma_n:=\alpha(G_n)\in\Sigma_n$. Then, if the series $F_1,\ldots,F_r$
are primitive elements for $\Delta$,
$$
G(1)*(F_1\cdots F_r) = (F_1\cdots F_r)\cdot\gamma_r \ .
$$
\end{lemma}
\cqfd

Using $\sigma(t)$ as reproducing kernel as in (\ref{REPRO}), one obtains:

\begin{corollary}\label{CORBCH}
The expression of $G(t)$ in the basis $\Psi^I$ is given by
$$
G(t)=\sum_{r\ge 0} \int_0^t dt_1\cdots\int_0^{t_{r-1}}dt_r
(\psi(t_r)\cdots \psi(t_1))\cdot\gamma_r \ .
$$
\end{corollary}
\cqfd

In particular, with $g(t)=(1+t)^x$, one finds
\begin{equation}
\sigma(t)^x = 1+\sum_{r\ge 1}\int_0^t dt_1\cdots\int_0^{t_{r-1}}dt_r
\sum_{k\ge 0} x^k\,
(\psi(t_r)\cdots \psi(t_1))\cdot e_r^{[k]}
\end{equation}
where the $e_r^{[k]}$ are the Eulerian idempotents. Using the expression
of Eulerian idempotents on the basis of descent classes and changing
the variables in the multiple integrals, one can as above get rid of the
summations over permutations by introducing an appropriate kernel.

As explained in \cite{BMP} and \cite{MP}, such a transformation is always
possible,
and to find the general expression of the kernel associated to an arbitrary
analytic function $g(t)$, we just have to find it for the particular
series
$$
f_z(t) ={1\over z-t} :=\sum_{n\ge 0} {1\over z^{n+1}} t^n
$$
since by the Cauchy integral formula
$$
G(t) = {1\over 2\pi i} \oint_{z=0} {g(z)\over z-(\sigma(t)-1)} dz \ .
$$

Using Corollary \ref{CORBCH}, we have
$$
F_z(t)=\sum_{r\ge 0}\int_0^t dt_1\cdots \int_0^{t_{r-1}} dt_r
(\psi(t_r)\cdots \psi(t_1))\cdot \gamma_r
$$
where the $\gamma_r=\alpha(F_{z,r})$ are given by
$$
\sum_{r\ge 0}F_{z,r}= {1\over z-(\sigma(1)-1)}
= {1\over z+1}\cdot {1\over 1-(z+1)^{-1}\sigma(1)}
=\sum_{n\ge 0}
{ {\cal A}_n \left( {1\over z+1} \right)
   \over
[1 - (z+1)^{-1}]^{n+1} }
$$
$$
=\sum_{n\ge 0} \left( {z+1\over z}\right)^{n+1} {\cal A}_n \left({1\over
z+1}\right)
=\sum_I {(z+1)^{ |I|-(\ell(I)-1) } \over z^{|I|+1} } \, R_I
$$
(by formula (\ref{GEN_A})).
Thus,
\begin{equation}\label{gamma_r}
\gamma_r = {1\over z^{r+1}} \sum_{\sigma\in\S_r} (z+1)^{r-d(\sigma)}\sigma
\end{equation}
and
$$
F_z(t) =
\sum_{r\ge 0} \int_0^t\cdots \int_0^t dt_1\cdots dt_r \
{1\over z^{r+1}} (z+1)^{\Theta_r}\  \psi(t_r)\cdots \psi(t_1)
$$
so that
$$
G(t)= {1\over 2\pi i} \oint_{z=0} g(z) F_z(t) dz
$$
$$
= \sum_{r\ge 0} \int_0^t\cdots \int_0^t dt_1\cdots dt_r \
K_r[g](t_1,\ldots,t_r)\  \psi(t_r)\cdots \psi(t_1) \ ,
$$
the kernels being given by
\begin{equation}
K_r[g](t_1,\ldots,t_r) =
{1\over 2\pi i} \oint_{z=0} {g(z)\over z^{r+1}} (z+1)^{\Theta_r} dz \ .
\end{equation}

One sees from equation (\ref{gamma_r}) that $\gamma_r$ is always
in the Eulerian subalgebra, so that the direct sum ${\cal E}=\bigoplus
{\cal E}_r$ appears as a natural operator algebra acting on Chen's series.
This observation, due to Mielnik and Pleba\'nski, has to be related to
the fact that Chen's complex for the loop space \cite{Chen35} can be decomposed
under the action of the Eulerian idempotents (see \cite{Pa2}, p. 74).


\def\QS{{\sl\widehat{Qsym}}}
\def\qs{{\sl Qsym}}
\def\+{\hat +}
\def\x{\hat\times}
\newpage
\section{Duality}\label{DUAL}

As recalled in section \ref{OUTLINE}, an important feature
of the algebra $\sym$ of commutative symmetric functions
is that it is a self-dual Hopf algebra (for the coproduct
$\Delta$ and the standard scalar product). Moreover, the
commutative internal product $*$ is dual to the second coproduct
$\delta : F\mapsto F(XY)$, and the two bialgebra structures are
intimately related ({\it e.g.} each coproduct is a morphism for
the product dual to the other one).

\smallskip
Such a situation cannot be expected for the noncommutative
Hopf algebra $\Sym$. A reason for this is that $\Delta$
is cocommutative, and  cannot be dual to a noncommutative
multiplication. One has thus to look for the dual bialgebra
$\Sym^*$ of $\Sym$. It follows from a recent
work by Malvenuto and Reutenauer (\cf \cite{MvR}) that this
dual can be identified with another interesting
generalisation of symmetric functions, that is, the
algebra $\QS$ of {\it quasi-symmetric functions},
whose definition is due to Gessel (see \cite{Ge}).


\subsection{Quasi-symmetric functions}\label{QSYF}

Let $X = \{ x_1 < x_2 < x_3 < \ldots\, \}$ be an infinite totally
ordered set of {\it commuting} indeterminates. A formal
series $f\in \Q[[X]]$ is said to be {\it quasi-symmetric}
if for any two finite sequences $y_1<y_2<\ldots <y_k$
and $z_1<z_2<\ldots <z_k$ of elements of $X$, and
any exponents $i_1,i_2,\ldots,i_k \in \N$, the monomials
$y_1^{i_1}\, y_2^{i_2}\,\cdots\, y_k^{i_k}$ and
$z_1^{i_1}\,z_2^{i_2}\, \cdots\, z_k^{i_k}$ have the same
coefficient in $f$.

\smallskip
The quasi-symmetric series (resp. polynomials) form
a subring denoted by $\QS$ (resp. $\qs$) of $\Q[[X]]$,
naturally graded by the graduation $(\QS_n)$ (resp. $(\qs_n)$)
inherited from $\Q[[X]]$. A natural basis of $\qs_n$ is then
provided by the {\it quasi-monomial functions}, defined by
$$
M_I = \sum_{y_1<y_2<\ldots <y_k} \,
y_1^{i_1}\, y_2^{i_2}\,\cdots\, y_k^{i_k}
$$
where $I=(i_1,\ldots,i_k)$ is any composition of $n$. In particular, the
dimension of $\qs_n$ is $2^{n-1}={\rm dim\, }\Sym_n$. Another convenient
basis, also introduced in \cite{Ge} is constituted by the functions
$$
F_I = \ \sum_{J\succeq I} \ M_J
$$
which we propose to call {\it quasi-ribbons}.

\smallskip
Let $Y$ be a second infinite totally ordered set of
indeterminates and denote by $X\+ Y$ the ordered
sum of $X$ and $Y$, {\it i.e.} their disjoint union
ordered by $x<y$ for $x\in X$ and $y\in Y$, and by
the previous orderings on $X$ and $Y$. Using the standard
identification $\QS\otimes\QS \equiv \QS(X,Y)$ (series
which are separately quasi-symmetric in $X$ and $Y$) defined by
$f\otimes g \equiv f(X)\, g(Y)$, Malvenuto and
Reutenauer (\cf \cite{MvR}) defined a coproduct $\gamma$
on $\QS$ by setting
\begin{equation}
\gamma (f) = f(X\+ Y)
\end{equation}
for $f \in \QS$. They show that $\QS$ becomes then a Hopf algebra,
with antipode $\nu$ given by
\begin{equation}
\nu (F_I) = (-1)^{|I|} \, F_{I^\sim}
\end{equation}
where $C^\sim$ is the conjugate composition. These operations, when
restricted to $\sym\subset\QS$, coincide with the usual ones. That is,
when $f$ is a symmetric function, $\gamma(f)=f(X+Y)$ and $\nu(f)=f(-X)$.

\smallskip
As shown in \cite{Ge},
the other coproduct $\delta$ of $\sym$ can also
be extended to $\QS$. Define $X\x Y$ to be
$XY$ endowed with the lexicographic ordering,
{\it i.e} $x_1y_1 < x_2y_2$ iff $x_1<x_2$ or
$x_1=x_2$ and $y_1<y_2$. The extension of $\delta$
to $\QS$ is then defined by setting
$$
\delta (f) = f(X\x Y) \
$$
for $f \in \QS$.


\subsection{The pairing between $\Sym$ and $\QS$}\label{PAIRING}

It has been shown in \cite{Ge} that the dual space
$\qs_n^*$ endowed with the product adjoint to $\delta$
is anti-isomorphic to the descent algebra $\Sigma_n$.
The dual Hopf algebra $\QS^*$ of the whole algebra
of quasi-symmetric series has been recently identified
in \cite{MvR}. If one introduces a pairing between
$\QS$ and $\Sym$ by setting
\begin{equation}\label{PAIR}
\< M_I\, ,\, S^J\> = \delta_{IJ} \
\end{equation}
for every compositions $I,J$, the results of \cite{Ge} and \cite{MvR}
which are relevant for our purposes can be summarized in the following
theorem.

\begin{theorem}
\begin{enumerate}
\item The pairing {\rm (\ref{PAIR})} induces an isomorphism of
Hopf algebras, given by $(S^J)^*\mapsto M_I$, between
the dual $\Sym^*$ of $\Sym$ and the Hopf algebra
$\QS$ of quasi-symmetric series (or equivalently,
an isomorphism between the graded dual $\Sym^{* gr}$
and the polynomial quasi-symmetric functions $\qs$).
More precisely, one has for $f,g\in\QS$ and $P,Q\in \Sym$
\begin{equation}
\<f\, ,\, PQ\> = \<\gamma f\, ,\, P\otimes Q\>
\end{equation}
\begin{equation}
\<fg\, ,\, P\> = \<f\otimes g\, ,\, \Delta P\>
\end{equation}
\begin{equation}
\<f\, ,\, \tilde\omega P\> = \<\nu f\, ,\, P\> \ .
\end{equation}

\item Moreover, the coproduct $\delta$ of $\QS$ is
dual to the internal product $*$ of $\Sym$:
\begin{equation}
\<\delta f\, ,\, P\otimes Q\> = \<f\, ,\, P*Q\> \ .
\end{equation}

\item The quasi-ribbons are dual to the ribbons,
{\it i.e.} $\<F_I\, ,\, R_J\> = \delta_{IJ}$.

\item The antipode $\nu$ of $\QS$ is given by
$\nu F_C=(-1)^{|C|}\, F_{C^\sim}$.

\item Let $\tau$ be any permutation with descent
composition $D(\tau):=c({\rm Des\, }\tau)=C$. Then,
\begin{equation}
\delta F_C =\ \sum_{\sigma\pi=\tau} \ F_{D(\pi)}\otimes F_{D(\sigma)} \ .
\end{equation}

\item If $g\in\QS$ is a symmetric function, then
\begin{equation}
\<g\, ,\, R_C\> = (g\, ,\, R_C)
\end{equation}
where, in the right-hand side, $R_C$ stands for the {\rm commutative}
ribbon Schur function. In other words, $g=\, \sum_C\, (g\, ,\, R_C)\, F_C$.
\end{enumerate}
\end{theorem}

The pairing (\ref{PAIR}) can be better understood by means of an
analog of the classical Cauchy formula of the
commutative theory. Let $A$ be a virtual noncommutative alphabet,
and $X$ a totally ordered commutative alphabet as above. Then,
one can define the symmetric functions of the noncommutative alphabet
$XA$ by means of the generating series
\begin{equation}
\sigma(XA,1) = \ \overrightarrow{\prod_{x\in X}} \ \sigma(A,x) \ ,
\end{equation}
the above product being taken with respect to the total order of $X$. The
expansion of this infinite product leads at once to the relation
\begin{equation}
\sigma(XA,1)= \ \sum_I \ M_I(X)\, S^I(A) \ .
\end{equation}
Expanding each $S^I$ on the ribbon functions, one obtains the following
identities :
\begin{equation}
\sigma(XA,1) = \ \sum_I \ M_I(X)\ \big( \ \sum_{J\preceq I} \ R_J(A) \ \big)
=\ \sum_J \ \big(\ \sum_{I\succeq J}\ M_I(X)\ \big) \ R_J(A)
=\ \sum_J \ F_J(X)\, R_J(A) \ .
\end{equation}
More generally, for any basis $(U_I)$ of $\Sym$ with dual
basis $(V_I)$ in $\qs$, one has
\begin{equation}
\sigma(XA,1) = \ \sum_I \ V_I(X)\, U_I(A) \ .
\end{equation}
This property can be used to describe the dual bases of the various
bases of $\Sym$. Also, the proof of Proposition \ref{FACTK} can be
interpreted as a computation of the specialization $X=\{1,q,q^2,\ldots \}$
of the quasi-symmetric functions $M_I$ and $F_I$.

\newpage
\section{Specializations}\label{SPECIA}

In this section, we study several interesting cases of specialization
of the symmetric functions defined in Section \ref{FORMAL}. In particular,
we exhibit two realizations of the specialization $\Lambda_k=0$ for $k>n$
by functions of $n$ noncommuting variables which are symmetric in
an appropriate sense. We also consider  extensions of the theory to skew
polynomial
rings, and another kind of specialization associated with a noncommutative
matrix. The use of these matrix symmetric functions is illustrated
by some computations in the universal enveloping algebra $U(gl(n,\C))$.

\smallskip
In some cases, it is of interest to consider specializations of general
quasi-Schur
functions. On these occasions, the word specialization means
`specialization of the free field
$K\!\not< \! S \!\not> := K\!\not< \! S_0,\S_1,\,S_2,\, \ldots \!\not>$',
that is, a ring homomorphism $\eta$ defined on a subring $R_\eta$ of $K\!\not<
\! S \!\not>$
containing $\Sym$ such that any element of $R_\eta$ not in the kernel of $\eta$
has an inverse in $R_\eta$. For more details on the category of fields and
specializations, see \cite{Co}.


\subsection{Rational symmetric functions of $n$ noncommutative variables}
\label{RATSYM}

We fix $n$ noncommutative indeterminates $x_1,\ x_2,\ \ldots ,\ x_n$, the
variable $t$ still being a commutative indeterminate. We set $x = t^{-1}$.
In the commutative case, the quasi-determinant
$$
g(x)=
\left|\matrix{
1      & \ldots & 1     & 1     \cr
x_1    & \ldots & x_n   & x     \cr
x_1^2  & \ldots & x_n^2 & x^2   \cr
\vdots &        &\vdots &\vdots \cr
x_1^n  & \ldots & x_n^n & \bo{x^n}   \cr
}\right|
$$
reduces to the polynomial
$$
(x-x_1)\, (x-x_2)\,\ldots\, (x-x_n)
=
t^{-n} \, (1-t\, x_1)\, (1-t\,x_2) \, \ldots\, (1-t\, x_n)
\ .
$$
In the noncommutative case, basic properties of  quasi-determinants
imply that $g(x)$ is again a monic (left) polynomial
of degree $n$, which vanishes under the substitution $x=x_i,\ i=1,\ldots ,\,n$.
In fact, according to a theorem of Bray and Whaples \cite{BW}, if the $x_i$
are specialized to $n$ pairwise nonconjugate elements $c_i$ of a division
ring, then the polynomial
$$
g(x)=
\left|\matrix{
1      & \ldots & 1     & 1     \cr
c_1    & \ldots & c_n   & x     \cr
c_1^2  & \ldots & c_n^2 & x^2   \cr
\vdots &        &\vdots &\vdots \cr
c_1^n  & \ldots & c_n^n & \bo{x^n}   \cr
}\right|
$$
is the only monic polynomial of degree $n$ such that $g(c_i)=0,\ i=1,\ldots
,\,n$.
Moreover, $g(x)$ has no other (right) roots, and any polynomial $h(x)$ having
all the
$c_i$ as roots is right divisible by $g(x)$, that is, $h(x) = q(x)\,g(x)$ for
some
polynomial $q$.

\medskip
Thus we are led in the noncommutative case to the following definition.

\begin{definition}
The elementary symmetric functions $\Lambda_k(x_1,\ldots ,x_n)$ are defined
by
$$
\sum_{k\geq 0}\ \Lambda_{k}(x_1,\ldots ,x_n)\ (-t)^k =
\left|\matrix{
1      & \ldots & 1     & t^n     \cr
x_1    & \ldots & x_n   & t^{n-1}     \cr
x_1^2  & \ldots & x_n^2 & t^{n-2}   \cr
\vdots &        &\vdots &\vdots \cr
x_1^n  & \ldots & x_n^n & \bo{1}   \cr
}\right|\ .
$$
\end{definition}

\medskip\noindent
In other words we specialize the symmetric functions of Section \ref{FORMAL} by
setting
\begin{equation}\label{LGENE}
\lambda(-t) =
\left|\matrix{
1      & \ldots & 1     & t^n     \cr
x_1    & \ldots & x_n   & t^{n-1}     \cr
x_1^2  & \ldots & x_n^2 & t^{n-2}   \cr
\vdots &        &\vdots &\vdots \cr
x_1^n  & \ldots & x_n^n & \bo{1}   \cr
}\right|\ .
\end{equation}

The expansion of the quasi-determinant $\lambda (-t)$ by its last row gives
the following result.

\begin{proposition}
For $0\leq k\leq n$, we have
$$
\Lambda_k(x_1,\ldots ,x_n)
=
(-1)^{k-1}
\left|\matrix{
1 & \ldots  & 1 \cr
\vdots &    & \vdots \cr
x_1^{n-k-1} &\ldots & x_n^{n-k-1} \cr
x_1^{n-k+1} &\ldots & x_n^{n-k+1} \cr
\vdots  &         & \vdots \cr
x_1^{n-1} &\ldots & \bo{x_n^{n-1}} \cr
}\right|
\left|\matrix{
1 & \ldots  & 1 \cr
\vdots &    & \vdots \cr
x_1^{n-k} &\ldots & \bo{x_n^{n-k}} \cr
\vdots  &         & \vdots \cr
x_1^{n-1} &\ldots & x_n^{n-1} \cr
}\right|^{-1} \ .
$$
and $\Lambda_k(x_1,\ldots ,x_n) = 0$ for  $k > n$.
\end{proposition}

\medskip
\begin{example}
{\rm For $n=2$, we obtain}
\begin{equation}
\Lambda_1(x_1,x_2)=
\left|\matrix{
1 & 1\cr
x_1^2 & \bo{x_2^2}\cr
}\right|
\left|\matrix{
1 & 1\cr
x_1 & \bo{x_2}\cr
}\right|^{-1}
=
(x_2^2-x_1^2)(x_2-x_1)^{-1} \ ,
\end{equation}
\begin{equation}\label{LAMBDA2}
\Lambda_2(x_1,x_2)=-
\left|\matrix{
x_1 & x_2\cr
x_1^2 & \bo{x_2^2}\cr
}\right|
\left|\matrix{
1 & \bo{1}\cr
x_1 & x_2\cr
}\right|^{-1}
=
(x_2^2-x_1x_2)(x_1^{-1}x_2-1)^{-1} \ .
\end{equation}
\end{example}

\medskip
It is not immediately clear on the expression (\ref{LAMBDA2}) that
$\Lambda_2(x_2,x_1) = \Lambda_2(x_1,x_2)$. However, one has the
following proposition.

\begin{proposition}
The $\Lambda_k(x_1,\ldots ,x_n)$ are symmetric functions of the noncommutative
variables $x_1,\ x_2,\ \ldots ,\ x_n$, that is, they are invariant under any
permutation of the $x_k$.
\end{proposition}

\Proof By Proposition \ref{PERMUT}, a quasi-determinant is invariant by any
permutation of its columns. Thus, the generating series (\ref{LGENE}) is a
symmetric function of $x_1,\, \ldots ,\,x_n$. Hence, its coefficients are
symmetric. \cqfd

\medskip
We shall now compute the complete symmetric functions $S_k(x_1,\ldots ,x_n)$.

\begin{proposition}\label{PSA}
For every $k,n\geq 0$, we have
\begin{equation}\label{SA}
S_k(x_1,\ldots ,x_n)
=
\left|\matrix{
1 & \ldots  & 1 \cr
\vdots &    & \vdots \cr
x_1^{n-2} &\ldots & x_n^{n-2} \cr
x_1^{n+k-1} &\ldots & \bo{x_n^{n+k-1}} \cr
}\right|
\left|\matrix{
1 & \ldots  & 1 \cr
\vdots  & \ddots  & \vdots \cr
x_1^{n-1} &\ldots & \bo{x_n^{n-1}} \cr
}\right|^{-1} \ .
\end{equation}
\end{proposition}

\Proof Denote temporarily the right-hand side of (\ref{SA}) by
$\overline{S}_k(x_1,\ldots ,x_n)$. To prove that
$S_k(x_1,\ldots ,x_n)=\overline{S}_k(x_1,\ldots ,x_n)$, it is sufficient
to check the identity
$$
\left|\matrix{
\overline{S}_1(x_1,\ldots ,x_n)&\overline{S}_2(x_1,\ldots ,x_n)&\ldots&
\overline{S}_{n-1}(x_1,\ldots ,x_n)&\bo{\overline{S}_n(x_1,\ldots ,x_n)}\cr
\overline{S}_0(x_1,\ldots ,x_n)&\overline{S}_1(x_1,\ldots ,x_n)&\ldots&
\overline{S}_{n-2}(x_1,\ldots ,x_n)&\overline{S}_{n-1}(x_1,\ldots ,x_n) \cr
0         & \overline{S}_0(x_1,\ldots ,x_n)& \ldots &
\overline{S}_{n-3}(x_1,\ldots ,x_n) &\overline{S}_{n-2}(x_1,\ldots ,x_n) \cr
\vdots    & \vdots    & \ddots & \vdots  &\vdots\cr
0         & 0         & \ldots & \overline{S}_0(x_1,\ldots ,x_n)
&\overline{S}_1(x_1,\ldots ,x_n)     \cr
}\right|
$$
$$
= (-1)^{n-1}\Lambda_n(x_1,\ldots ,x_n)\ .
$$
This will follow from a slightly generalized form of Bazin's theorem for
quasi-determinants. The computation is illustrated on an
example. Take $n=k=3$ and denote for short by $|i_1i_2\bo{i_3}|$ the
quasi-minor
$$
|i_1i_2\bo{i_3}| =
\left|\matrix{
x_1^{i_1} &x_2^{i_1} &x_3^{i_1}\cr
x_1^{i_2} &x_2^{i_2} &x_3^{i_2}\cr
x_1^{i_3} &x_2^{i_3} &\bo{x_3^{i_3}}\cr
}\right| \ .
$$
Then one has
$$
\left|\matrix{
\overline{S}_1(x_1,x_2,x_3)&\overline{S}_2(x_1,x_2 ,x_3)&
\bo{\overline{S}_3(x_1,x_2,x_3)}\cr
\overline{S}_0(x_1,x_2,x_3)&\overline{S}_1(x_1,x_2,x_3)&
\overline{S}_2(x_1,x_2,x_3) \cr
0         & \overline{S}_0(x_1,x_2,x_3)     &\overline{S}_1(x_1,x_2,x_3)
\cr
}\right|
=
\left|\matrix{
|01\bo{3}| & |01\bo{4}| & \bo{|01\bo{5}|}\cr
|01\bo{2}| & |01\bo{3}| & |01\bo{4}| \cr
0          & |01\bo{2}| & |01\bo{3}|\cr
}\right|\
|01\bo{2}|^{-1}
$$
$$
=
\left|\matrix{
|23\bo{5}| & |12\bo{5}| & \bo{|01\bo{5}|}\cr
|23\bo{4}| & |12\bo{4}| & |01\bo{4}| \cr
|23\bo{3}| & |12\bo{3}| & |01\bo{3}|\cr
}\right|\
|01\bo{2}|^{-1}
=
|34\bo{5}||\bo{2}34|^{-1}|01\bo{2}||01\bo{2}|^{-1}\ ,
$$
\smallskip
$$
=|12\bo{3}||\bo{0}12|^{-1}=\Lambda_3(x_1,x_2,x_3)\ .
$$
Here, the second equality is obtained by multiplying the columns of the
quasi-minors from the right by suitable powers of $x_1,\, x_2,\, x_3$. The
third equality follows from Theorem~\ref{BAZINPLUS}.
\cqfd

\medskip
More generally, one can express a quasi-Schur function
$\check{S}_I(x_1,\ldots ,x_n)$ as the ratio of two quasi-minors of the
Vandermonde matrix. This may be seen as a noncommutative analog of the
classical expression of a Schur function as a ratio of two alternants.

\begin{sloppypar}
\begin{proposition} {\rm (Cauchy-Jacobi formula for quasi-Schur functions)}\\
\label{CAUJAC}
Let $I=$ $(i_1,\ldots ,i_m)$ be a partition of length $m\leq n$. Set
$$
(s_1,\ldots ,s_n)=(0,1,\ldots ,n-1)+(0,\ldots ,0,i_1,\ldots ,i_m)\ ,
$$
$$
(t_1,\ldots ,t_n)=(0,1,\ldots ,n-1)+(0,\ldots ,0,i_1-1,\ldots ,i_{m-1}-1)\ .
$$
Then,
$$
\check{S}_I(x_1,\ldots ,x_n)=
(-1)^{m-1}\,
\left|\matrix{
x_1^{s_1} &\ldots & x_n^{s_1} \cr
\vdots &\ddots    & \vdots \cr
x_1^{s_n} &\ldots & \bo{x_n^{s_n}} \cr
}\right|
\left|\matrix{
x_1^{t_1} & \ldots  & x_n^{t_1} \cr
\vdots  &   & \vdots \cr
x_1^{t_{n-m+1}} &\ldots &\bo{x_n^{t_{n-m+1}}} \cr
\vdots  &   & \vdots \cr
x_1^{t_n} & \ldots  & x_n^{t_n} \cr
}\right|^{-1} \ .
$$
\end{proposition}
\end{sloppypar}

\Proof The proof is similar to the proof of Proposition~\ref{PSA}. \cqfd

\medskip
Of course all the formulas of Section \ref{FORMAL} may be applied to the
symmetric functions of $x_1,\ldots ,x_n$ defined in this section. This is
worth noting since a direct computation may be quite difficult even in the
simplest cases. For example it is a good exercise to check by hand the
formula
$$
(x_2^2-x_1^2)(x_2-x_1)^{-1}(x_2^2-x_1^2)(x_2-x_1)^{-1}
=
(x_2^3-x_1^3)(x_2-x_1)^{-1} +(x_2^2-x_1x_2)(x_1^{-1}x_2-1)^{-1}\ ,
$$
that is, $S_1(x_1,x_2)^2 = S_2(x_1,x_2)+\Lambda_2(x_1,x_2)$.


\subsection{Rational $(s,d)$-symmetric functions of $n$ noncommutative
variables}
\label{SD}

We present here a natural extension of the case studied in
Section~\ref{RATSYM}. We fix in the same way $n$ noncommutative
variables $x_1,\dots, x_n$. Let $K$ be the skew field generated by
these variables, $s$ an automorphism of $K$ and $d$ a $s$-derivation
of $K$. Consider an indeterminate $X$ such that
$$
X \, k = s(k) \, X + d(k)
$$
for every $k \in K$. The polynomial algebra obtained in this way is
denoted $K[X,s,d]$. Lam and Leroy have defined in this framework a notion
of Vandermonde matrix associated with the family $(x_i)$ which allows to
extend the results of Section~\ref{RATSYM} to this more general context.
Let us first give the following definition.

\begin{definition} {\rm (\cite{LaLe})} Let $k$ be an element
of $K$. Its $(s,d)$-power $P_n(k)$ of order $n$ is then inductively
defined by
$$
P_0(k) = 1 \quad \hbox{and} \quad P_{n+1}(k) = s(P_n(k))\, k + d(P_n(k)) \ .
$$
\end{definition}

This definition is motivated by the following result.
Let $f(X) = \sum_i \ f_i \, X^i$ be a polynomial
of $K[X,s,d]$. Then there exists a unique polynomial $q(t)$ such that
$$
f(X) = q(X) \, (X-a) + \sum_i \ f_i \, P_i(a) \ .
$$
This  shows in particular that $\sum_i \, f_i \, P_i(a)$ is the
good ``evaluation" of $f(X)$ for $X = a$ and hence that $P_n(a)$ really
plays the role in the $(s,d)$-context of the $n$-th power of $a$.

\vskip 0.5mm
We can now introduce the elementary $(s,d)$-symmetric functions
(defined in \cite{LaLe}) which are given by the following
specialization of the series $\lambda(t)$.

\begin{definition}
The elementary symmetric functions $\Lambda_k^{(s,d)}(x_1,\dots,x_n)$ are
defined by
$$
\sum_{k\geq 0} \ \Lambda_{k}^{(s,d)}(x_1,\dots,x_n) \, (-t)^k
=
\left|
\matrix{
1        & \dots  & 1        & t^n \cr
x_1      & \dots  & x_n      & t^{n-1} \cr
P_2(x_1) & \dots  & P_2(x_n) & t^{n-2} \cr
\vdots   &  & \vdots   & \vdots \cr
P_n(x_1) & \dots  & P_n(x_n) & \bo{1} \cr
}
\right| \ .
$$
\end{definition}

\vskip 0.5mm
Expanding the quasi-determinant by its last column, we get the
following result.

\begin{proposition}
For every $k \in [0,n]$, we have
$$\label{LAMQUAS}
\Lambda_{k}^{(s,d)}(x_1,\dots,x_n) =
(-1)^{k-1} \,
\left|
\matrix{
1            &\!\! \dots\!\!     & 1 \cr
\vdots       &\!\!      & \vdots \cr
P_{n-k-1}(x_1) &\!\! \dots\!\!      & P_{n-k-1}(x_n) \cr
P_{n-k+1}(x_1) & \!\!\dots \!\!     & P_{n-k+1}(x_n) \cr
\vdots       &  \!\!    & \vdots \cr
P_{n}(x_1)   & \!\!\dots\!\!      & \bo{P_{n}(x_n)} \cr
}
\right| \,
\left|
\matrix{
1            & \!\!\dots\!\!  & 1 \cr
\vdots       & \!\! & \vdots \cr
P_{n-k}(x_1) & \!\!\dots\!\!  & \bo{P_{n-k}(x_n)} \cr
\vdots       & \!\! & \vdots \cr
P_{n-1}(x_1) &\!\!\dots\!\!  & P_{n-1}(x_n) \cr
}
\right|^{-1}
$$
and $\Lambda^{(s,d)}_k(x_1,\dots,x_n) = 0$ for every $k > n$.
\end{proposition}

\begin{example}
{\rm For $n = 2$, we obtain the following elementary symmetric functions}
$$
\Lambda^{(s,d)}_1(x_1,x_2) =
\left|
\matrix{
1 & 1 \cr
s(x_1)\, x_1 + d(x_1) & \bo{s(x_2) \, x_2 + d(x_2)} \cr
}
\right| \
\left|
\matrix{
1   & 1 \cr
x_1 & \bo{x_2}
}
\right|^{-1}
$$
\vskip 0.5mm
$$
= (s(x_2) \, x_2 - s(x_1) \, x_1 + d(x_2-x_1)) \, (x_2 - x_1)^{-1} \ ,
$$
\vskip 0.5mm
$$
\Lambda^{(s,d)}_2(x_1,x_2) =
\left|
\matrix{
x_1 & x_2 \cr
s(x_1)\, x_1 + d(x_1) & \bo{s(x_2) \, x_2 + d(x_2)} \cr
}
\right| \
\left|
\matrix{
1    & \bo{1} \cr
x_1  & x_2
}
\right|^{-1}
$$
\vskip 0.5mm
$$
= (s(x_2)\, x_2 + d(x_2) - s(x_1)\, x_2 - d(x_1)\, x_1^{-1}\, x_2) \,
(1 - x_1^{-1}\, x_2)^{-1}
\ .
$$
\end{example}

Again, Proposition \ref{PERMUT} shows immediately that the
$\Lambda_k^{(s,d)}(x_1,\dots,x_n)$ are invariant under any permutation
of the noncommutative variables $x_i$.

\smallskip
Using the same method as in Section~\ref{RATSYM}, it is also possible to
derive the following quasi-determinantal expression of the complete
$(s,d)$-symmetric functions.

\begin{proposition}
For every $k \geq 0$, we have
$$
S_k^{(s,d)}(x_1,\dots, x_n) =
\left|
\matrix{
1              & \dots  & 1 \cr
x_1            & \dots  & x_n \cr
\vdots         &  & \vdots \cr
P_{n-2}(x_1)   & \dots  & P_{n-2}(x_n) \cr
P_{n+k-1}(x_1) & \dots  & \bo{P_{n+k-1}(x_n)} \cr
}
\right| \,
\left|
\matrix{
1            & \dots  & 1 \cr
x_1          & \dots  & x_n \cr
\vdots       &  & \vdots \cr
P_{n-1}(x_1) & \dots  & \bo{P_{n-1}(x_n)} \cr
}
\right|^{-1} \ .
$$
\end{proposition}

\vskip 0.5mm
We mention that there also exists a $(s,d)$-version of the Cauchy-Jacobi
formula~\ref{CAUJAC}, obtained by substituting the $(s,d)$-powers $P_j(x_i)$
to the ordinary powers $x_i^j$.

\smallskip
Let us now define the $(s,d)$-conjugate $k^l$ of $k\in K$ by $l\in K$
as follows
$$
k^l = s(l) \, k \, l^{-1} + d(l) \, l^{-1} \ .
$$
We then have the following result ({\it cf.} \cite{LaLe}) which shows that
the defining series of the elementary $(s,d)$-symmetric functions can be
factorized in $K[X,s,d]$.

\begin{proposition} \label{PROPSDL}
Let $(y_i)$ and $(z_i)$ be the two families of elements of $K$ defined by
$$
z_k =
\left|
\matrix{
1        & \dots  & 1 \cr
x_1      & \dots  & x_k \cr
\vdots   &  & \vdots \cr
P_{k-1}(x_1) & \dots  & \bo{P_{k-1}(x_k)}
}
\right| \ , \quad
y_k =
\left|
\matrix{
1        & \dots  & 1 \cr
x_1      & \dots  & x_k \cr
\vdots   &  & \vdots \cr
\bo{P_{k-1}(x_1)} & \dots  & P_{k-1}(x_k)
}
\right| \ .
$$
Then the following relations hold in $K[X,s,d]$
$$
\sum_{k\geq 0} \ \Lambda_{n-k}^{(s,d)}(x_1,\dots,x_n) \, (-X)^k
=
(X - x_n^{z_n}) \ \dots \ (X - x_1^{z_1})
=
(X - x_1^{y_1}) \ \dots \ (X - x_n^{y_n})
\ .
$$
\end{proposition}

\Proof We shall only establish the first identity, the second one being
proved in the same way. Before going further, let us define the polynomial
$V(x_1,\dots,x_n,X)$ of $K[X,s,d]$ as follows
$$
V(x_1,\dots,x_n,X)
=
\left|
\matrix{
1        & \dots  & 1        & 1   \cr
x_1      & \dots  & x_n      & X   \cr
\vdots   &  & \vdots   & \vdots \cr
P_n(x_1) & \dots  & P_n(x_n) & \bo{X^n} \cr
}
\right| \ .
$$
The desired relation will then follow from the following more
explicit lemma.

\begin{lemma}
The following relation holds in $K[X,s,d]$ :
$$
V(x_1,\dots,x_n,X)
=
V(x_2^{x_2-x_1},\dots,x_n^{x_n-x_1},X) \, (X - x_1) \ .
$$
\end{lemma}

\Proof Let us first recall the following formula which holds in $K[X,s,d]$
\begin{equation} \label{DIFFXP}
X^n - P_n(k) = (X^{n-1} + \dots + X^{n-i} s(P_i(k)) + \dots + s(P_n(k))) \,
(X-k) \ ,
\end{equation}
for every $k \in K$. We also need the easily checked  formula
\begin{equation} \label{DIFFSDP}
(P_{n+1}(k) - P_{n+1}(l))\, (k-l)^{-1} =
k^{P_n(k)-P_n(l)} \, (P_{n}(k)-P_{n}(l)) \, (k-l)^{-1} + s(P_n(l)) \ ,
\end{equation}
which holds for every $k,l \in K$.
Let us now apply Sylvester's noncommutative identity to the
quasi-determinant $V(x_1,\dots,x_n,X)$ with the entry $(1,1)$
as pivot. We get
$$
V(x_1,\dots,x_n,X)
=
\left|
\matrix{
\left|\matrix{
1   & 1 \cr
x_1 & \bo{x_2}
}\right|
& \ldots &
\left|\matrix{
1   & 1 \cr
x_1 & \bo{X}
}\right| \cr
\vdots & \ddots & \vdots \vtr{3} \cr
\left|\matrix{
1        & 1 \cr
P_n(x_1) & \bo{P_n(x_2)}
}\right|
& \ldots &
\bo{
\left|\matrix{
1   & 1 \cr
P_n(x_1) & \bo{X^n}
}\right|}
}\right| \ ,
$$
which can be rewritten as
$$
V(x_1,\dots,x_n,X)
=
\left|
\matrix{
x_2 - x_1            & \ldots & x_n-x_1           & X-x_1 \cr
P_2(x_2) - P_2(x_1)  & \ldots & P_2(x_n)-P_2(x_1) & X^2 -P_2(x_1) \cr
\vdots               &  & \vdots            & \vdots \vtr{2} \cr
P_n(x_2) - P_n(x_1)  & \ldots & P_n(x_n)-P_n(x_1) & \bo{X^n - P-n(x_1)} \cr
}
\right| \ .
$$
Using now relations (\ref{DIFFXP}) and (\ref{DIFFSDP}) and
basic properties of quasi-de\-ter\-mi\-nants, we obtain that the last
quasi-determinant is equal to
$$
\left|
\matrix{
1                      & \ldots & 1                      & 1 \cr
x_2^{x_2-x_1} + s(x_1) & \ldots & x_n^{x_n-x_1} + s(x_1) & X+s(x_1) \cr
r_2(x_2)               & \ldots & r_2(x_n)  & X^2+X\, s(x_1) + s(P_2(x_1)) \cr
\vdots                 & & \vdots                 & \vdots \vtr{2} \cr
r_{n-1}(x_2)           & \ldots & r_{n-1}(x_n)           &
\bo{X^{n-1}+\ldots+s(P_{n-1}(x_1))}
}
\right| \  (X-x_1) \ ,
$$
where we set $r_i(x_j) = x_j^{P_i(x_j)-P_i(x_1)} \, (P_i(x_j)-P_i(x_1)) \,
(x_j-x_1)^{-1} + s(P(P_i(x_1))$ for every $i,j$. Now, appropriate
linear combinations of columns will cast this expression into the
required form. \cqfd

Turning back to the proof of Proposition \ref{PROPSDL},
we see that it will follow from the last lemma by induction on $n$.
Indeed, the identity to
be shown is a simple consequence of the fact that $(k^l)^m = k^{ml}$
for every $k,l,m \in K$. \cqfd

\begin{example}
{\rm For $n = 2$, we have in $K[X,s,d]$
$$
X^2 - \Lambda_1^{(s,d)}(x_1,x_2) \, X + \Lambda_2^{(s,d)}(x_1,x_2)
=
(X - x_1^{x_1-x_2})\, (X - x_2) = (X - x_2^{x_2-x_1}) \, (X - x_1) \ .
$$
Thus, by expanding the right-hand side, one has
$$
\Lambda_1^{(s,d)}(x_1,x_2)
=
x_1^{x_1-x_2} + s(x_2) + d(x_2)
=
x_2^{x_2-x_1} + s(x_1) + d(x_1)\, ,
$$
$$
\Lambda_2^{(s,d)}(x_1,x_2) = x_1^{x_1-x_2}\, x_2 = x_2^{x_2-x_1}\, x_1\ .
$$
More generally, one obtains in this way expansions
of the quasi-determinantal formula of Proposition~\ref{LAMQUAS}
which look like the familiar commutative expression of the elementary
symmetric functions, especially in the case when $d=0$.}
\end{example}


\subsection{Polynomial symmetric functions of $n$ noncommutative va\-ria\-bles}
\label{POLSYM}

In this section, we fix $n$ noncommutative indeterminates
$x_1,\ \ldots ,\ x_n$ and we specialize the formal symmetric
functions of Section \ref{FORMAL} by setting
$$
\Lambda_k(x_1,\ldots ,x_n)
=
\sum_{i_1>i_2>\ldots >i_k}\, x_{i_1}\, x_{i_2}\, \ldots\, x_{i_k} \ .
$$
That is, we take,
\begin{equation}
\lambda(t)
=
\ \overleftarrow{\prod_{1\le k\le n}}\ (1+tx_k)
=
(1+tx_n)\, (1+tx_{n-1})\, (1+tx_{n-2})\,\cdots\, (1+tx_1) \ ,
\end{equation}
$t$ being a commutative indeterminate.
The generating series for the complete functions is thus
\begin{equation}
\sigma(t) = \lambda(-t)^{-1}
=
\ \overrightarrow{\prod_{1\le k\le n}}\ (1-tx_k)^{-1}
=
(1-tx_1)^{-1}\, (1-tx_2)^{-1}\, \cdots\, (1-tx_n)^{-1}
\end{equation}
so that the complete symmetric functions specialize to
$$
S_k(x_1,\ldots ,x_n)
=
\sum_{i_1\leq i_2\leq \ldots \leq i_k}
\, x_{i_1}\, x_{i_2}\, \ldots\, x_{i_k} \ .
$$

The ribbon Schur functions then specialize according to the next proposition.
We first introduce some vocabulary. Let $w=x_{i_1}\ldots x_{i_k}$ be a word.
An integer $m$ is called a {\it descent} of $w$ if $1\leq m\leq k-1$ and
$i_m > i_{m+1}$.

\begin{proposition}
Let $J=(j_1,\ldots ,j_n)$ be a composition of $m$. The ribbon Schur function
$R_J$ specializes to
\begin{equation}\label{RB}
R_J(x_1,\ldots ,x_n) = \ \sum \ x_{i_1}\,\ldots\, x_{i_m} \ ,
\end{equation}
the sum running over all words $w=x_{i_1}\ldots x_{i_m}$ whose descent set
is exactly equal to $\{j_1,\, j_1+j_2,\, \ldots \,,\, j_1+\ldots +j_{k-1}\}$.
\end{proposition}

\Proof Denote temporarily by $\overline{R}_J$ the polynomials defined by
(\ref{RB}). It is clear that they satisfy the two relations
$$
\overline{R}_{1^k} = \Lambda_k(x_1,\ldots ,x_n) \ ,
$$
$$
\overline{R}_J \,  \overline{R}_K = \overline{R}_{J\triangleright K}
+ \overline{R}_{J\cdot K} \ .
$$
But these relations characterize the ribbon Schur functions (see Section
\ref{RIBBONS}). Therefore
$\overline{R}_J =  R_J(x_1,\ldots ,x_n)$ for any composition $J$. \cqfd

\begin{example}
Let $X=\{x_1,\,x_2,\,x_3\}$. Then
$$
\Lambda_2(X)=x_2 \, x_1 + x_3\, x_1 + x_3\, x_2\ ,
$$
$$
S_2(X) = x_1^2 + x_1\, x_2 + x_1\, x_3 + x_2^2 + x_2\, x_3 + x_3^2 \ ,
$$
$$
R_{12}(X) = x_2\, x_1^2 + x_2\, x_1\, x_2 + x_2\, x_1\, x_3 +
x_3\, x_1^2 + x_3\, x_1\, x_2 + x_3\, x_1\, x_3
+ x_3\, x_2^2 + x_3\, x_2\, x_3 \ ,
$$
$$
R_{21}(X) =
x_1\, x_2\, x_1 + x_2^2\, x_1 + x_1\, x_3\, x_1 + x_2\, x_3\, x_1
+ x_3^2 \, x_1 + x_1\, x_3\, x_2 + x_2\, x_3\, x_2 + x_3^2\, x_2 \ .
$$
\end{example}

The functions defined in this section are not invariant under permutation
of the va\-ria\-bles $x_i$. However, they are still symmetric, {\it i.e.}
invariant under an action of the symmetric group on the free algebra
$Z\<X\>$ different from the usual one. This action, defined in \cite{LS2}, is
compatible with Young tableaux considered as words in the free algebra. For
various applications such as standard bases or Demazure's character formula
see \cite{LS3}. We recall  the algorithmic description of the
action of simple transpositions.

\smallskip
Consider first the case of a two-letter alphabet $X=\{x_1,x_2\}$. Let $w$ be
a word on $X$. Bracket every factor $x_2\, x_1$ of $w$. The letters which are
not bracketed constitute a subword $w_1$ of $w$. Then, bracket every factor
$x_2\, x_1$ of $w_1$. There remains a subword $w_2$. Continue this procedure
until it stops, giving a word $w_k$ of type $w_k=x_1^r\, x_2^s$. The image of
$w$ under the transposition $\sigma = \sigma_{12}$ is by definition the word
$w^{\sigma}$ in which $w_k$ is replaced by $w_k^{\sigma}=x_1^s\, x_2^r$,
and the bracketed letters remain unchanged.

\smallskip
Let us give an example, denoting for short $w = x_{i_1} \ldots x_{i_k}$
by $i_1 \ldots i_k$. Choose
$$
w = 1\, 2\,  2\, 2\, 1\, 2\, 1\, 1\, 1\, 1\, 2\, 1\, 2\, 2\, 2\ .
$$
The successive bracketings of $w$ give
$$
1\, (2\,  (2\, (2\, 1)\, (2\, 1)\, 1)\, 1)\, 1\, (2\, 1)\, 2\, 2\, 2 \ ,
$$
and $w_3 = 1\, 1\, 2\, 2\, 2$. Then $w_3^{\sigma} = 1\, 1\, 1\, 2\, 2$ and
$w^{\sigma} = 1\, 2\,  2\, 2\, 1\, 2\, 1\, 1\, 1\, 1\, 2\, 1\, 1\, 2\, 2$.
Returning now to a general alphabet $X=\{x_1,x_2\ldots ,x_n\}$, one defines
the action of the simple transposition $\sigma_i$ of $x_i$ and $x_{i+1}$ on
the word $w$, by the preceding rule applied to the subword $w$ restricted to
$\{x_i,x_{i+1}\}$, the remaining letters being unchanged. For example the
image by $\sigma_2$ of the word $w=2\,1\,3\,1\,4\,2\,1\,3\,4\,3$
is $w^{\sigma_2}=2\,1\,3\,1\,4\,2\,1\,2\,4\,3$.

\smallskip
It is proven in \cite{LS2} that $w\rightarrow w^{\sigma_i}$ extends to an
action of the symmetric group on $Z\<X\>$, \it i.e. \rm that given a
permutation $\mu$ and a word $w$, all factorizations of
$\mu = \sigma\sigma'\ldots \sigma''$ into simple transpositions
produce the same word $((w^{\sigma})^{\sigma'}\ldots )^{\sigma''}$
denoted $w^{\mu}$. We can now state the following proposition.

\begin{proposition}
The polynomial symmetric functions defined above are invariant under
the previous action of the symmetric group.
\end{proposition}

\Proof By definition, for any word $w$ and any permutation $\mu$,
$w^{\mu}$ and $w$ have the same descents. Therefore the ribbon Schur
functions are invariant under this action, and the result follows from
the fact that these functions constitute a linear basis of {\bf Sym}. \cqfd

Denoting by $SYM(X_n)$ the algebra of polynomial symmetric functions
of $X_n=\{x_1,\ldots,x_n\}$, one can see that the algebra
$\Sym$ of formal symmetric functions can be realized as the
inverse limit in the category of graded algebras
$$
\Sym\  \simeq\  \limproj_n \ SYM(X_n)
$$
with respect to the projections $F(x_1,\ldots,x_p,x_{p+1},\ldots,x_{p+q})
\longmapsto F(x_1,\ldots,x_p,0,\ldots,0)$. One can thus realize
$\Sym$ as $SYM(X)$, where $X$ is an infinite alphabet. Note also that
the homogeneous component of weight $k$ of $SYM(X_n)$ has for dimension
$$
{\rm dim\, }SYM_k(X_n)= \, \sum_{1\le i\le n} \, {k-1\choose i}
$$
which is equal to the dimension of the space of quasi-symmetric
functions of weight $k$ in $n$ variables, and the duality
between quasi-symmetric functions and noncommutative symmetric functions
still holds in the case of a finite alphabet.

\begin{note}
{\rm The ribbon Schur functions defined in this section are particular cases
of the noncommutative Schur polynomials defined in \cite{LS2} as sums of
tableaux in the free algebra. However these noncommutative Schur polynomials
do not belong in general to the algebra (or skew field) generated by the
elementary symmetric functions $\Lambda_k(X)$.}
\end{note}


\subsection{Symmetric functions associated with a matrix}\label{MATSYM}

In this section, we fix a matrix $A$ of order $n$, with entries in a
noncommutative ring. Recall
first that in the commutative case, one has
\begin{equation}\label{ELEMA}
\hbox{det} (I + tA) = \ \sum_{k=0}^n \, \Lambda_k ({\bf\alpha}) \, t^{k} \ ,
\end{equation}
\begin{equation}\label{MCMH}
\hbox{det} (I - tA)^{-1} = \ \sum_{k\geq 0} \, S_k ({\bf\alpha}) \, t^{k} \ ,
\end{equation}
\begin{equation}
-{d\over dt}\left(\hbox{log}\left(\hbox{det}(I-tA)\right)\right)
= \ \sum_{k\geq 1} \, \Psi_k  ({\bf\alpha}) \, t^{k-1} \ ,
\end{equation}
where ${\bf\alpha}$ is the set of eigenvalues of $A$. Formula~(\ref{MCMH})
is commonly known as MacMahon's master theorem.

\smallskip
Replacing in (\ref{ELEMA}) the determinant by a quasi-determinant, we arrive
at
\begin{equation}\label{QELEM}
| I + tA |_{ii} = \hbox{det}(I+tA)\left(\hbox{det}(I+tA^{ii})\right)^{-1}=
\ \sum_{k=0}^{+\infty} \, \Lambda_k ({\bf\alpha}-{\bf\alpha}^i) \,
t^k
\end{equation}
where ${\bf\alpha} - {\bf\alpha}^i$ denotes the difference in the sense
of $\lambda$-rings of $\alpha$ and the set $\alpha^i$ of eigenvalues of
$A^{ii}$. We decide to take identity~(\ref{QELEM}) as a definition when
the entries of $A$ no longer commute.

\begin{definition}\label{SYMMATRIX}
Let $A$ be a matrix of order $n$ (with entries in an arbitrary ring), and
$i$ a fixed integer between 1 and $n$. The elementary
symmetric functions $\Lambda_k (\alpha_i)$ associated with $A$ (and $i$)
are defined by
$$
| I + tA|_{ii} = \ \sum_{k=0}^{+\infty} \ \Lambda_k (\alpha_i) \, t^k \ .
$$
\end{definition}

The others families of noncommutative symmetric functions of $A$ are defined
accordingly by their generating functions
$$
\sum_{k\geq 0} \, S_k (\alpha_i) \, t^{k} = |I-tA|_{ii}^{-1} \ ,
$$
$$
\sum_{k\geq 1} \, \Phi_k  (\alpha_i) \, t^{k-1} =
-{d\over dt}\, \hbox{log}\left(|I-tA|_{ii}\right) \ ,
$$
$$
\sum_{k\geq 1} \, \Psi_k  (\alpha_i) \, t^{k-1} =
|I-tA|_{ii} \, {d\over dt}\, |I-tA|_{ii}^{-1} \ .
$$

\medskip
The specializations of the elementary, complete, power sums (of first or
second kind) and ribbon functions are polynomials in the entries of $A$,
which can be combinatorially interpreted in terms of graphs.

\begin{proposition}\label{SFGR}
Let ${\cal A}$ be the complete oriented graph associated with the matrix $A$
(\cf Section \ref{QUASIDET}). Then,
\begin{enumerate}
\item $S_k (\alpha_i)$ is the sum of all words labelling paths of length $k$
going from $i$ to $i$ in ${\cal A}$.

\item $(-1)^{k-1}\, \Lambda_k (\alpha_i)$ is the sum of all words labelling
simple paths of length $k$ going from $i$ to $i$ in ${\cal A}$.

\item $\Psi_k (\alpha_i)$ is the sum of all words labelling paths of
length $k$ going from $i$
to $i$ in ${\cal A}$, the coefficient of each path being the length
of the first return to $i$.

\item $\Phi_k (\alpha_i)$ is the sum of all words labelling paths of
length $k$ going from $i$ to $i$ in ${\cal A}$, the coefficient of each
path being the ratio of $k$ to the number of returns to $i$.

\item Let $I$ be a composition of $k$. Then $(-1)^{\ell(I)-1}\, R_I(\alpha_i)$
is the sum of all words labelling paths of length $k$ going from $i$ to $i$
in ${\cal A}$ such that every return to $i$ occurs at some length
$\bar I^\sim_1 + \dots + \bar I^\sim_j$ with $j \in [1,\dots,r]$
(where we set $\bar I^\sim = (\bar I^\sim_j)_{j=1,r}$).
\end{enumerate}
\end{proposition}

\Proof The graphical interpretations of $\Lambda_k$ and $S_k$ follow
from Proposition~\ref{AUTO}. The interpretation of $\Psi_n$
results then from the formula
$$
\Psi_k = \ \sum_{j=1}^{k} \ (-1)^{j-1} \, j \, \Lambda_j \, S_{k-j} \ .
$$
Similarly, the interpretation of $\Phi_k$ follows from the formula
$$
\Phi_k = \ \sum_{j=1}^k \ (-1)^{k-j} \ {k \over j} \
( \, \sum_{i_1+\dots+i_j=k} \ \Lambda_{i_1} \, \dots \, \Lambda_{i_j} \ )
$$
which is a particular case of Proposition~\ref{LAMPHI}.
Finally, the interpretation of $R_I$ follows from (\ref{R2LA}).
\cqfd

\medskip
\begin{example}
{\rm For a matrix $A=(a_{ij})$ of order $n=3$, we have}
$$
\Lambda_1(\alpha_1)=S_1(\alpha_1)=\Psi_1(\alpha_1)=\Phi_1(\alpha_1)=a_{11} \ ,
$$
$$
\Lambda_2(\alpha_1) = R_{11}(\alpha_1) = -a_{12}\, a_{21} - a_{13}\, a_{31} \ ,
$$
$$
S_2(\alpha_1) = R_2(\alpha_1) = a_{11}^2 + a_{12}\, a_{21} + a_{13}\, a_{31} \
,
$$
$$
\Psi_2(\alpha_1) = \Phi_2(\alpha_1) = a_{11}^2 + 2\, a_{12}\, a_{21} +
2\, a_{13}\, a_{31} \ ,
$$
$$
\Lambda_3(\alpha_1)=R_{111}(\alpha_1) =
a_{12}\, a_{22}\, a_{21} + a_{12}\, a_{23}\, a_{31} + a_{13}\, a_{32}\, a_{21}
+ a_{13}\, a_{33}\, a_{31} \ ,
$$
$$
S_3(\alpha_1) = R_3(\alpha_1) = a_{11}^3 + a_{11}\, a_{12}\, a_{21} +
a_{11}\, a_{13}\, a_{31} + a_{12}\, a_{21}\, a_{11} +
a_{12}\, a_{22}\, a_{21}
$$
$$
+ a_{12}\, a_{23}\, a_{31} + a_{13}\, a_{31}\, a_{11} +
a_{13}\, a_{32}\, a_{21} + a_{13}\, a_{33}\, a_{31} \ ,
$$
$$
\Psi_3(\alpha_1) = a_{11}^3 + a_{11}\, a_{12}\, a_{21} +
a_{11}\, a_{13}\, a_{31} + 2\, a_{12}\, a_{21}\, a_{11} +
3\, a_{12}\, a_{22}\, a_{21}
$$
$$
+ 3\, a_{12}\, a_{23}\, a_{31} + 2\, a_{13}\, a_{31}\, a_{11} +
3\, a_{13}\, a_{32}\, a_{21} + 3\, a_{13}\, a_{33}\, a_{31} \ ,
$$
$$
\Phi_3(\alpha_1) = a_{11}^3 + {3\over 2}\, a_{11}\, a_{12}\, a_{21} +
{3\over 2}\, a_{11}\, a_{13}\, a_{31} + {3\over 2}\, a_{12}\, a_{21}\, a_{11}
+ 3\, a_{12}\, a_{22}\, a_{21}
$$
$$
+ 3\, a_{12}\, a_{23}\, a_{31} + {3\over 2}\, a_{13}\, a_{31}\, a_{11} +
3\, a_{13}\, a_{32}\, a_{21} + 3\, a_{13}\, a_{33}\, a_{31} \ ,
$$
$$
R_{12}(\alpha_1) = - a_{12}\, a_{21}\, a_{11} -
a_{12}\, a_{22}\, a_{21} - a_{12}\, a_{23}\, a_{31} - a_{13}\, a_{31}\, a_{11}
- a_{13}\, a_{32}\, a_{21} - a_{13}\, a_{33}\, a_{31} \ ,
$$
$$
R_{21}(\alpha_1) =  - a_{11}\, a_{12}\, a_{21} - a_{11}\, a_{13}\, a_{31} -
a_{12}\, a_{22}\, a_{21} - a_{12}\, a_{23}\, a_{31}  - a_{13}\, a_{32}\, a_{21}
- a_{13}\, a_{33}\, a_{31} \ .
$$
\end{example}

\medskip
\begin{note}{\rm
It follows from Proposition~\ref{SFGR} that the generating series of the
functions $S_k(\alpha_i)$, $\Lambda_k(\alpha_i)$ or $\Psi_k(\alpha_i)$
are all rational in the sense of automata theory (\cf Section~\ref{AUTOMA}).
Moreover the automata
which recognize these series can be explicitely described and have no
multiplicities.

For example, the minimal automaton which recognizes the generating
series of the functions $\Psi_k(\alpha_1)$ is given, for $n=2$, by

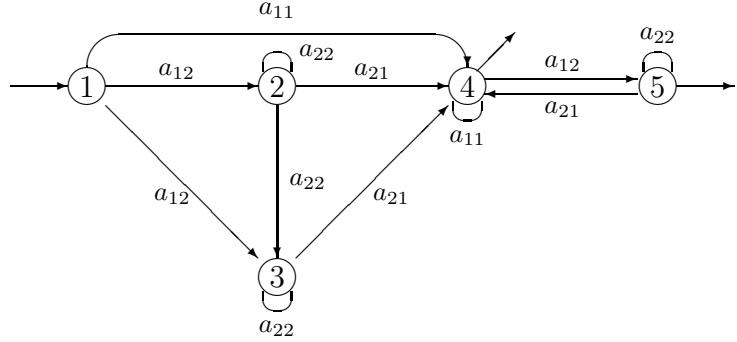
\setlength{\unitlength}{0.72pt}
\centerline{
\begin{picture}(330,190)(0,-140)
 \put(0,0){\circle{20}}
 \put(0,0){\makebox(0,0){1}}
 \put(100,0){\circle{20}}
 \put(100,0){\makebox(0,0){2}}
 \put(200,0){\circle{20}}
 \put(200,0){\makebox(0,0){4}}
 \put(300,0){\circle{20}}
 \put(300,0){\makebox(0,0){5}}
 \put(100,-100){\circle{20}}
 \put(100,-100){\makebox(0,0){3}}
 \put(10,0){\vector(1,0){80}}
 \put(110,0){\vector(1,0){80}}
 \put(209,4){\vector(1,0){80}}
 \put(100,-10){\vector(0,-1){80}}
 \put(10,-10){\vector(1,-1){80}}
 \put(110,-90){\vector(1,1){80}}
 \put(100,8){\oval(15,20)[t]}
 \put(200,-8){\oval(15,20)[b]}
 \put(300,8){\oval(15,20)[t]}
 \put(100,-108){\oval(15,20)[b]}
 \put(-40,0){\vector(1,0){30}}
 \put(310,0){\vector(1,0){30}}
 \put(205,8){\vector(1,1){20}}
 \put(200,12){\vector(0,-1){2}}
 \put(100,10){\oval(200,35)[t]}
 \put(289,-4){\vector(-1,0){80}}
 \put(50,8){\makebox(0,0){\hbox{\footnotesize $a_{12}$} }}
 \put(150,8){\makebox(0,0){\hbox{\footnotesize $a_{21}$}}}
 \put(250,12){\makebox(0,0){\hbox{\footnotesize $a_{12}$}}}
 \put(250,-12){\makebox(0,0){\hbox{\footnotesize $a_{21}$}}}
 \put(300,27){\makebox(0,0){\hbox{\footnotesize $a_{22}$}}}
 \put(116,-50){\makebox(0,0){\hbox{\footnotesize $a_{22}$}}}
 \put(45,-58){\makebox(0,0){\hbox{\footnotesize $a_{12}$}}}
 \put(160,-58){\makebox(0,0){\hbox{\footnotesize $a_{21}$}}}
 \put(100,40){\makebox(0,0){\hbox{\footnotesize $a_{11}$}}}
 \put(121,18){\makebox(0,0){\hbox{\footnotesize $a_{22}$}}}
 \put(100,-127){\makebox(0,0){\hbox{\footnotesize $a_{22}$}}}
 \put(200,-27){\makebox(0,0){\hbox{\footnotesize $a_{11}$}}}
\end{picture}
}

On the other hand, Proposition~\ref{SFGR} also shows
that the generating series of the family $\Phi_k(\alpha_i)$ is not rational
(\cf \cite{BR} or \cite{Eil} for more details).
}
\end{note}


\subsection{Noncommutative symmetric functions and the center
of $U({\goth gl}_n)$} \label{NCUGL}

In this section, we illustrate the results of Section~\ref{MATSYM} in the
particular case when $A = E_n = (e_{ij})_{i,j=1,n}$, the matrix of the
standard generators of the universal envelopping algebra $U({\goth gl}_n)$.
Recall that these generators satisfy the relations
$$
[e_{ij},e_{kl}] = \delta_{jk}e_{il}-\delta_{li}e_{kj} \ .
$$
The Capelli identity of classical invariant theory indicates that the
determinant of the noncommutative matrix $E_n$ makes sense, provided that
one subtracts $0,\, 1,\,2,\, \ldots \, n-1\,$ to its leading diagonal
(see \cite{Tu} for instance). Thus one defines
$$
{\rm det}\, E_n =
\left|
\matrix{
e_{11} & e_{12}   &  \ldots  & e_{1n}  \cr
e_{21} & e_{22}-1 &  \ldots  & e_{2n} \cr
\vdots & \vdots   &  \ddots  & \vdots   \cr
e_{n1} & e_{n2} & \ldots   & e_{nn}-n+1\cr
}\right|
$$
$$
:=
\sum_{\sigma\in {\goth S}_n}\ (-1)^{\ell(\sigma)}\
e_{\sigma(1)\,1}\, (e_{\sigma(2)\,2}-\delta_{\sigma(2)\,2})\, \ldots
\, (e_{\sigma(n)\,n}-(n-1)\delta_{\sigma(n)\,n}) \ .
$$
This is better understood
from the following well-known result (see \cite{Ho}).

\begin{theorem}\label{ZGL}
The coefficients of the polynomial in the variable $t$
$$
{\rm det}\, (I+tE_n) =
\left|
\matrix{
1+te_{11} & te_{12}   &  \ldots  & te_{1n}  \cr
te_{21} & 1+t(e_{22}-1) & \ldots & te_{2n}   \cr
\vdots &  \vdots     & \ddots  & \vdots \cr
te_{n1} & te_{n2} & \ldots   & 1+t(e_{nn}-n+1)\cr
}\right|
$$
$$
:=
\sum_{\sigma\in {\goth S}_n}\ (-1)^{\ell(\sigma)}\,
(\delta_{\sigma(1)\,1}+te_{\sigma(1)\,1})\, \ldots \,
(\delta_{\sigma(n)\,n}+t(e_{\sigma(n)\,n}-(n-1)\delta_{\sigma(n)\,n})) \ .
$$
generate the center $Z(U({\goth gl}_n))$ of $U({\goth gl}_n)$.
\end{theorem}

It is shown in \cite{GR1} and \cite{GR2} that the Capelli
determinant can be expressed as a product of quasi-determinants. More
precisely, one has

\begin{theorem} \label{CAP}
Keeping the notation of Theorem~{\rm\ref{ZGL}}, ${\rm det}\, (I+tE_n)$ can be
factorized in the algebra of formal power series in $t$ with coefficients in
$U({\goth gl}_n)$
$$
{\rm det}(I+tE_n) =
$$
$$
(1+te_{11})
\left|
\matrix{
1+t(e_{11}-1) & \!\! te_{12} \cr
te_{21} & \!\! \bo{1+t(e_{22}-1)} \cr
}\right|
\ldots
\left|
\matrix{
1+t(e_{11}-n+1)    &  \ldots \!\!  & te_{1n}  \cr
\vdots &       \!\!  & \vdots \cr
te_{n1} & \ldots \!\!  & \bo{1+t(e_{nn}-n+1)}\cr
}\right|
$$
and the factors in the right-hand side commute with each other.
\end{theorem}

\begin{note}
{\rm Theorem~\ref{CAP} is stated in \cite{GR2} in its formal differential
operator version. The above version is obtained by using the classical
embedding of $U({\goth gl}_n)$ into the Weyl algebra
${\displaystyle K[x_{ij},
\ {\partial \over \partial x_{ij}} \ ,1\leq i,\,j \leq n]}$
of order $n^2$ defined by
$$
e_{ij} \hookrightarrow \ \sum_{k=1}^n \ x_{ik} \, {\partial\over \partial
x_{jk}} \ .
$$
}
\end{note}

We adopt the following notations. We write for short
$F_m = E_m-(m-1)I = (f_{ij})_{1\leq i,j\leq m}$, and we put $A=F_m$, $i=m$ in
Definition~\ref{SYMMATRIX}, the symmetric functions thus obtained being denoted
by $\Lambda_k(\epsilon_m)$. In other words, by Proposition \ref{SFGR},
the polynomials  $\Lambda_k(\epsilon_m)$ are given by
$$
\Lambda_k(\epsilon_m)
=
(-1)^{k-1}\ \sum_{1\leq i_1,\ldots ,i_{k-1}\leq m-1}
\, f_{m\,i_1}\, f_{i_1\,i_2}\, \ldots\, f_{i_{k-1}m} \ .
$$
Combining now the two previous theorems we arrive at the following result,
concerning the symmetric functions associated with the matrices
$E_1,\,E_2-I,\, \ldots ,\,E_n-(n-1)I$.

\begin{theorem}\label{ZSF}
The symmetric functions $\Lambda_k(\epsilon_m)$ for $k\geq 0$ and
$1\leq m\leq n$ generate a commutative subalgebra in $U({\goth gl}_n)$.
This algebra is the smallest subalgebra of $U({\goth gl}_n)$ containing
the centers
$Z(U({\goth gl}_1)),\ Z(U({\goth gl}_2)), \ldots,\ Z(U({\goth gl}_n))$.
\end{theorem}

\Proof Let ${\cal Z}$ denote the subalgebra of $U({\goth gl_n})$ generated by
$Z(U({\goth gl}_1)),\,\ldots,\ Z(U({\goth gl}_n))$. This is of
course a commutative algebra. Set
$$
\lambda(t,\epsilon_m)= \sum_k \ t^k\, \Lambda_k(\epsilon_m)
=
\left|
\matrix{
1+t(e_{11}-m+1)    &  \ldots  & te_{1m}  \cr
\vdots  & \ddots   & \vdots \cr
te_{m1} & \ldots   & \bo{1+t(e_{mm}-m+1)}\cr
}\right|
\ .
$$
By Theorem~\ref{ZGL} and Theorem~\ref{CAP}, one sees
that the coefficients of the formal power series in $t$
$$
\lambda_m(t)=\lambda(t,\epsilon_1)\,\lambda(t,\epsilon_2)\,\ldots
\,\lambda(t,\epsilon_m)
$$
generate the center $Z(U({\goth gl}_m))$. Therefore the symmetric
functions $\Lambda_k(\epsilon_m)$ with $k\geq 0$ and $1\leq m\leq n$
generate a subalgebra of $U({\goth gl_n})$ containing ${\cal Z}$.
Conversely, the polynomial
$\lambda_{m-1}(t)$ being invertible in ${\cal Z}[[t]]$, there holds
$$
\lambda(t,\epsilon_m)=\lambda_{m-1}(t)^{-1}\, \lambda_m(t) \ ,
$$
Hence the symmetric functions  $\Lambda_k(\epsilon_m)$ belong to ${\cal Z}$,
which proves the theorem. \cqfd

\smallskip
As a consequence, the center of $U({\goth gl_n})$ may be described in
various ways in terms of the different types of symmetric functions associated
with the matrices $E_1,\,E_2-I,\, \ldots \,, \ E_n-(n-1)I$.

\begin{corollary}\label{ZSF2}
The center $Z(U({\goth gl}_n))$ is generated by the scalars and either of
the following families of formal symmetric functions
\begin{enumerate}
\item $\displaystyle\Lambda_k^{(n)} = \sum_{i_1+\ldots +i_n=k} \
\Lambda_{i_1}(\epsilon_1) \, \ldots\,  \Lambda_{i_n}(\epsilon_n) \ , \ \
\hbox{for} \ 1\leq k\leq n,$
\item $\displaystyle
S_k^{(n)} = \sum_{i_1+\ldots +i_n=k}\
S_{i_1}(\epsilon_1)\, \ldots\, S_{i_n}(\epsilon_n) \ , \ \
\hbox{for} \ 1\leq k\leq n,$
\item $\displaystyle\Psi_k^{(n)} = \sum_{1\leq m\leq n} \ \Psi_k(\epsilon_m) \
,
\ \ \hbox{for} \ 1\leq k \leq n.$
\end{enumerate}
\end{corollary}

\Proof The functions $\Lambda_k^{(n)}$ are nothing else but the coefficients of
the polynomial in $t$ ${\rm det}\, (I+tE_n)$, which generate
$Z(U({\goth gl}_n))$ according to Theorem~\ref{ZGL}. Since we are in fact in a
commutative situation, the functions $S_k^{(n)}$ and $\Psi_k^{(n)}$ are
related to the functions $\Lambda_k^{(n)}$ by the basic identities (\ref{RSL})
and (\ref{RPL}), and thus also generate  $Z(U({\goth gl}_n))$. \cqfd

\begin{example}
{\rm
The center of $U({\goth gl}_3)$ is generated by
$1,\,\Psi_1^{(3)},\,\Psi_2^{(3)},\,\Psi_3^{(3)}$, where
}
$$
\Psi_1^{(3)} = e_{11} + e_{22} -1 + e_{33} - 2\ ,
$$
$$
\Psi_2^{(3)} = e_{11}^2 + (e_{22} -1)^2 + 2\, e_{21}\, e_{12} +
(e_{33} - 2)^2 + 2\, e_{31}\, e_{13} + 2\, e_{32}\, e_{23}\ ,
$$
$$
\Psi_3^{(3)} = e_{11}^3 + (e_{22} -1)^3 + (e_{22} -1)\, e_{21}\, e_{12} +
2\, e_{21}\, e_{12}\, (e_{22} -1) +3\, e_{21}\, (e_{11} -1)\, e_{12}
$$
$$
+ (e_{33} - 2)^3 + (e_{33} - 2)\, e_{31}\, e_{13}
+ (e_{33} - 2)\, e_{32}\, e_{23} +2\, e_{31}\, e_{13}\, (e_{33} - 2)
+ 2\, e_{32}\, e_{23}\, (e_{33} - 2)
$$
$$
+ 3 \, e_{31}\, e_{12}\, e_{23} + 3\, e_{32}\, e_{21}\, e_{13}
+ 3\, e_{31}\, (e_{11} -2)\, e_{13} + 3\, e_{32}\, (e_{22} -2)\, e_{23}\ .
$$
\end{example}

\medskip
The results of this section are strongly connected with Gelfand-Zetlin
bases for the complex Lie algebra $gl_n$. Recall that each irreducible
finite-dimensional $gl_n$-module $V$ has a canonical decomposition into
a direct sum of one-dimensional subspaces associated with the chain of
subalgebras
$$
gl_1 \subset gl_2 \subset \ldots \subset gl_n \ .
$$
These subspaces $V_{\mu}$ are parametrized by Gelfand-Zetlin schemes
$\mu = (\mu_{ij})_{1\leq j \leq i \leq n}$ in such a way that for every
$m=1,\,2,\,\ldots ,\,n$, the module $V_{\mu}$ is contained in an irreducible
$gl_m$-module with highest weight $(\mu_{m1},\,\mu_{m2},\, \ldots
,\,\mu_{mm})$.
Since each irreducible representation of $gl_{m-1}$ appears at most once in
the restriction to $gl_{m-1}$ of an irreducible representation of $gl_m$,
this set of conditions defines $V_{\mu}$ uniquely (\cf \cite{GZ}).
Moreover, the integers $\mu_{ij}$ must satisfy the condition
$\mu_{ij}\ge \mu_{i-1\, j}\ge \mu_{i\, j+1}$ for all $i,j$.

\smallskip
Another characterization of the subspaces $V_{\mu}$ given by
Cherednik \cite{Ch} (see also
Nazarov and Tarasov \cite{NT}) is the following. The $V_{\mu}$ are
exactly the eigenspaces of the smallest subalgebra ${\cal Z}$ of $U(gl_n)$
containing the centers $Z(U(gl_1)),\,\ldots ,\,Z(U(gl_n))$. Taking into
account Theorem~\ref{ZSF}, this description translates into the following
proposition.

\begin{proposition}
Let $x_{\mu}$ be a non-zero vector of the one-dimensional subspace $V_{\mu}$
para\-me\-trized by the Gelfand-Zetlin scheme
$\mu = (\mu_{ij})_{1\leq j \leq i \leq n}$. Set $\nu_{ij} = \mu_{ij}-j+1$ and
$\nu_i=\{ \nu_{i1},\,\ldots ,\,\nu_{ii}\}$ (with $\nu_0 = \emptyset$). For
every $F\in {\bf Sym}$ and every $m=1,\,2,\,\ldots ,\,n$, $x_{\mu}$ is an
eigenvector of the noncommutative symmetric function $F(\epsilon_m)$. The
corresponding eigenvalue is the specialization of $F$ to the alphabet of
commutative variables $\nu_m-\nu_{m-1}$ (in the sense of $\lambda$-rings) :
$$
F(\epsilon_m)\,x_{\mu} = F(\nu_m-\nu_{m-1})\,x_{\mu} \ .
$$
The eigenvalues associated with different $\mu$ being pairwise distinct, this
characterizes completely the subspaces $V_{\mu}$.
\end{proposition}

\Proof Since the $\Psi_k$ generate {\bf Sym}, it is enough to prove the
proposition in the case when $F=\Psi_k$. It results from
Proposition~\ref{SFGR} that
$$
\Psi_k(\epsilon_m) =
$$
$$
\sum_{1\leq i_1,\ldots ,i_{k-1}\leq m} \alpha_{i_1\ldots i_{k-1}}\,
(e_{m{i_1}} - (m-1)\delta_{m{i_1}})\,
(e_{{i_1}{i_2}} - (m-1)\delta_{{i_1}{i_2}}) \ldots
(e_{i_{k-1}m} - (m-1)\delta_{i_{k-1}m}) \, ,
$$
where the $\alpha_{i_1\ldots i_{k-1}}$ are integers and $\alpha_{m\ldots m}=1$.
Thus, denoting by ${\cal N}_m$ the subalgebra of $gl_m$ spanned by the
elements $e_{ij},\ 1\leq i<j \leq m$, we see that
\begin{equation}\label{MODN}
\sum_{1\leq j\leq m} \, \Psi_k(\epsilon_j) \, -
\sum_{1\leq j\leq m}\,  (e_{jj}-j+1)^k \ \in U(gl_m)\, {\cal N}_m \ .
\end{equation}
By Corollary~\ref{ZSF2}, $\sum_{1\leq j\leq m} \Psi_k(\epsilon_j)$
belongs to $Z(U(gl_m))$, and therefore acts as a scalar on any irreducible
$gl_m$-module. By definition, $x_{\mu}$ belongs to the irreducible
$gl_m$-module with highest weight
$(\mu_{m1},\,\mu_{m2},\, \ldots ,\,\mu_{mm})$, and we can compute this scalar
by acting on the highest weight vector, which is annihilated
by ${\cal N}_m$. Therefore, we deduce from (\ref{MODN}) that
$$
\sum_{1\leq j\leq m}\, \Psi_k(\epsilon_j) \, x_{\mu} =
\sum_{1\leq j\leq m} \, (\mu_{mj}-j+1)^k \, x_{\mu}
= \Psi_k(\nu_m) \, x_{\mu} \ ,
$$
and by difference,
$$
\Psi_k(\epsilon_m) \, x_{\mu}
=
\left(\Psi_k(\nu_m) - \Psi_k(\nu_{m-1})\right) \, x_{\mu}
=  \Psi_k(\nu_m-\nu_{m-1})\,x_{\mu} \ .
$$
\cqfd

\begin{example} {\rm Let $V$ be the irreducible $gl_3$-module with
highest weight $(5,3,2)$. Consider the GZ-scheme
$$
\mu=\ \matrix{
5 &   & 3 &   & 2  \cr
  & 4 &   & 2 &    \cr
  &   & 3 &   &
}
$$
The action of the operator
$$
\Psi_3(\epsilon_3) =
(e_{33} - 2)^3 + (e_{33} - 2)\, e_{31}\, e_{13} +
(e_{33} - 2)\, e_{32}\, e_{23} + 2\, e_{31}\, e_{13}\, (e_{33} - 2)
$$
$$
+ 2\, e_{32}\, e_{23}\, (e_{33} - 2) + 3\, e_{31}\, e_{12}\, e_{23} +
3\, e_{32}\, e_{21}\, e_{13} + 3\, e_{31}\, (e_{11} -2)\, e_{13} +
3\, e_{32}\, (e_{22} -2)\, e_{23}
$$
on the weight vector $x_\mu$ is given by
$$
\Psi_3(\epsilon_3)\, x_\mu
=
\left[ \, 5^3+(3-1)^3+(2-2)^3-4^3-(2-1)^3\, \right]x_\mu
=
68\, x_\mu \ .
$$
}
\end{example}


\subsection{Twisted symmetric functions}\label{SKEW}

Let $K$ be a skew field, let $s$ be an automorphism of $K$ and let
$(S_i^{(s)})_{i\geq 0}$ be a family of $K$. Let then $K[X,s]$ be the
$K$-algebra of twisted polynomials over $K$, i.e. of polynomials with
coefficients in $K$ where the variable $X$ satisfies to
\begin{equation}\label{sX}
X \, k = s(k) \, X
\end{equation}
for every $k \in K$. Let now $\sigma (t)$ be the formal power series
defined by
$$
\sigma (t) = \ \sum_{i=0}^{+\infty} \ (S_i^{(s)} \, X^i) \, t^i \ .
$$
The symmetric functions $S_n = S_i^{(s)} \, X^i$ defined using this generating
function belong to $K[X,s]$. It is however more interesting to define twisted
symmetric functions which belong to the smallest $s$-stable algebra generated
by the family $S_i^{(s)}$.

\begin{definition}
The {\rm twisted symmetric functions} $S_n^{(s)},\,\Lambda_n^{(s)},\,
\Phi_n^{(s)},\,\Psi_n^{(s)},\,R_I^{(s)}$ are the elements of $K$ defined by
$$
S_n^{(s)} = S_n\, X^{-n} \, , \ \Lambda_n^{(s)} = \Lambda_n \, X^{-n} \, , \
\Phi_n^{(s)} = \Phi_n \, X^{-n} \, , \
\Psi_n^{(s)} = \Psi_n \, X^{-n} \, , \
R_I^{(s)} = R_I \, X^{-|I|} \, .
$$
\end{definition}

The quasi-determinantal formulas of Section \ref{ELEM} can be easily adapted
to take care of the present situation.

\begin{proposition}
\begin{equation}\label{LSD}
\Lambda_n^{(s)} = (-1)^{n-1} \,
\left|
\matrix{
S_1^{(s)} & S_2^{(s)}      & \dots  & S_{n-1}^{(s)} & \bo{S_n^{(s)}} \vtr{3}\cr
s(S_0^{(s)}) & s(S_1^{(s)}) &\dots & s(S_{n-2}^{(s)}) &
s(S_{n-1}^{(s)})\vtr{3} \cr
0       & s^2(S_0^{(s)}) & \dots  & s^2(S_{n-3}^{(s)}) &
s^2(S_{n-2}^{(s)}) \vtr{3} \cr
\vdots  & \vdots  & \ddots & \vdots                & \vdots \vtr{3}  \cr
0       &   0     & \dots  & s^{n-1}(S_{0}^{(s)})  & s^{n-1}(S_1^{(s)})  \cr
}
\right| \ ,
\end{equation}
\medskip
\begin{equation}
S_n^{(s)} = (-1)^{n-1} \,
\left|
\matrix{
\Lambda_1^{(s)}    & \Lambda_2^{(s)}      & \dots  & \Lambda_{n-1}^{(s)}
& \bo{\Lambda_n^{(s)}} \vtr{3} \cr
s(\Lambda_0^{(s)}) & s(\Lambda_1^{(s)})   & \dots  & s(\Lambda_{n-2}^{(s)})
& s(\Lambda_{n-1}^{(s)}) \vtr{3} \cr
0       & s^2(\Lambda_0^{(s)}) & \dots  & s^2(\Lambda_{n-3}^{(s)})
& s^2(\Lambda_{n-2}^{(s)}) \vtr{3} \cr
\vdots  & \vdots  & \ddots & \vdots                & \vdots \vtr{3}  \cr
0       &   0     & \dots  & s^{n-1}(\Lambda_{0}^{(s)})
& s^{n-1}(\Lambda_1^{(s)})  \cr
}
\right| \ ,
\end{equation}
\medskip
\begin{equation}
\Psi_n^{(s)} = (-1)^{n-1} \,
\left|
\matrix{
s^{n-1}(S_1^{(s)})     & s^{n-1}(S_0^{(s)})  & 0                 & \dots  &
0 \vtr{3} \cr
s^{n-2}(2\,S_2^{(s)}) & s^{n-2}(S_1^{(s)})  & s^{n-2}(S_0^{(s)}) & \dots  &
0 \vtr{3} \cr
s^{n-3}(3\, S_3^{(s)}) & s^{n-3}(S_2^{(s)}) & s^{n-3}(S_{1}^{(s)}) &\dots &
0\vtr{3} \cr
\vdots            & \vdots         & \vdots           & \ddots &
\vdots \vtr{3} \cr
\bo{n\,S_n^{(s)}} & S_{n-1}^{(s)} & S_{n-2}^{(s)} & \dots & S_1^{(s)} \cr
}
\right| \ .
\end{equation}
\end{proposition}

\Proof We only give the proof of the first formula, the other relations
being proved in the same way. According to Corollary \ref{DETFORM}, we have
$$
\Lambda_n^{(s)} \, X^{n} = (-1)^{n-1} \
\left|
\matrix{
S_1^{(s)}\, X  & S_2^{(s)}\, X^2 & \dots  & S_{n-1}^{(s)}\, X^{n-1} &
\bo{S_n^{(s)}\, X^n} \vtr{3} \cr
S_0^{(s)}      & S_1^{(s)}\, X   & \dots  & S_{n-2}^{(s)}\, X^{n-2} &
S_{n-1}^{(s)}\, X^{n-1} \vtr{3} \cr
0      & S_0^{(s)} & \dots  & S_{n-3}^{(s)}\, X^{n-3} &
S_{n-2}^{(s)}\, X^{n-2} \vtr{3} \cr
\vdots & \vdots    & \ddots & \vdots       & \vdots   \vtr{3} \cr
0      &   0       & \dots  & S_{0}^{(s)}  & S_1^{(s)}\, X  \cr
}
\right| \ .
$$
Multiplying on the left the $k$-th row by $X^{k-1}$ and using
relation (\ref{sX}), we get
$$
\Lambda_n^{(s)} \, X^{n} = (-1)^{n-1} \
\left|
\matrix{
S_1^{(s)}\, X  & S_2^{(s)}\, X^2 & \dots  & S_{n-1}^{(s)}\, X^{n-1} &
\bo{S_n^{(s)}\, X^n} \vtr{3} \cr
s(S_0^{(s)})\, X   & s(S_1^{(s)})\, X^2  & \dots  & s(S_{n-2}^{(s)})\, X^{n-1}
&
s(S_{n-1}^{(s)})\, X^{n} \vtr{3} \cr
0      & s^2(S_0^{(s)})\, X^2 & \dots  & s^2(S_{n-3}^{(s)})\, X^{n-1} &
s^2(S_{n-2}^{(s)})\, X^{n} \vtr{3} \cr
\vdots & \vdots    & \ddots & \vdots       & \vdots   \vtr{3} \cr
0      &   0       & \dots  & s^{n-1}(S_{0}^{(s)})\, X^{n-1}
& s^{n-1}(S_1^{(s)})\, X^n  \cr
}                                                             :
\right| \ .
$$
Multiplying on the right the $k$-th column by $X^{-k}$,
we obtain (\ref{LSD}). \cqfd

Using the same method, one
can also give suited versions of the other determinantal identities
given in Section \ref{FORMAL}. It is also interesting to notice that one can
generalize these results to the case when the variable $t$ satisfies
a relation of the form
$$
t \, k = s(k) \, t + d(k)
$$
for every scalar $k \in K$, $s$ being an automorphism of $K$ and
$d$ a $s$-derivation. In such a case, one must consider that the
$S_i$ are the coefficients of the Laurent series
$\sum_{i\geq 0} \ S_i \, t^{-i}$. In this framework, it is then easy to
adapt the results given here, using  formulas which can be found in
\cite{LaLe} or \cite{Kr} for instance.

\newpage
\section{Noncommutative rational power series}\label{EXPAN}

In this section, we demonstrate that noncommutative symmetric functions
provide an efficient tool for handling formal power series in one variable
over a skew field. Special attention is paid to rational power series,
and to the problem of approximating a given noncommutative power series $F(t)$
by a rational one, the so-called Pad\'e approximation (Section~\ref{PADE}).
Particular Pad\'e approximants can be obtained by expanding $F(t)$ into a
noncommutative
continued fraction of $J$-type or of $S$-type (Sections~\ref{CONFRAC} and
\ref{NCJFRAC}).
The sequence of denominators
of the partial quotients of a $J$-fraction is orthogonal with respect to
an appropriate noncommutative scalar product, and it satisfies a three-term
recurrence relation (Section~\ref{ORTHPOL}). The systematic use of quasi-Schur
functions enables one to derive all these results in a straightforward and
unified way.

\smallskip
These topics gave rise to an enormous literature in the past twenty years,
as shown by the 250 titles compiled by Draux in his commented bibliography
\cite{Dr}. We shall indicate only a few of them, the reader being referred
to this source for precise attributions and bibliographical informations.

\smallskip
Interesting examples of noncommutative rational power series are provided
by the generating series
$\sigma(t,\alpha_i)$ of the complete symmetric functions associated with
the generic matrix of order $n$ defined in Section~\ref{MATSYM}. Their
denominators appear as noncommutative analogs of the characteristic
polynomial, for which a version of the Cayley-Hamilton theorem can be
established.
In particular, the generic matrix posesses $n$
{\it pseudo-determinants}, which are true noncommutative analogs of
the determinant. These pseudo-determinants reduce in the case of
$U(gl_n)$ to the Capelli determinant, and in the case of the quantum
group $GL_q(n)$, to the quantum determinant (up to a power of $q$).


\subsection{Noncommutative continued $S$-fractions}\label{CONFRAC}

Continued fractions with coefficients in a noncommutative algebra
have been considered by several authors, and especially by Wynn
(\cf \cite{Wy}). The convergence theory of these noncommutative
expansions is discussed for example in \cite{Fa}.
As in the commutative case, many formulas can
be expressed in terms of quasi-determinants. In this section,
we give the noncommutative analog of Stieltjes continued
fractions in terms of quasi-Schur functions,
and specialize them to the noncommutative tangent defined in
Section~\ref{TRIG}.

\smallskip
Let $a = (a_i)_{i\geq 0}$ be a sequence of noncommutative indeterminates.
We can associate with it two types of continued $S$-fractions,
which are the series defined by
\begin{equation}\label{LCONT}
a_0  \,
{ 1 \over 1 + a_1\, t \, \displaystyle
{ 1 \over \, \ddots \, \displaystyle
{ \hbox{ } \atop \displaystyle
{ \hbox{ } \atop \displaystyle
{ 1 \over 1 + a_n t \, \displaystyle
{ 1 \over \, \ddots \,
}}}}}}
= a_0 \,
(1+a_1 \, t ( 1 + \dots + a_{n-1} \, t ( 1 + a_n\, t ( 1 + \dots )^{-1} )^{-1}
\dots )^{-1} \, )^{-1}
\end{equation}
and by
$$
{ 1 \over \displaystyle
{ 1 \over \displaystyle
{ \hbox{ } \atop \displaystyle
{ \hbox{ } \atop \displaystyle
{ 1 \over \displaystyle
{ 1 \over \, \adots \, } \, a_n\, t + 1 }}} \, \adots \, }
\, a_1\, t + 1 } \, a_0
= (( \dots ((\dots + 1 )^{-1} a_n\, t + 1 )^{-1} a_{n-1}\, t + \dots + 1 )^{-1}
a_1\, t + 1 )^{-1} \, a_0 \ .
$$

\smallskip
The partial fractions of these infinite continued fractions admit
quasi-determinantal
expressions \cite{GR2}. Thus, the $n$-th convergent of (\ref{LCONT}) is equal
to
$$a_0\,
\left|\matrix{
\bo{1}  & a_1t  & 0     & \ldots & 0 \cr
-1      &    1  & a_2t  & \ldots & 0  \cr
\vdots  &\ddots &\ddots & \ddots & \vdots   \cr
0       & \ldots &  -1   &   1    &a_nt    \cr
0       &  \ldots     &   0    &   - 1  & 1   \cr
}\right|^{-1}
= a_0\,
\left|\matrix{
1  & a_2t  & 0     & \ldots & \bo{0} \cr
-1      &    1  & a_3t  & \ldots & 0  \cr
\vdots  &\ddots &\ddots & \ddots & \vdots   \cr
0       & \ldots &  -1   &   1    &a_nt    \cr
0       &  \ldots     &   0    &   - 1  & 1   \cr
}\right|
\left|\matrix{
1  & a_1t  & 0     & \ldots & \bo{0} \cr
-1      &    1  & a_2t  & \ldots & 0  \cr
\vdots  &\ddots &\ddots & \ddots & \vdots   \cr
0       & \ldots &  -1   &   1    &a_nt    \cr
0       &  \ldots     &   0    &   - 1  & 1   \cr
}\right|^{-1}
$$
\smallskip\noindent
where the two quasi-determinants of the right-hand side are polynomials
in $t$ whose expansion is given by Proposition~\ref{SSTRI}.

\medskip
The following result expresses the
coefficients of  the noncommutative
continued fraction expansion of a formal power series as
{\it Hankel quasi-determinants}, that is, from the viewpoint of noncommutative
symmetric functions, as rectangular quasi-Schur functions.
Indeed, as illustrated in Section~\ref{SPECIA}, any noncommutative
formal power series $F(t)$ may be seen as a specialization of the
generating series  $\sigma(t) = \sum_{k\geq 0} \ S_k \, t^k$
of the complete symmetric functions $S_k$. With this in mind,
the Hankel quasi-determinants associated as in the commutative
case with $F(t)$ appear as the specializations of the quasi-Schur functions
indexed by rectangular partitions of the form $m^n$.

\begin{proposition}\label{CONTFRAC}
Let $\sigma(t) = \sum_{k\geq 0} \ S_k \, t^k$ be a noncommutative formal
power series. Then one has the following expansion of $\sigma(t)$ into a left
$S$-fraction:
$$
\sigma(t) = \check{S}_0 \
{ 1 \over 1 -  \check{S}_0^{-1}\, \check{S}_1 \,  t \ \displaystyle
{ 1 \over 1 + \check{S}_1^{-1}\, \check{S}_{11} \,  t \ \displaystyle
{ 1 \over 1 - \check{S}_{11}^{-1}\, \check{S}_{22} \, t \ \displaystyle
{ 1 \over \ \ddots \ \displaystyle
{ \hbox{ } \atop \displaystyle
{ \hbox{ } \atop \displaystyle
{ 1 \over 1 + (-1)^{n} \, \check{S}_{I_n}^{-1} \, \check{S}_{I_{n+1}} \,  t \
\displaystyle
{ 1 \over \ \ddots \
}}}}}}}}
$$
where $I_{2\,n} = (n^n)$ and $I_{2n+1} = (n^{n+1})$.
\end{proposition}
\begin{note}{\rm
Throughout this section, we relax for convenience the assumption
$S_0 = 1$ made above, and we only suppose that $S_0 \not = 0$.}
\end{note}
\Proof Let us introduce the series $\rho(t)$ defined by
$$
\sigma (t) = S_0 \ \displaystyle { 1 \over 1 - t \, \rho (t) } \ .
$$
That is,
$$
\rho (t) = \displaystyle {1 - \lambda(-t) \, S_0 \over t }
=
\ \sum_{n\geq 0} \ (-1)^{n-1} \, \Lambda_{n+1} \, S_0 \, t^n \ .
$$
Denote by $a_n(\sigma)$ the $n$-th coefficient of the Stieltjes-type
development in right continued fraction of $\sigma(t)$.
According to the definition of $\rho$,
\begin{equation} \label{STFC}
a_{n+1}(\sigma) = a_n(\rho)
\end{equation}
for $n \geq 0$. We prove by induction on $n$ that
$$
a_n(\sigma) = (-1)^n \, \check{S}_{I_{n}}^{-1} \, \check{S}_{I_{n+1}}
$$
for every $n \geq 0$. This  is clear for $n = 0,\ 1$, and
the induction step is as follows.
We only indicate  the
case $n = 2m$, the odd case being similar. According to
relation (\ref{STFC}) and to the induction hypothesis, we have
\begin{equation} \label{STFC2}
a_{2m+1}(\sigma)
=
a_{2m}(\rho)
=
\ \Delta_m^{-1} \, \Delta_{m+1}
\end{equation}
where we set
$$
\Delta_m
=
\left|
\matrix{
(-1)^{m-1} \, \Lambda_{m+1} \, S_0 & (-1)^m \, \Lambda^{m+2} \, S_0 &
\dots & \bo{\Lambda_{2m} \, S_0} \cr
(-1)^m \, \Lambda_{m} \, S_0 & (-1)^{m-1} \, \Lambda_{m+1} \, S_0
& \dots & - \Lambda_{2m-1} \, S_0 \cr
\vdots & \vdots & \ddots & \vdots \cr
- \Lambda_1 \, S_0 & \Lambda_2 \, S_0 & \dots &
(-1)^{m-1} \, \Lambda_{m+1} \, S_0
}
\right|
$$
and
$$
\Delta_{m+1}
=
\left|
\matrix{
(-1)^{m-1} \, \Lambda_{m+1} \, S_0 & (-1)^m \, \Lambda^{m+2} \, S_0 &
\dots & \bo{- \Lambda_{2m+1} \, S_0} \cr
(-1)^m \, \Lambda_{m} \, S_0 & (-1)^{m-1} \, \Lambda_{m+1} \, S_0
& \dots & \Lambda_{2m} \, S_0 \cr
\vdots & \vdots & \ddots & \vdots \cr
0 & - \Lambda_1 \, S_0 & \dots & (-1)^{m-1} \, \Lambda_{m+1} \, S_0
}
\right|
\ .
$$
Using now basic properties of quasi-determinants and
Naegelbasch's formula for quasi-Schur functions, we get
$$
\Delta_m
=
\left|
\matrix{
\Lambda_{m+1} & \Lambda_{m+2} & \dots  & \bo{\Lambda_{2m}} \cr
\Lambda_{m}   & \Lambda_{m+1} & \dots  & \Lambda_{2m-1} \cr
\vdots        & \vdots        & \ddots & \vdots \cr
\Lambda_1     & \Lambda_2     & \dots  & \Lambda_{m+1}
}
\right|
\ S_0
=
\check{S}_{m^{m+1}}
=
\check{S}_{I_{2m+1}} \ .
$$
Arguing in the same way, we have
$$
\Delta_{m+1}
=
- \,
\left|
\matrix{
\Lambda_{m+1} & \Lambda_{m+2} & \dots & \bo{\Lambda_{2m+1}} \cr
\Lambda_{m}   & \Lambda_{m+1} & \dots & \Lambda_{2m} \cr
\vdots & \vdots & \ddots & \vdots \cr
0 & \Lambda_1 & \dots & \Lambda_{m+1}
}
\right| \
S_0
=
\check{S}_{(m+1)^{m+1}}
=
\check{S}_{I_{2m+2}}
\ .
$$
The conclusion follows from these last two formulas
and from formula (\ref{STFC2}).  \cqfd

\medskip
Applying $\omega$ to the formula given by Proposition~\ref{CONTFRAC}, we get
immediately the following one.

\begin{corollary}
Let $\sigma(t) = \sum_{k\geq 0} \ S_k \, t^k$ be a noncommutative formal
power series. Then one has the following expansion :
$$
\sigma(t) =
{ 1 \over \displaystyle
{ 1 \over \displaystyle
{ 1 \over \displaystyle
{ 1 \over \displaystyle
{ \hbox{ } \atop \displaystyle
{ \hbox{ } \atop \displaystyle
{ 1 \over \displaystyle
{ 1 \over \ \adots \ }
\check{S}_{I_{n+1}} \, \check{S}_{I_n}^{-1}\,  (-1)^n \, t + 1 } } }
\ \adots \ }
\check{S}_{22} \, \check{S}_{11}^{-1} \, (-t) +1}
\check{S}_{11} \, \check{S}_1^{-1}\,  t + 1}
\check{S}_1 \, \check{S}_0^{-1}\, (-t) + 1 }
\ \check{S}_0
$$
where $I_{2\,n} = (n^n)$ and $I_{2n+1} = (n^{n+1})$.
\end{corollary}

The action of $\omega$ may also be interpreted in another way in order to
obtain the de\-ve\-lop\-ment in continued fraction of the inverse of a formal
power series.

\begin{corollary}
Let $\sigma(t) = \sum_{k\geq 0} \ S_k \, t^k$ be a noncommutative formal
power series. Then one has the following expansion :
$$
\sigma(-t)^{-1} = \check{S}_0 \
{ 1 \over 1 - \check{S}_0^{-1}\, \check{S}_1 \,  t \ \displaystyle
{ 1 \over 1 + \check{S}_1^{-1}\, \check{S}_{2} \, t \ \displaystyle
{ 1 \over 1 - \check{S}_{2}^{-1}\, \check{S}_{22} \, t \ \displaystyle
{ 1 \over \ \ddots \ \displaystyle
{ \hbox{ } \atop \displaystyle
{ \hbox{ } \atop \displaystyle
{ 1 \over 1 + (-1)^{n} \, \check{S}_{J_n}^{-1} \,\check{S}_{J_{n+1}} \,  t \
\displaystyle
{ 1 \over \ \ddots \
}}}}}}}}
$$
where $J_{2\,n} = (n^n)$ and $J_{2n+1} = ((n+1)^n)$.
\end{corollary}

\medskip
We apply now these results to the noncommutative tangent defined in \ref{TRIG}.
If we set $S_i(B) = T_{2i+1}^{(r)}(A)$ ($A$ and $B$ being merely labels to
distinguish between two families of formal symmetric functions), so that
$$
t\, \sigma(B;t^2) = TAN_r(A;t)
=
\ \sum_{n\ge 0}\ T_{2n+1}^{(r)}(A)\, t^{2n+1} \ ,
$$
we can, using Propositions \ref{CONTFRAC} and \ref {LASPRAG}, identify the
coefficients of the continued fraction expansion of the
tangent in terms of staircase quasi-Schur functions. For example, one has
$$
\check{S}_{333}(B) =
\left|\matrix{
T^{(r)}_7 & T^{(r)}_9 & \bo{T^{(r)}_{11}} \cr
T^{(r)}_5 & T^{(r)}_7 & T^{(r)}_9 \cr
T^{(r)}_3 & T^{(r)}_5 & T^{(r)}_7 \cr
}\right|
=
\left|\matrix{T^{(r)}_3 & T^{(r)}_5 & T^{(r)}_7 \cr
T^{(r)}_5 & T^{(r)}_7 & T^{(r)}_9 \cr
T^{(r)}_7 & T^{(r)}_9 & \bo{T^{(r)}_{11}} \cr
}\right|
$$
$$
=
\left|\matrix{R_{12}&R_{122}&R_{1222}\cr
R_{122}&R_{1222}&R_{12222}\cr
R_{1222}&R_{12222}&\bo{R_{122222}}\cr
}\right|
= \check{S}_{123456}(A) \ ,
$$
\medskip
\noindent
the last equality following from Proposition \ref{LASPRAG}.

\medskip
This identity may be seen as a particular case of the following proposition,
which expresses more general quasi-determinants in the $T_{2n+1}^{(r)}$ as
quasi Schur-functions (indexed by skew partitions). Let $\rho_k$ denote the
staircase partition $(1,2,\ldots ,k)$.

\begin{proposition}
Let $I=(i_1,\ldots ,i_n)$ be a partition, such that $i_1\ge n-1$. Set
$N=i_n+n$.
Then one has
\begin{equation}
\check{S}_I(B) = \check{S}_{\rho_N/J}(A) \ ,
\end{equation}
where $J$ is the partition whose diagram is obtained by removing from
the diagram of $\rho_N$, $n$ successive ribbon shapes indexed by the
compositions
$$
12^{i_n+n-1},\ 12^{i_{n-1}+n-3},\ \ldots\ , 12^{i_1-n+1} \ .
$$
\end{proposition}

\Proof This follows from Bazin's theorem for quasi-determinants. \cqfd

\begin{example} {\rm Take $I = (2,3,4)$ so that $N=7$. The ribbon shapes
to be extracted from the diagram of $\rho_7$ are $12^6,\ 12^3,\ 1$. Hence,
$J=(2,5)$, as illustrated by the following picture.

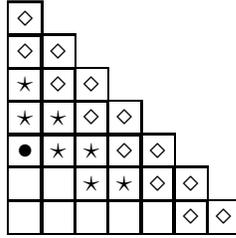
\setlength{\unitlength}{0.25pt}
\centerline{
\begin{picture}(400,400)(0,-50)
\put(0,0){\framebox(50,50){}}
\put(50,0){\framebox(50,50){}}
\put(100,0){\framebox(50,50){}}
\put(150,0){\framebox(50,50){}}
\put(200,0){\framebox(50,50){}}
\put(250,0){\framebox(50,50){$\diamond$}}
\put(300,0){\framebox(50,50){$\diamond$}}
\put(0,50){\framebox(50,50){$$}}
\put(0,100){\framebox(50,50){$\bullet$}}
\put(0,150){\framebox(50,50){$\star$}}
\put(0,200){\framebox(50,50){$\star$}}
\put(0,250){\framebox(50,50){$\diamond$}}
\put(0,300){\framebox(50,50){$\diamond$}}
\put(50,50){\framebox(50,50){$$}}
\put(100,50){\framebox(50,50){$\star$}}
\put(150,50){\framebox(50,50){$\star$}}
\put(200,50){\framebox(50,50){$\diamond$}}
\put(250,50){\framebox(50,50){$\diamond$}}
\put(50,100){\framebox(50,50){$\star$}}
\put(50,150){\framebox(50,50){$\star$}}
\put(50,200){\framebox(50,50){$\diamond$}}
\put(50,250){\framebox(50,50){$\diamond$}}
\put(100,100){\framebox(50,50){$\star$}}
\put(150,100){\framebox(50,50){$\diamond$}}
\put(200,100){\framebox(50,50){$\diamond$}}
\put(100,150){\framebox(50,50){$\diamond$}}
\put(100,200){\framebox(50,50){$\diamond$}}
\put(150,150){\framebox(50,50){$\diamond$}}
\end{picture}
}

\noindent
Thus, $\check{S}_{234}(B) = \check{S}_{1234567/25}(A)$.}
\end{example}

We end this Section by considering the specialization
$$
S_i \longrightarrow 0 \ ,\ \  \hbox{for} \ i > n \ .
$$
In this case the continued fraction for $TAN_r(A,t)$ terminates, and
by means of the recurrence formulas given in \cite{GR2} for computing
its numerator and denominator, one obtains

\begin{proposition}\label{IMPAIR}
Suppose that $S_i=0$ for $i>n$. Then the complete functions $S_i,\ i \leq n$
may be expressed as rational functions of the staircase quasi Schur functions
$S_{\rho_i},\ i \leq n$. Namely, writing for short
$c_i=\check{S}_{\rho_i}^{-1} \, \check{S}_{\rho_{i+1}}$ for $1\leq i < n$
and $c_i=0$  for $i\geq n$, one has
$$
S_2=\ \sum_{1\leq i} \ c_i\ ,\ \ S_4 =\ \sum_{1\leq i\leq j}\ c_i\, c_{j+2}\ ,
\ \ S_6 =\ \sum_{1\leq i\leq j\leq k}\ c_i\, c_{j+2}\, c_{k+4}\ ,\ \ldots
$$
$$
S_1^{-1}\, S_3 = \ \sum_{2\leq i} \ c_i\ ,\
\ S_1^{-1}\, S_5 = \ \sum_{2\leq i\leq j} \ c_i\, c_{j+2}\ ,\ \
\ S_1^{-1}\, S_7 =\ \sum_{2\leq i\leq j\leq k} \ c_i\, c_{j+2}\, c_{k+4}\ ,\
\ldots
$$
\end{proposition}

Proposition~\ref{IMPAIR} may be regarded as a noncommutative analog of
a classical question investigated by Laguerre and Brioschi. The problem was
to express any symmetric polynomial of $n$ indeterminates as a rational
function of the power sums of odd degree $\Psi_{2k+1}$ (\cf \cite{Lag},
\cite{Po}). Foulkes gave a solution by means of Schur functions. Indeed,
in the commutative case, the only Schur functions belonging to the ring
$\Q[\Psi_1,\Psi_3,\Psi_5,\ldots]$ are the one indexed by staircase
partitions $\rho_n$, so that Laguerre's problem is equivalent to
expressing a symmetric polynomial of $n$ indeterminates as a rational function
of the staircase Schur functions (\cf \cite{F3}). Note that in the
noncommutative setting, this no longer holds. For example
$$
\check{S}_{12} = R_{12} = -1/3\, \Psi^3 - 1/6\, \Psi^{12} + 1/6 \, \Psi^{21} +
1/3\, \Psi^{111} \ ,
$$
does not belong to $\Q[\Psi_1,\Psi_3,\Psi_5,\ldots]$.
The fact that the symmetric functions depend
only on $n$ variables is replaced in Proposition~\ref{IMPAIR} by the
condition $S_i=0$ for $i>n$.


\subsection{Noncommutative Pad\'e approximants}\label{PADE}

The classical Pad\'e approximant is the ratio of two determinants, in which
all columns, except for the first ones, are identical. In the commutative
case, these determinants can be interpreted as {\it multi Schur functions}
(see \cite{La}). One obtains a noncommutative analog by replacing determinants
by quasi-determinants. Here the problem is to approximate a noncommutative
formal series
\begin{equation}\label{SERIE}
F(t) = S_0 + S_1\, t + S_2\, t^2 + \ \ldots + S_n\, t^n+\ \ldots\ \ ,
\end{equation}
where the parameter $t$ commutes with the $S_i$, by a rational fraction
$Q(t)^{-1}P(t)$ up to a fixed order in $t$. This kind of computation appears
for instance in Quantum Field Theory (\cite{Bessis}, \cite{GG}),
$F$ being in this case a perturbation series, or in electric networks
computations \cite{BB}.

\begin{proposition}\label{PAD}
Let $S_0, S_1,\ldots,S_{m+n}$ be noncommutative
indeterminates, and let $t$ be an indeterminate commuting with the $S_i$.
We set $C_k(t) = S_0 + S_1\, t + \cdots + S_k\, t^k$ ($C_k(t) = 0$ if $k<0$),
and
\begin{equation}\label{NUMPAD}
P_m(t)
=
\left|\matrix{
\bo{C_m(t)}  & S_{m+1} & \cdots & S_{m+n} \cr
t\, C_{m-1}(t) & S_m & \cdots & S_{m+n-1}\cr
\vdots & \vdots & \ddots & \vdots \cr
t^n\, C_{m-n}(t)& S_{m-n+1}& \cdots & S_m\cr
}\right| \ ,
\end{equation}
\begin{equation}\label{DENPAD}
Q_n(t)=
\left|\matrix{
\bo{1} & S_{m+1} & \cdots & S_{m+n} \cr
t & S_m & \cdots & S_{m+n-1} \cr
\vdots & \vdots & \ddots & \vdots \cr
t^n & S_{m-n+1} &\cdots & S_m \cr
}\right| \ .
\end{equation}
Then we have
$$
Q_n(t)\cdot (S_0 + S_1\, t+\cdots +S_{m+n}\, t^{m+n}) =
P_m(t)+ O(t^{m+n+1}) \ .
$$
\end{proposition}

\Proof Expanding $Q_n(t)$ by its first column and multiplying to the right
by $S_0 + S_1\, t + \cdots + S_{m+n}\, t^{m+n}$, one obtains for the terms
in $t^k$ with $k\le m+n$ the following expression :
$$
\left|\matrix{
\bo{S_{m+n}} & S_{m+1} & \cdots & S_{m+n} \cr
S_{m+n-1}&S_m&\cdots&S_{m+n-1}\cr
\vdots & \vdots & \ddots &\vdots \cr
S_m & S_1 & \cdots & S_m \cr
}\right|\ t^{m+n}
+
\left|\matrix{
\bo{S_{m+n-1}} & S_{m+1} & \cdots & S_{m+n} \cr
S_{m+n-2}&S_m&\cdots&S_{m+n-1}\cr
\vdots & \vdots & \ddots &\vdots \cr
S_{m-1} & S_1 & \cdots & S_m \cr
}\right|\ t^{m+n-1}
+\cdots
$$
$$
+\, \cdots\, +
\left|\matrix{
\bo{S_{0}} & S_{m+1} & \cdots & S_{m+n} \cr
0&S_m&\cdots&S_{m+n-1}\cr
\vdots & \vdots & \ddots &\vdots \cr
0 & S_1 & \cdots & S_m \cr
}\right|\ t^{0}
= P_m(t) \ ,
$$
since the coefficients of the $t^k$ with $k>m$ are zero, the corresponding
quasi-determinants having two identical columns. \cqfd

\begin{definition} The Pad\'e approximants of the noncommutative
series {\rm (\ref{SERIE})} are the rational functions
$$
[m/n] = Q_n(t)^{-1}\, P_m(t) \ .
$$
\end{definition}

We have for instance
$$
[1/1] = (1-S_2\, S_1^{-1}\, t)^{-1}\,
(S_0+(S_1\! -\! S_2\, S_1^{-1}S_0)\,t) = S_0 + S_1\, t + S_2\, t^2 + O(t^3)
\ .
$$
Applying the $*$ involution (\cf \ref{ELEM}, \ref{QUASI}, and \ref{ORTHPOL}
below),
one obtains the equivalent expression $[m/n] = R_m(t) \, T_n(t)^{-1}$, where
\begin{equation}\label{NUMPAD2}
R_m(t) =
\left|\matrix{
\bo{C_m(t)} & tC_{m-1}(t) & \ldots & t^nC_{m-n}(t) \cr
S_{m+1}     & S_m         & \ldots & S_{m-n+1}     \cr
\vdots      &\vdots       & \ddots & \vdots        \cr
S_{m+n}     & S_{m+n-1}   & \ldots & S_m           \cr
}\right| \ ,
\end{equation}
\begin{equation}\label{DENPAD2}
T_n(t) =
\left|\matrix{
\bo{1} & t& \ldots & t^n \cr
S_{m+1}     & S_m         & \ldots & S_{m-n+1}     \cr
\vdots      &\vdots       & \ddots & \vdots        \cr
S_{m+n}     & S_{m+n-1}   & \ldots & S_m           \cr
}\right| \ .
\end{equation}
Thus one also has
$$
[1/1] = (S_0+(S_1-S_0S_1^{-1}S_2)t)(1-S_1^{-1}S_2t)^{-1} \ .
$$
This last remark may be interpreted as an effective proof
of the fact that the ring of polynomials in one indeterminate with coefficients
in a skew field is an Ore domain. For, given two polynomials
$P(t)$ and $Q(t)$ of respective degrees $m$ and $n$ with no common left
factor, and assuming $Q(0)\not = 0$, one has
$$
P(t)\,T(t) = Q(t)\,R(t) \ ,
$$
where $R(t)\,T(t)^{-1}$ is the Pad\'e approximant $[m/n]$ of the
power series $Q(t)^{-1}\,P(t)$.


\subsection{Noncommutative orthogonal polynomials} \label{ORTHPOL}

Motivated by formula~(\ref{DENPAD}), we consider in this section the
sequence of polynomials
\begin{equation}\label{ORTHPOLY}
\pi_n(x)=
\left|\matrix{
S_{n} & \cdots & S_{2n-1} &\bo{x^n}\cr
S_{n-1} & \cdots & S_{2n-2} & x^{n-1}\cr
\vdots & \ddots & \vdots & \vdots \cr
S_0 & \cdots & S_{n-1} & 1 \cr
}\right| \ ,
\end{equation}
in the commutative variable $x$. In the case when the $S_i$ commute
with each other, $\pi_n(x)$ is none other than the sequence of
orthogonal polynomials associated with the moments $S_i$. Thus,
when the $S_i$ belong to a skew field, these polynomials may be
regarded as natural noncommutative analogs of orthogonal polynomials.
They are indeed orthogonal with respect to the noncommutative
scalar product to be defined now. This scalar product is a formal
analog of the matrix-valued ones used for example in \cite{GG} and
\cite{BB}.

\smallskip
Let $R$ be the ring of polynomials in the commutative indeterminate
$x$, with coefficients in the free field
$K\!\not< \! S_0,\,S_1,\,S_2,\, \ldots \!\not>$. Recall that this
field is endowed with a natural involution $F \longrightarrow F^*$
defined as the {\it anti}-automorphism such that $S_i^*=S_i$ for all $i$.
This involution is extended to $R$ by putting
$$\left( \sum_i a_i\,x^i\right)^* = \sum_i a_i^* \, x^i \ . $$
In particular, we have
$$
\pi_n(x)^* =
\left|\matrix{
S_{n-1} & S_n & \cdots & S_{2n-1}\cr
\vdots  & \vdots & \ddots & \vdots \cr
S_0 & S_1 & \cdots & S_n \cr
1 & x & \cdots & \bo{x^n} \cr
}\right| \ .
$$
We define a scalar product $\<.\, ,\, . \>$ in $R$ by
$$
\< \sum_i a_i\,x^i\, , \, \sum_j b_j\,x^j \> := \sum_{i,j} a_i\,S_{i+j}\,b_j^*
\ .
$$
This product satisfies the following obvious properties
$$
\< \alpha_1\,p_1(x) + \alpha_2\,p_2(x)\, ,\, q(x) \>
 = \alpha_1\, \< p_1(x) \, , \, q(x) \>
  + \alpha_2\, \< p_2(x) \, , \, q(x) \> \ ,
$$
$$
\< p(x) \, , \, \beta_1\,q_1(x) + \beta_2\,q_2(x) \>
 = \<p(x)\, ,\, q_1(x) \>\, \beta_1^* +
 \<p(x)\, ,\, q_2(x) \>\, \beta_2^* \ ,
$$
$$
\< q(x) \, , \, p(x) \> = \< p(x) \, , \, q(x) \>^* \ ,
$$
$$
\<x\, p(x) \, , \, q(x) \> = \< p(x) \, , \, x\,q(x) \> \ ,
$$
from which one deduces, by means of \ref{DEVQDET}, that
$$
\< \pi_n(x) \, ,\, x^i \> = \check S_{n^n i} \ , \ \ \ \ \ \
\< x^i \, ,\, \pi_n(x) \> = \check S_{n^n i}^* \ ,
$$
for $i \ge n$, and that
$$
\< \pi_n(x) \, ,\, x^i \> = \< x^i \, ,\, \pi_n(x) \> = 0
$$
for $i=0,\,1,\, \ldots \, , \, n-1 $. Hence we obtain

\begin{proposition}\label{ORTH}
The sequence of polynomials $\pi_n(x)$ satisfies
$
\< \pi_n(x) \, ,\, \pi_m(x) \> = 0
$
for $n \not = m$.
\cqfd
\end{proposition}

\smallskip
As in the commutative case, the sequence $\pi_n(x)$ follows
a typical three term recurrence relation that we shall now
explicit. Set
$$
x\, \pi_n(x) = \sum_{0\le k \le n+1} a_k^{(n)} \, \pi_k(x) \ .
$$
It follows from \ref{ORTH} that
$$
\<x\, \pi_n(x)\, , \, \pi_k(x) \> =
a_k^{(n)} \<\pi_k(x)\, , \, \pi_k(x) \> =
a_k^{(n)} \<\pi_k(x)\, , \, x^k \> =
a_k^{(n)} \check S_{k^{k+1}}\ .
$$
On the other hand, if $k\le n-2$,
$$
\<x\, \pi_n(x)\, , \, \pi_k(x) \> =
\<\pi_n(x)\, , \, x\,\pi_k(x) \> = 0
$$
and $a_k^{(n)} = 0$ for $k =0,\,1,\, \ldots \, , \, n-2 $.
The polynomials $\pi_k(x)$ being monic, it is clear that
$a_{n+1}^{(n)} = 1 $, and it remains only to compute
$a_{n}^{(n)}$ and $a_{n-1}^{(n)}$. We have
$$
\<x\, \pi_n(x)\, , \, \pi_{n-1}(x) \> =
\<\pi_n(x)\, , \, x\,\pi_{n-1}(x) \> =
\<\pi_n(x)\, , \, x^n \> = \check S_{n^{n+1}} \ ,
$$
and, also, by expanding the quasi-determinant $\pi_n(x)$
by its last column,
$$
\<x\, \pi_n(x)\, , \, \pi_{n}(x) \> =
\< x^{n+1}, \pi_n(x) \>
- \check S_{(n-1)^{n-1}n}^*\, \check S_{(n-1)^n}^{-1} \< x^n, \pi_n(x) \>
$$
$$
=
\check S_{n^n(n+1)}^* - \check S_{(n-1)^{n-1}n}^*\, \check S_{(n-1)^n}^{-1}
\, \check S_{n^{n+1}} \ .
$$
Hence we obtain

\begin{proposition}
The noncommutative orthogonal polynomials $\pi_n(x)$ satisfy the
three term recurrence relation
\begin{equation}\label{RECUR}
\pi_{n+1}(x) - (x - \check S_{n^n(n+1)}^* \check S_{n^{n+1}}^{-1}
+ \check S_{(n-1)^{n-1}n}^*\, \check S_{(n-1)^n}^{-1}) \pi_n(x)
+\check S_{n^{n+1}}\,\check S_{(n-1)^n}^{-1} \, \pi_{n-1}(x) = 0
\end{equation}
for $n\ge 1$. \cqfd
\end{proposition}

\smallskip
Applying the $*$ involution to (\ref{RECUR}), we get a similar
recurrence for the polynomials $\pi_n^*(x)$, namely
\begin{equation}\label{RECUR2}
\pi_{n+1}^*(x) - \pi_n^*(x)\,(x - \check S_{n^{n+1}}^{-1}\,\check S_{n^n(n+1)}
+ \check S_{(n-1)^n}^{-1}\,\check S_{(n-1)^{n-1}n})
+ \check S_{(n-1)^n}^{-1}\,\check S_{n^{n+1}} \, \pi_{n-1}(x) = 0
\end{equation}
Note also that the symmetric function
$\check S_{n^n(n+1)}^* - \check S_{(n-1)^{n-1}n}^*\, \check S_{(n-1)^n}^{-1}
\, \check S_{n^{n+1}}$
being equal to the scalar product $\<x\, \pi_n(x)\, , \, \pi_n(x) \>$,
it is invariant under the $*$ involution, which gives the following
interesting identity
$$
\check S_{n^n(n+1)}^* - \check S_{(n-1)^{n-1}n}^*\, \check S_{(n-1)^n}^{-1}
\, \check S_{n^{n+1}}
=
\check S_{n^n(n+1)} - \check S_{n^{n+1}}\, \check S_{(n-1)^n}^{-1}
\check S_{(n-1)^{n-1}n}\ .
$$


\subsection{Noncommutative continued $J$-fractions}\label{NCJFRAC}

In this section we consider continued fractions of the type
\begin{equation}\label{JFRAC}
b_1  \,
{ 1 \over x - a_1 + b_2\, \displaystyle
{ 1 \over x - a_2 + b_3\, \displaystyle
{ 1 \over \, \ddots \, \displaystyle
{ \hbox{ } \atop \displaystyle
{ \hbox{ } \atop \displaystyle
{ 1 \over x - a_n + b_{n+1} \, \displaystyle
{ 1 \over \, \ddots \,
}}}}}}}
\end{equation}
In the commutative case, these fractions are called $J$-fractions,
and their connexion with orthogonal polynomials and Pad\'e
approximants is well-known (\cf \cite{Wa}). As observed by
several authors (see {\it e.g.} \cite{GG} \cite{BB}, and \cite{Dr}
for other references)
similar properties hold true when $(a_i)$ and $(b_i)$ are
replaced by two sequences of noncommutative indeterminates.

\medskip
Denote by $p_n(x)\, q_n(x)^{-1}$ the $n$-th partial quotient of
(\ref{JFRAC}):
$$
p_n(x)\, q_n(x)^{-1} =
b_1  \,
{ 1 \over x - a_1 + b_2\, \displaystyle
{ 1 \over x - a_2 + b_3\, \displaystyle
{ 1 \over \, \ddots \, \displaystyle
{ \hbox{ } \atop \displaystyle
{ \hbox{ } \atop \displaystyle
{ 1 \over x - a_n
}}}}}}
$$
It is convenient to put also $p_0 = 0$, $q_0 = 1$. The polynomials
$p_n(x)$ and $q_n(x)$ then satisfy the three term recurrence relation
\cite{GR2}, \cite{BR}
\begin{equation}\label{JREC}
p_{n+1} = p_n\,(x-a_{n+1}) + p_{n-1}\, b_{n+1}\ , \ \ \ \ \
q_{n+1} = q_n\,(x-a_{n+1}) + q_{n-1}\, b_{n+1}
\end{equation}
Comparing with (\ref{RECUR2}), we are led to the following result
\begin{theorem}
Let $S_i$ be a sequence of noncommutative indeterminates. The expansion
of the formal power series $\sum_{k\ge 0}S_k\,x^{-k-1}$ into a
noncommutative $J$-fraction is equal to
\begin{equation}\label{JEXPAN}
b_1  \,
{ 1 \over x - a_1 + b_2\, \displaystyle
{ 1 \over \, \ddots \, \displaystyle
{ \hbox{ } \atop \displaystyle
{ \hbox{ } \atop \displaystyle
{ 1 \over x - a_n + b_{n+1} \, \displaystyle
{ 1 \over \, \ddots \,
}}}}}}
\end{equation}
where $a_1=S_1$, $b_1=S_0$, and
$$a_n=\check S_{(n-1)^n}^{-1}\,\check S_{(n-1)^{n-1}n}
-\check S_{(n-2)^{n-1}}^{-1}\,\check S_{(n-2)^{n-2}(n-1)}\ ,$$
$$b_n = \check S_{(n-2)^{n-1}}^{-1}\,\check S_{(n-1)^n}$$
for $n\ge 2$.
The $n$-th partial quotient $p_n(x)\, q_n(x)^{-1}$ is given by
\begin{equation}\label{NUMEXP}
p_n(x)=
\left|\matrix{
S_{n-1} & \cdots & S_{2n-2} & S_{2n-1} \cr
\vdots  & \ddots & \vdots   & \vdots   \cr
S_0     & \cdots & S_{n-1}  & S_n      \cr
0       & \cdots & \sum_{k=0}^{n-2} S_k\, x^{n-k-2} &
\bo{\sum_{k=0}^{n-1} S_k\, x^{n-k-1}} \cr
}\right| \ ,
\end{equation}
\begin{equation}\label{DENEXP}
q_n(x)=
\left|\matrix{
S_{n-1} & \cdots & S_{2n-2} & S_{2n-1} \cr
\vdots  & \ddots & \vdots   & \vdots   \cr
S_0     & \cdots & S_{n-1}  & S_n      \cr
1       & \cdots & x^{n-1}  & \bo{x^n} \cr
}\right| \ ,
\end{equation}
and one has
$$
p_n(x)\, q_n(x)^{-1} = S_0\,x^{-1} + \cdots + S_{2n-1}\,x^{-2n} + O(x^{-2n-1})
\ .
$$
\end{theorem}

\Proof This is just a matter of properly relating the results of the previous
sections. We first prove that the numerator and denominator of the $n$-th
convergent of (\ref{JEXPAN}) admit the quasi-determinantal expressions
(\ref{NUMEXP})
and (\ref{DENEXP}). To this aim we denote temporarily by $\overline{p}_n(x)$,
$\overline{q}_n(x)$ the right-hand sides of (\ref{NUMEXP}) and (\ref{DENEXP}),
and we check that these polynomials satisfy the recurrence relations
(\ref{JREC}).
The relation for $\overline{q}_n(x)$ is none other than (\ref{RECUR2}) (with
the
necessary change of notations). Since $\overline{q}_0 = q_0$ and
$\overline{q}_1 = q_1$, we deduce that $\overline{q}_n = q_n$ for all $n$.
The same argument is applied to  $\overline{p}_n(x)$. Indeed, it results from
(\ref{NUMPAD2}), (\ref{DENPAD2}) and Proposition~\ref{PAD} with $t=x^{-1}$,
$m=n-1$, that
$$ \overline{p}_n(x) = (\sum_{k\ge 0}S_k\,x^{-k-1})\,q_n(x) + O(x^{-2n-1}) \
,$$
which shows that $\overline{p}_n(x)$ satisfies the same recurrence relation
as $q_n(x)$. Since $\overline{p}_1 = p_1$, and $\overline{p}_2=p_2$ as one
easily checks, we get that $\overline{p}_n= p_n$ for all $n$, and the rest
of the theorem follows. \cqfd


\subsection{Rational power series and Hankel matrices} \label{RPS}

Denote by $K[[t]]$ the ring of formal power series in $t$ with coefficients
in the skew field $K$.
A series $F(t)=\ \sum_{k\geq 0}\, S_k\, t^k$ in $K[[t]]$ is said to be {\em
rational}
iff it is a rational fraction, that is, iff there exists polynomials $P$, $Q$
in $K[t]$ such that $F(t) = Q(t)^{-1}\, P(t)$.

\smallskip
We begin by extending to the noncommutative case the classical characterization
of rational series by means of Hankel determinants.
With every series $F(t) = \sum_{k\geq 0} \ S_k \, t^k$ in $K[[t]]$ is
associated an infinite Toeplitz matrix ${\bf S} = (S_{j-i})_{i,j\geq 0}$.
The quasi-minors of ${\bf S}$ are conveniently seen as specializations
of quasi-Schur functions and we shall still denote them by $\check S_I$,
the series $F$ being understood. In particular, the quasi-Schur functions
indexed by rectangular partitions correspond to Hankel quasi-determinants.

\begin{proposition}\label{RATIO}
The series $F(t)$ is rational iff ${\bf S}$ is of finite rank. More
explicitely,
$F(t)=Q^{-1}(t)\,P(t)$ where $P(t)$ and $Q(t)$ have respective degrees $m$ and
$n$, and no common left factor, iff the submatrix
$$
R=
\left(\matrix{
S_m & S_{m+1} & \ldots & S_{m+n-1} \cr
S_{m-1} & S_m & \ldots & S_{m+n-2} \cr
\vdots  & \vdots & \ddots & \vdots \cr
S_{m-n+1} & S_{m-n+2} & \ldots & S_m \cr
}\right)
$$
is invertible and $\check{S}_{(m+1)^n(m+p)} = 0 $ for every $p\geq 1$.
$Q$ and $P$ are then given by formulas {\rm (\ref{DENPAD}) and
(\ref{NUMPAD})}.
\end{proposition}

\Proof Suppose that $R$ is invertible. Then one can define $Q$ and $P$
by formulas (\ref{DENPAD}) (\ref{NUMPAD}), and the same computation as in
the proof of Proposition~\ref{PAD} shows that
$$
Q(t)\,F(t) = P(t) + \ \sum_{p\geq 1} \ \check{S}_{(m+1)^n(m+p)}\, t^{m+n+p} \ .
$$
Hence, if $\check S_{(m+1)^n(m+p)} = 0$ for $p\geq 1$, one obtains
$F(t) = Q^{-1}(t)\,P(t)$. Conversely, if $F(t)$ is rational, one can
write $F(t) = Q^{-1}(t)\,P(t)$ with $Q(0)=1$, $P$ and $Q$ having no common
left factor, that is, $P$ and $Q$ of minimal degree. Therefore, if
${\rm deg}Q = n,\ {\rm deg}P = m$,
we have, setting $Q(t) = \sum_{0\le i \le n} b_i\, t^i$,
\begin{equation}
\sum_{0\le i \le n} b_i \, S_{r-i} = 0 \ , \ \ \ r\ge m+1 \ ,\label{RELAT}
\end{equation}
which shows that the matrix $R$ has rank less or equal to $n$. This rank
cannot be less than $n$, otherwise $P$ and $Q$ would not be of minimal degree.
Therefore $R$ is invertible, and the relation $\check{S}_{(m+1)^n(m+p)} = 0 $
follows from (\ref{RELAT}). \cqfd

\medskip
The coefficients $S_i$ of a rational series $F(t) = Q^{-1}(t)\,P(t)$ satisfy
many remarkable identities. The following result may be seen as a natural
analog of the factorization formulas for Schur polynomials in two finite sets
of commutative variables given in \cite{BeRe}. The reader is referred to
\cite{Pr} for algebro-geometric applications of these important relations
in the commutative case.

\medskip
We denote by $(I,\, m^n+J)$ the partition whose Young diagram is

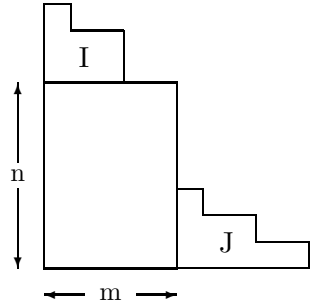
\setlength{\unitlength}{1pt}

\centerline{
\begin{picture}(120,140)(0,-20)
\put(0,0){\framebox(50,70){}}
\put(0,70){\line(0,1){30}}
\put(0,100){\line(1,0){10}}
\put(10,100){\line(0,-1){10}}
\put(10,90){\line(1,0){20}}
\put(30,90){\line(0,-1){20}}
\put(50,0){\line(1,0){50}}
\put(100,0){\line(0,1){10}}
\put(80,10){\line(1,0){20}}
\put(80,10){\line(0,1){10}}
\put(60,20){\line(1,0){20}}
\put(60,20){\line(0,1){10}}
\put(60,30){\line(-1,0){10}}
\put(69,10){\makebox(0,0){\hbox{\small J}}}
\put(15,80){\makebox(0,0){\hbox{\small I}}}
\put(15,-10){\vector(-1,0){15}}
\put(35,-10){\vector(1,0){15}}
\put(25,-10){\makebox(0,0){\hbox{\footnotesize m}}}
\put(-10,30){\vector(0,-1){30}}
\put(-10,40){\vector(0,1){30}}
\put(-10,35){\makebox(0,0){\hbox{\footnotesize n}}}
\end{picture}
}

\begin{proposition}\label{FACT}
Let $F(t) = Q^{-1}(t)\,P(t)$ be a rational series, and assume that
$P(t)$ and $Q(t)$ have respective degrees $m$ and $n$, and no common left
factor. Let $I$ be a partition whose largest part is less or equal to $m$,
$J$ be a partition whose length is less or equal to $n$, and $i$, $j$ be
integers. There holds
\begin{equation}\label{FACT1}
\check S_{(1^i,\,m^n+J)} = \check S_{(1^i,\,m^n)}\, \check S_{m^n} ^{-1}\,
\check S_{(m^n+J)} \ ,
\end{equation}
\begin{equation}\label{FACT2}
\check S_{(I,\,m^n+j)} = \check S_{(I,\,m^n)}\, \check S_{m^n} ^{-1} \,
\check S_{(m^n+j)} \ .
\end{equation}
\end{proposition}

Replacing each quasi-Schur function $\check S_H$ by the Young diagram of $H$,
relations (\ref{FACT1}) and (\ref{FACT2}) are pictured as follows

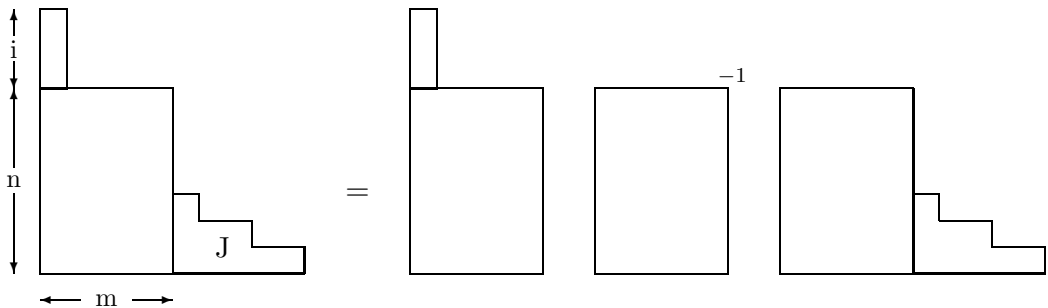
\centerline{
\begin{picture}(350,140)(0,-20)
\put(0,0){\framebox(50,70){}}
\put(0,70){\framebox(10,30){}}
\put(50,0){\line(1,0){50}}
\put(100,0){\line(0,1){10}}
\put(80,10){\line(1,0){20}}
\put(80,10){\line(0,1){10}}
\put(60,20){\line(1,0){20}}
\put(60,20){\line(0,1){10}}
\put(60,30){\line(-1,0){10}}
\put(69,10){\makebox(0,0){\hbox{\small J}}}
\put(15,-10){\vector(-1,0){15}}
\put(35,-10){\vector(1,0){15}}
\put(25,-10){\makebox(0,0){\hbox{\footnotesize m}}}
\put(-10,30){\vector(0,-1){30}}
\put(-10,40){\vector(0,1){30}}
\put(-10,35){\makebox(0,0){\hbox{\footnotesize n}}}
\put(-10,80){\vector(0,-1){10}}
\put(-10,90){\vector(0,1){10}}
\put(-10,85){\makebox(0,0){\hbox{\footnotesize i}}}
\put(120,30){\makebox(0,0){\hbox{=}}}
\put(140,0){\framebox(50,70){}}
\put(140,70){\framebox(10,30){}}
\put(210,0){\framebox(50,70){}}
\put(262,73){\makebox(0,0){${}^{-1}$}}
\put(280,0){\framebox(50,70){}}
\put(330,0){\line(1,0){50}}
\put(380,0){\line(0,1){10}}
\put(360,10){\line(1,0){20}}
\put(360,10){\line(0,1){10}}
\put(340,20){\line(1,0){20}}
\put(340,20){\line(0,1){10}}
\put(340,30){\line(-1,0){10}}
\end{picture}
}

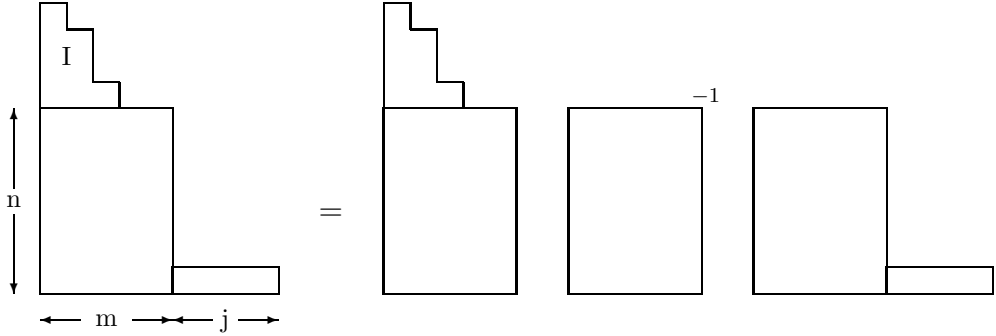
\centerline{
\begin{picture}(350,145)(0,-20)
\put(0,0){\framebox(50,70){}}
\put(50,0){\framebox(40,10){}}
\put(0,70){\line(0,1){40}}
\put(0,110){\line(1,0){10}}
\put(10,100){\line(0,1){10}}
\put(10,100){\line(1,0){10}}
\put(20,80){\line(0,1){20}}
\put(20,80){\line(1,0){10}}
\put(30,80){\line(0,-1){10}}
\put(10,90){\makebox(0,0){\hbox{\footnotesize I}}}
\put(15,-10){\vector(-1,0){15}}
\put(35,-10){\vector(1,0){15}}
\put(25,-10){\makebox(0,0){\hbox{\footnotesize m}}}
\put(-10,30){\vector(0,-1){30}}
\put(-10,40){\vector(0,1){30}}
\put(-10,35){\makebox(0,0){\hbox{\footnotesize n}}}
\put(65,-10){\vector(-1,0){15}}
\put(75,-10){\vector(1,0){15}}
\put(70,-10){\makebox(0,0){\hbox{\footnotesize j}}}
\put(110,30){\makebox(0,0){\hbox{=}}}
\put(130,0){\framebox(50,70){}}
\put(130,70){\line(0,1){40}}
\put(130,110){\line(1,0){10}}
\put(140,100){\line(0,1){10}}
\put(140,100){\line(1,0){10}}
\put(150,80){\line(0,1){20}}
\put(150,80){\line(1,0){10}}
\put(160,80){\line(0,-1){10}}
\put(200,0){\framebox(50,70){}}
\put(252,73){\makebox(0,0){${}^{-1}$}}
\put(270,0){\framebox(50,70){}}
\put(320,0){\framebox(40,10){}}
\end{picture}
}

The proof of Proposition~\ref{FACT} is a consequence of the next Lemma,
valid for generic quasi-Schur functions and of independent interest.

\begin{lemma}
Let $H=(h_1,\, \ldots \,,\,h_r)$ be a partition, and $j,\,k$ be two
integers with $k \leq h_1$. Then,
\begin{equation}\label{LEM}
\left|\matrix{
\bo{\check S_{(k,\,H+j)}}  & \check S_{(k,\,H)}\cr
 \check S_{H+j}            & \check S_H       \cr
}\right|
=
\check S_{(h_1+1,\,h_2+1,\,\ldots \,h_r+1,\,h_r+j)}
\ .
\end{equation}
\end{lemma}

\Proof This is a simple consequence of Bazin's theorem. \cqfd

\noindent
\it Proof of Proposition \rm\ref{FACT}\ --- \
Let $K$ denote the partition obtained from $I$ by removing its
first part. The hypothesis imply that, in the
specialization to the series $F$, the quasi-Schur functions
$\check S_{(L,\,(m+1)^n(m+p))}$ are sent to $0$, for any $p\geq 1$ and any
partition $L$ whose largest part is less or equal to $m+1$.
Therefore, putting
$H=(K,\,m^n)$ in (\ref{LEM}), we get
$$
\check S_{(I,\,m^n+j)} = \check S_{(I,\,m^n)} \, \check S_{(K,\,m^n)}^{-1} \,
\check S_{(K,\,m^n+j)}
\ ,
$$
and formula (\ref{FACT2}) follows by induction on the length of $I$. Formula
(\ref{FACT1}) may be proved similarly, or derived directly from (\ref{FACT2})
by means of the involution $\omega$. \cqfd


\subsection{The characteristic polynomial of the generic matrix, and a
non\-com\-mu\-ta\-ti\-ve Cayley-Hamilton theorem} \label{CHAR}

In this section, we illustrate the results of Section~\ref{RPS} on a
fundamental example, namely we take $F(t)$ to be the generating series
$\sigma(t,\alpha_i) =\, \sum_{k\geq 0}\, S_k(\alpha_i)\,t^k= |I-tA|_{ii}^{-1}$
of the complete symmetric functions associated with the generic $n\times n$
matrix $A$, and its $i$-th diagonal entry $a_{ii}$
({\it cf.} Section~\ref{MATSYM}). We first prove that this series is rational.

\begin{proposition} \label{RATSIG}
For the generic $n\times n$ matrix $A=(a_{ij})$, there holds
$$
\sigma(t,\alpha_i) = Q_i(t)^{-1}\, P_i(t) \ ,
$$
$Q_i$ and $P_i$ being polynomials of degree $n$ and $n-1$ without left common
factor, with coefficients in the free field generated by the entries $a_{ij}$
of $A$. Moreover, assuming the normalizing condition $Q_i(0)=P_i(0)=1$, the
polynomials $Q_i(t)$, $P_i(t)$ are unique.
\end{proposition}

\Proof
Let $v_k$ denote the $i$-th row vector of $A^k$. In other words, the sequence
$v_k$ is defined by $v_0 = (0,\,\ldots ,\,1,\,\ldots ,\,0)$ (the entry $1$ is
at the $i$-th place) and, for $k\geq 0$, by $v_{k+1}=v_k\,A$.
The $v_k$ may be regarded as vectors in $\fsf^n$, which is at the same time a
left and right $n$-dimensional vector space on the free field $\fsf$. Hence,
there exists scalars $\lambda_0,\,\lambda_1,\,\ldots ,\,\lambda_n$ in $\fsf$
such that one has
\begin{equation}\label{CL}
\lambda_0\,v_0 + \lambda_1\,v_1 + \ldots +\lambda_n\,v_n = 0\ .
\end{equation}
Multiplying (\ref{CL}) from the right by $A^j$, and using the fact that the
$i$-th component of $v_k$ is equal to $S_k(\alpha_i)$, one obtains the
relations
\begin{equation}\label{LINSYS}
\sum_{0\leq p \leq n} \ \lambda_p \,S_{p+j}(\alpha_i) = 0\ , \ \ j\geq 0 \ ,
\end{equation}
which shows that the rank of the Hankel matrix ${\bf S}(\alpha_i)$ is finite,
equal to $n$. More precisely, we have,
\begin{equation}
\left|\matrix{
S_n(\alpha_i)  & \ldots & S_{2n-1}(\alpha_i) & \bo{S_{j+n}(\alpha_i)}\cr
\vdots & \ddots      & \vdots   & \vdots    \cr
S_1(\alpha_i) &  \ldots & S_n(\alpha_i) & S_{j+1}(\alpha_i) \cr
S_0(\alpha_i)   & \ldots & S_{n-1}(\alpha_i) & S_j(\alpha_i)   \cr
}\right|
= 0 \ , \ \ j\geq 0 \ ,
\end{equation}
and the conclusion follows from Proposition~\ref{RATIO}. \cqfd

Let $A^{ii}$ denote the $(n-1) \times (n-1)$
matrix obtained by removing the $i$-th row and column of $A$.
As was explained in the introduction of Section~\ref{MATSYM}, the series
$\sigma(t,\alpha_i)$ is a noncommutative analog of
${\rm det}(I-tA^{ii})/{\rm det}(I-tA)$, that is, of the ratio of the
characteristic polynomial of $A^{ii}$ to the characteristic polynomial of $A$
(up to the change of variable $u=t^{-1}$).
Therefore, we can regard the polynomials $Q_i(t)$ of Proposition~\ref{RATSIG}
as natural analogs of ${\rm det}(I-tA)$. This is supported by the fact that
these polynomials give rise to a form of the Cayley-Hamilton theorem
for the generic noncommutative matrix.

\smallskip
For instance, when $n=2$,
the two polynomials $Q_1(t)$, $Q_2(t)$ are given by
\begin{equation}
Q_1(t) = 1 - (a_{11} + a_{12}\,a_{22}\,a_{12}^{-1})\,t
             + (a_{12}\,a_{22}\,a_{12}^{-1}\,a_{11} - a_{12}\,a_{21})\,t^2 \ ,
\end{equation}
\begin{equation}
Q_2(t) = 1 - (a_{22} + a_{21}\,a_{11}\,a_{21}^{-1})\,t
             + (a_{21}\,a_{11}\,a_{21}^{-1}\,a_{22} - a_{21}\,a_{12})\,t^2  \ .
\end{equation}

\smallskip\noindent
Writing for short $Q_i(t) = 1 - {\rm tr}_i(A)\,t + {\rm det}_i(A)\,t^2$
for $i=1,\,2$, one can check that
\begin{equation}
A^2 -
\left(\matrix{
{\rm tr}_1(A)& 0 \cr
\noalign{\vskip 1mm}
0 &  {\rm tr}_2(A)\cr
}\right)
\,A
+
\left(\matrix{
{\rm det}_1(A)& 0 \cr
\noalign{\vskip 1mm}
0 &  {\rm det}_2(A)\cr
}\right)
= 0 \ ,
\end{equation}
the Cayley-Hamilton theorem for the generic matrix of order 2. The general
result is as follows. Set
\begin{equation}
Q_i(t) = \ \sum_{0\leq j \leq n} \ (-1)^j\, L_j^{(i)}(A)\,t^j \ ,
\end{equation}
\begin{equation}
L_j(A) =
\left(\matrix{
L_j^{(1)}(A) & \ldots & 0 \cr
\vdots       & \ddots & \vdots \cr
\noalign{\vskip 1mm}
0            & \ldots & L_j^{(n)}(A)
}\right)
\ ,
\end{equation}
for every $1 \leq i \leq n$.

\begin{theorem} \label{CAHA}
The generic noncommutative matrix $A$ satisfy the
following polynomial equation, the coefficients of which are diagonal
matrices with entries in $\fsf$
\begin{equation}\label{CH}
\sum_{0\leq j \leq n} \ (-1)^j\, L_j(A)\,A^{n-j} = 0 \ .
\end{equation}
\end{theorem}

\Proof
One has to check that the $i$-th row vector of the left-hand side of (\ref{CH})
is zero. But this is exactly
what is done in the proof of Proposition~\ref{RATSIG}. Indeed, keeping the
notations therein, the coefficients $L_j^{(i)}(A)$ of
\begin{equation}\label{ORTHCHAR}
Q_i(t)
=
\left|\matrix{
S_n(\alpha_i) & \ldots & S_{2n-1}(\alpha_i) & \bo{1}\cr
\vdots  & \ddots      & \vdots   & \vdots    \cr
S_1(\alpha_i) & \ldots & S_n(\alpha_i) & t^{n-1} \cr
S_0(\alpha_i) & \ldots & S_{n-1}(\alpha_i) & t^n  \cr
}\right|
\ ,
\end{equation}
form a system of solutions $\lambda_j$ for the linear system~(\ref{LINSYS}),
and therefore satisfy (\ref{CL}). \cqfd

\medskip
Due to the significance of the polynomials $Q_i(t)$, we shall give different
expressions of them. The first expression,
already encountered, is formula~(\ref{ORTHCHAR}). Expanding this
quasi-determinant by its last column, we compute the coefficients
$$
L_{n-k}^{(i)}(A) = (-1)^{n-k-1}
\left|\matrix{
S_n(\alpha_i)  & \!\!\ldots\!\! & \bo{S_{2n-1}(\alpha_i)}\cr
\vdots  & \!\!      & \vdots       \cr
S_{k+1}(\alpha_i)  & \!\!\ldots\!\! & S_{k+n}(\alpha_i)  \cr
S_{k-1}(\alpha_i)  & \!\!\ldots\!\! & S_{k+n-2}(\alpha_i)  \cr
\vdots  & \!\!      & \vdots       \cr
S_0(\alpha_i)   & \!\!\ldots\!\! & S_{n-1}(\alpha_i)    \cr
}\right|
\,
\left|\matrix{
S_{n-1}(\alpha_i)  & \!\!\ldots\!\! & S_{2n-2}(\alpha_i)\cr
\vdots  & \!\!      & \vdots       \cr
S_{k}(\alpha_i) & \!\!\ldots\!\! & \bo{S_{k+n-1}(\alpha_i)}  \cr
\vdots  & \!\!      & \vdots       \cr
S_0(\alpha_i)   & \!\!\ldots\!\! & S_{n-1}(\alpha_i)    \cr
}\right|^{-1}
\ ,\ i\geq 1 \ .
$$

Recalling that $\sigma(t,\alpha_i)=\lambda(-t,\alpha_i)^{-1}$,
we obtain by means of (\ref{NUMPAD})
the following expressions in terms of the elementary symmetric functions
$\Lambda_j(\alpha_i)$
\begin{equation}
Q_i(t) =
\left|\matrix{
\bo{\Lambda_0(\alpha_i) - t\, \Lambda_1(\alpha_i)
+\ldots +(-t)^n\,\Lambda_n(\alpha_i)} & \Lambda_{n+1}(\alpha_i)
& \ldots &\Lambda_{2n-1}(\alpha_i) \cr
-t\,\Lambda_0(\alpha_i) + \ldots +(-t)^n\,\Lambda_{n-1}(\alpha_i) &
\Lambda_n(\alpha_i) & \ldots &\Lambda_{2n-2}(\alpha_i) \cr
\vdots                                              &  \vdots   & \ddots
& \vdots         \cr
(-t)^{n-1}\,\Lambda_0(\alpha_i)+(-t)^n\,\Lambda_1(\alpha_i)
      & \Lambda_2(\alpha_i) & \ldots & \Lambda_n(\alpha_i) \cr
}\right| \ ,
\end{equation}
\smallskip
\begin{equation}\label{CHARLAMB}
L_k^{(i)}(A) =
\left|\matrix{
\bo{\Lambda_k(\alpha_i)} & \Lambda_{n+1}(\alpha_i) & \ldots &
\Lambda_{2n-1}(\alpha_i) \cr
\Lambda_{k-1}(\alpha_i) & \Lambda_n(\alpha_i) & \ldots &
\Lambda_{2n-2}(\alpha_i) \cr
\vdots        &  \vdots   & \ddots       & \vdots         \cr
\Lambda_{k-n+1}(\alpha_i)  & \Lambda_2(\alpha_i) & \ldots &
\Lambda_n(\alpha_i) \cr
}\right| \ .
\end{equation}
\smallskip
\noindent
Also, as shown in the Proof of \ref{RATSIG}, $Q_i(t)$ can be expressed in
terms of the entries of $A^k$, that we denote by $a_{ij}^{(k)}$. Indeed,
one has
\begin{equation}\label{KRYLOV}
Q_i(t)=
\left|\matrix{
a_{i1}^{(n)} & a_{i2}^{(n)}& \ldots & a_{in}^{(n)} & \bo{1}\cr
\vdots &\vdots &       & \vdots   & \vdots    \cr
\noalign{\vskip 1,5mm}
a_{i1}^{(1)} & a_{i2}^{(1)} & \ldots & a_{in}^{(1)} & t^{n-1} \cr
\noalign{\vskip 1,5mm}
a_{i1}^{(0)}   & a_{i2}^{(0)}   & \ldots & a_{in}^{(0)} & t^n  \cr
}\right|
\ .
\end{equation}

\smallskip
\noindent
One can recognize in (\ref{KRYLOV}) the exact noncommutative analog of
Krylov's expression for the characteristic polynomial of a matrix ({\it cf.}
\cite{Gan}). Thus we find again that, when specialized to the commutative
generic matrix, all the polynomials $Q_i(t)$ reduce to the familiar
characteristic polynomial (up to the change of variable $u=t^{-1}$).
In particular, the leading coefficients of the $Q_i(t)$ provide $n$
noncommutative analogs of the determinant, that we shall call
the $n$ {\it pseudo-determinants} of the generic $n\times n$ matrix A, and
denote by ${\rm det}_i(A)\,,\, i=1,\ldots\,,\,n$. They admit the following
quasi-determinantal expressions, obtained from (\ref{CHARLAMB}) and
(\ref{KRYLOV})
\begin{equation}\label{PSEUDET1}
{\rm det}_i(A)
=
\left|\matrix{
\bo{\Lambda_n(\alpha_i)} & \Lambda_{n+1}(\alpha_i) & \ldots &
\Lambda_{2n-1}(\alpha_i) \cr
\Lambda_{n-1}(\alpha_i) & \Lambda_n(\alpha_i) & \ldots &
\Lambda_{2n-2}(\alpha_i) \cr
\vdots        &  \vdots   & \ddots       & \vdots         \cr
\Lambda_{1}(\alpha_i)  & \Lambda_2(\alpha_i) & \ldots &
\Lambda_n(\alpha_i) \cr
}\right| \ ,
\end{equation}
\smallskip
\begin{equation}\label{PSEUDET2}
{\rm det}_i(A) = (-1)^{n-1}
\left|\matrix{
a_{i1}^{(n)} & \ldots & \bo{a_{ii}^{(n)}}& \ldots & a_{in}^{(n)}\cr
\vdots       &        &\vdots            &        &  \vdots    \cr
\noalign{\vskip 1,5mm}
a_{i1}^{(2)} & \ldots & a_{ii}^{(2)} & \ldots & a_{in}^{(2)}  \cr
\noalign{\vskip 1,5mm}
a_{i1}^{(1)}   & \ldots & a_{ii}^{(1)}   & \ldots & a_{in}^{(1)} \cr
}\right|
\ .
\end{equation}

\medskip
We shall illustrate the results of this section on two important examples,
namely, the case of the Lie algebra $gl_n$, and the case of the quantum
group $GL_q(n)$.

\begin{example} {\rm
We take $A$ to be the matrix $E_n$ of standard generators of $U(gl_n)$
({\it cf.} Section~\ref{NCUGL}). In this specialization, the
pseudo-determinants ${\rm det}_i(E_n)$ are all equal to the Capelli
determinant. More precisely, we have, using the notations of \ref{NCUGL}
$$
Q_i(t)={\rm det}(I-t(E_n+(n-1)I))
=
\left|
\matrix{
1\!-\! t(e_{11}\!+\! n\!-\! 1) & \!\! -te_{12}   &  \ldots  &-te_{1n}  \cr
-te_{21} & \!\! 1\!-\! t(e_{22}\!+\! n\!-\! 2) & \ldots &-te_{2n}   \cr
\vdots & \!\! \vdots     & \ddots  & \vdots \cr
-te_{n1} &\!\! -te_{n2} & \ldots   & 1\!-\! te_{nn}\cr
}\right| \ ,
$$
for all $i=1,\ldots ,\,n$. Let
${\overline Q}(u) = u^n\,Q_i(u^{-1}) = {\rm det}(uI-(E_n+(n-1)I))$
denote the characteristic polynomial of $E_n$. By Theorem~\ref{CAHA},
${\overline Q}(E_n)=0$. Consider an irreducible finite-dimensional
$gl_n$-module $V$ with highest weight
$\lambda = (\lambda_1,\,\ldots \,,\,\lambda_n)$. It follows from
Section~\ref{NCUGL} that the coefficients of ${\overline Q}(u)$ belong
to the center of $U(gl_n)$, and act upon $V$ as the elementary symmetric
fonctions of the commutative variables
$\lambda_1 + n -1,\, \lambda_2 +n-2,\, \ldots \,,\,\lambda_n$. Therefore,
if we denote by $E_n^{\lambda} = (e_{ij}^{(\lambda)})$ the matrix whose
$(i,j)$-entry is the image of $e_{ij}$ in the module $V$, we find that
$$
\prod_{1\le k \le n} \ (E_n^{\lambda} - (\lambda_k+n-k)) = 0 \ ,
$$
the so-called characteristic identity satisfied by the generators of $gl_n$.
}
\end{example}

\smallskip
\begin{example}
{\rm Here, $A$ is taken to be the matrix $A_q=(a_{ij})$ of the generators of
the quantum group $GL_q(n)$. Recall that the $a_{ij}$ are subject to the
following relations

\medskip
\vskip 0.5mm
\centerline{$
a_{ik}a_{il}=q^{-1}a_{il}a_{ik} \ \ \ {\rm for} \ k<l, \quad
a_{ik}a_{jk}=q^{-1}a_{jk}a_{ik} \ \ \ {\rm for} \ i<j,
$}

\medskip
\centerline{$
a_{il}a_{jk} = a_{jk}a_{il} \ \ \ {\rm for} \ i<j,\ k<l,
$}

\medskip
\centerline{$
a_{ik}a_{jl}-a_{jl}a_{ik}= (q^{-1}\!-\! q)\, a_{il}a_{jk} \ \ \ {\rm for}
\ i<j,\ k<l\ .
$}

\medskip\noindent
In this specialization, the pseudo-determinants ${\rm det}_i(A_q)$ are
all equal, up to a power of $q$, to the quantum determinant
$$
\left|\matrix{
a_{11} & a_{12} &\ldots & a_{1n} \cr
a_{21} & a_{22} &\ldots & a_{2n} \cr
\vdots &\vdots  &\ddots & \vdots \cr
a_{n1} & a_{n2} &\ldots & a_{nn} \cr
}\right|_q
:=
\ \displaystyle\sum_{\sigma\in{\goth{S}}_n} \
(-q)^{-\ell(\sigma)} \, a_{1\,j_{\sigma(1)}}\,\ldots\, a_{n\,j_{\sigma(n)}}
\ .
$$
The other coefficients of the $Q_i(t)$ also specialize to
polynomials in the $a_{ij}$, and one recovers the quantum Cayley-Hamilton
theorem proved by Zhang  in \cite{Zh}.
For instance, when $n=3$, one obtains
$$
Q_1(t)=1-(a_{11}+q^{-1}a_{22}+q^{-1}a_{33})\,t
$$
$$
+\left(q^{-1}
\left|\matrix{
a_{11} & a_{12} \cr
a_{21} & a_{22} \cr
}\right|_q
+
q^{-1}
\left|\matrix{
a_{11} & a_{13} \cr
a_{31} & a_{33} \cr
}\right|_q
+
q^{-2}
\left|\matrix{
a_{22} & a_{23} \cr
a_{32} & a_{33} \cr
}\right|_q\right)\,t^2
-q^{-2}\,
\left|\matrix{
a_{11} & a_{12}  & a_{13} \cr
a_{21} & a_{22}  & a_{23} \cr
a_{31} & a_{32}  & a_{33} \cr
}\right|_q
t^3
\ ,
$$
and for $n=2$, the Cayley-Hamilton theorem assumes the form
$$
A_q^2 -(q^{1/2}a_{11} + q^{-1/2}a_{22})
\left(\matrix{
q^{-1/2} & 0 \cr
\noalign{\vskip 1mm}
0  & q^{1/2}   \cr
}\right)\,A_q
+
(a_{11}a_{22}-q^{-1}a_{12}a_{21})
\left(\matrix{
q^{-1} & 0 \cr
\noalign{\vskip 1mm}
0  & q \cr
}\right)
=0 \ .
$$
}\end{example}

\medskip
In view of these examples, some questions arise naturally. The coefficients
$L_j^{(i)}(A)$ involved in the Cayley-Hamilton theorem for the generic
noncommutative matrix $A$ are fairly complicated expressions in terms
of the entries $a_{ij}$ of $A$, involving inverses $a_{ij}^{-1}$, and
thus belonging to the skew field generated by the $a_{ij}$. Moreover, the
$L_j^{(i)}(A)$ depend on the index $i$, that is, the coefficients of
the noncommutative characteristic polynomial are no longer scalars
but diagonal matrices. It seems to be an interesting problem to investigate
the following classes of matrices.
\begin{enumerate}
\item The matrices $A$ for which the coefficients $L_j^{(i)}(A)$ are
polynomials in the entries $a_{ij}$.
\item The matrices $A$ for which the coefficients $L_j^{(i)}(A)$ do
not depend on $i$.
\end{enumerate}
As shown by the previous examples, the matrix $E_n$ of generators of
$U(gl_n)$ belongs to both classes, while the matrix $A_q$ of generators
of $GL_q(n)$ belongs to the first one.

\newpage
\section{Appendix : automata and matrices over noncommutative rings}
\label{AUTOMA}

Identities between quasi-determinants can often be interpreted in terms
of {\it automata}. We present here the basic notions of this theory. More
information can be found for instance in the classical
references \cite{BR} or \cite{Eil}.

\smallskip
Let $K$ be a field and let $A$ be a noncommutative alphabet.
A {\it $K$-automaton} of order $n$ over $A$ is a triple
${\cal A} = (I,\mu,F)$ where
\begin{enumerate}
\item $I$ is a row vector of $K^n$, called the {\it initial vector},
\item $\mu$ is a monoid morphism from the free monoid $A^*$ into $M_n(K)$,
\item $F$ a column vector of $K^n$, called the {\it final vector}.
\end{enumerate}
\noindent
One can represent graphically an automaton by a labeled graph (also denoted
${\cal A}$) which is defined by
\begin{itemize}
\item the set of vertices of ${\cal A}$ is $[1,n]$ (they are called
      the states of ${\cal A}$),
\item for every $i,j \in [1,n]$ and for every $a \in A$, there is an arrow
      going from $i$ to $j$ and labeled by $\mu(a)_{ij}\, a$,
\item for every vertex $i \in [1,n]$, there is an arrow pointing on $i$
      and labeled by $I_i$,
\item for every vertex $i \in [1,n]$, there is an arrow issued
      from $i$ and labeled by $F_i$,
\end{itemize}
with the convention that one does not represent an arrow  labelled
by $0$. In other words, the generic form of the graph ${\cal A}$ is the
following

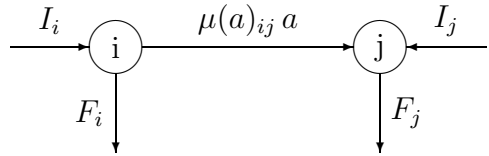
\centerline{
\begin{picture}(200,90)(-50,-55)
\put(0,0){\circle{20}}
\put(0,0){\makebox(0,0){i}}
\put(100,0){\circle{20}}
\put(100,0){\makebox(0,0){j}}
\put(10,0){\vector(1,0){80}}
\put(50,10){\makebox(0,0){$\mu(a)_{ij}\, a$}}
\put(0,-10){\vector(0,-1){30}}
\put(100,-10){\vector(0,-1){30}}
\put(-40,0){\vector(1,0){30}}
\put(140,0){\vector(-1,0){30}}
\put(-25,10){\makebox(0,0){$I_i$}}
\put(-10,-25){\makebox(0,0){$F_i$}}
\put(125,10){\makebox(0,0){$I_j$}}
\put(110,-25){\makebox(0,0){$F_j$}}
\end{picture}
}

\noindent
Note that the graph can be identified with the automaton since it encodes
both the vectors $I, F$ and the matrices $(\mu(a))_{a\in A}$ which
completely define the morphism $\mu$. The {\it behaviour} of ${\cal A}$
is the formal noncommutative power series over $A$ defined by
$$
\underline{{\cal A}} = \ \sum_{w\in A^*} \ (\, I \, \mu(w) \, F\, ) \, w \
\ \in K\! <\! <\! A \! >\! > \ .
$$
The behaviour of ${\cal A}$ has a simple graphical interpre\-ta\-tion. For
every path $\pi\! = (i_1,\dots,i_{n+1})$ in the graph ${\cal A}$ going from
$i_1$ to $i_{n+1}$ and indexed by the word $w = a_1 \, \dots \, a_n$, we
define the cost $c(\pi)$ by
$$
c(\pi) = I_{i_1}\, \mu(a_1)_{i_1,i_2}  \dots
\mu(a_n)_{i_n,i_{n+1}}\,  F_{i_{n+1}} \ .
$$
In other words, the cost of a path is just the product of the elements of
$k$ that index the different arrows it encounters on its way.
Defining the {\it cost} $c_{{\cal A}}(w)$ of a word $w$ relatively to
${\cal A}$, as the sum of the costs of all paths indexed by $w$ in
${\cal A}$, we have by definition,
$$
\underline{{\cal A}}
=
\ \sum_{w\in A^*} \ c_{{\cal A}}(w) \, w \ .
$$
\smallskip
It is not difficult to show that
$$
\underline{{\cal A}}
=
I \cdot \left( \ \sum_{a\in A} \ \mu(a) \, a \, \right)^* \!\! \cdot F \ ,
$$
where the {\it star} $M^*$ of a matrix $M$ is defined by
\begin{equation}\label{STAR}
M^* = (\, I - M \, )^{-1} = \ \sum_{n\geq 0} \ M^n \ .
\end{equation}
It follows  that
$$
I \cdot \left( \ \sum_{a\in A} \ \mu(a) \, a \, \right)^* \!\! \cdot F
=
\ \sum_{w\in A^*} \ c_{{\cal A}}(w) \, w \ .
$$

\smallskip
A series $f \in K\! <\! <\! A \! >\! >$ is said to be {\it recognizable}
if there exists a $K$-automaton ${\cal A}$ over $A$ for which
$f=\underline{{\cal A}}$. On the other hand, one can also define a notion
of rational noncommutative series in $K\! <\! <\! A \! >\! >$. A series
$f \in K\! <\! <\! A \! >\! >$ is {\it rational} in the sense of automata
theory if it belongs to the
subalgebra $K_{rat}\! <\! <\! A \! >\! >$, which is the smallest subring
of $K\! <\! <\! A \! >\! >$ which contains $K\! <\!  A \! >$ and which
is closed under  inversion of formal series. One then has the following
fundamental theorem of Sch\"{u}tzenberger (1961) (see \cite{BR} or
\cite{Eil} for a proof).

\begin{theorem}
A series  $f \in K\! <\! <\! A \! >\! >$ is rational if and only if
it is recognizable.
\end{theorem}

The link between this notion of rationality and the one considered in
Section~\ref{EXPAN} of the present paper is provided by a result of
Fliess (\cite{Fl}, see also \cite{Ca}):
when the free field is realized as the subfield of the
field of Malcev-Neumann series over the free group $F(A)$
generated by the $K$-alpgebra of $F(A)$, one has
$K_{rat}\! <\! <\! A \! >\! > =  K\! <\! <\! A \! >\! > \cap \ \fsf$.

\medskip
For our purposes, we need only the following particular case.
We take $A = \{\, a_{ij},\, 1 \leq i,j \leq n\, \}$ and $\mu(a_{ij})=E_{ij}$,
where $E_{ij}$ denotes the $n\times n$ matrix with only one nonzero entry
equal to $1$ in position $i,\,j$. The graph ${\cal A}$ is then the complete
oriented graph on $n$ vertices, the arrow from $i$ to $j$ being labelled by
$a_{ij}$. Thus, for $n=2$, the graph ${\cal A}$ is for instance

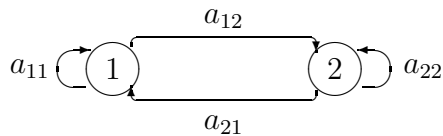
\setlength{\unitlength}{0.7pt}

\centerline{
\begin{picture}(170,92)(-25,-40)
\put(0,0){\circle{29}}
\put(0,0){\makebox(0,0){{\rm 1}}}
\put(120,0){\circle{29}}
\put(120,0){\makebox(0,0){{\rm 2}}}
\put(60,12){\oval(100,10)[t]}
\put(60,27){\makebox(0,0){$a_{12}$}}
\put(60,-12){\oval(100,10)[b]}
\put(60,-30){\makebox(0,0){$a_{21}$}}
\put(-15,0){\oval(30,20)[l]}
\put(-45,0){\makebox(0,0){$a_{11}$}}
\put(135,0){\oval(30,20)[r]}
\put(168,0){\makebox(0,0){$a_{22}$}}
\put(-13,10){\vector(1,0){1}}
\put(133,10){\vector(-1,0){1}}
\put(10,-10){\vector(0,1){1}}
\put(110,10){\vector(0,-1){1}}
\end{picture}
}

\noindent
Denoting also by $A$ the $n \times n$ matrix $(a_{ij})$, we then have
$$
\underline{{\cal A}}
=
I\, A^*\, F \ .
$$
In particular, taking all the entries of $I$ and $F$ equal to zero with
the exception of one equal to $1$, one obtains a graphical interpretation
for any entry of the star of a matrix. That is, denoting by ${\cal P}_{ij}$
the set of words associated with a path from $i$ to $j$, one has
$$
(A^*)_{ij} = \sum_{w \in {\cal P}_{ij}} w \ .
$$
This leads to the classical formula (see {\it e.g. } \cite{BR})
\begin{equation} \label{AUTOMATE}
\left(\,
\matrix{
a_{11} & a_{12} \cr
a_{21} & a_{22}
}
\,\right)^*
=
\left(\,
\matrix{
(a_{11} + a_{12}\, a_{22}^*\, a_{21})^* &
a_{11}^*\, a_{12}\, (a_{22} + a_{21}\, a_{11}^*\, a_{12})^* \cr
a_{22}^*\, a_{21}\, (a_{11} + a_{12}\, a_{22}^*\, a_{21})^* &
(a_{22} + a_{21}\, a_{11}^*\, a_{12})^*
}
\,\right) \ .
\end{equation}
Indeed, the entry $(1,1)$ of the matrix in the right-hand side
of (\ref{AUTOMATE}) represents the set of words labelling a path from $1$
to $1$ in the automaton ${\cal A}$ associated with the generic matrix of
order $2$. Here, the star of a series $s$ in the $a_{ij}$ with zero constant
coefficient is defined, as in (\ref{STAR}), by setting
$s^* = \, \sum_{n\ge 0}\, s^n$.

\medskip
Observe that formulas (\ref{QD2})
are exactly the images
of (\ref{AUTOMATE}) under the involutive field automorphism $\iota$
of $\fsf$ defined by
\begin{equation}
\iota(a_{ij}) = \left\{ \
\matrix{
1 - a_{ii} & \hbox{if} \ \ i = j \cr
- a_{ij} & \hbox{if} \ \ i \not= j \cr
}
\right.
\end{equation}
(see \cite{Co} p. 89), which maps the generic matrix
$A$ on $I - A$, so that $\iota(A^*) = A^{-1}$.

\medskip
As an illustration, let us sketch an automata theoretic proof of
Proposition~\ref{SSTRI}.
The quasi-determinant in the left-hand side of (\ref{POLENT}) can be written as
\begin{equation} \label{AUTD}
D =
a_{1n} - (a_{11} \ a_{12} \ \dots \ a_{1,n-1}\, ) \
(I_{n-1} - M)^{-1}
\
\left(\matrix{
a_{2n} \vtr{2} \cr a_{3n} \cr \vdots \vtr{2} \cr a_{nn} }\right) \ ,
\end{equation}
where $M$ denotes the strictly upper triangular matrix defined by
$$
M =
\left(
\matrix{
0      & a_{22} & \ldots & a_{2,n-1} \vtr{2} \cr
0      & 0      & \ldots & a_{3,n-1} \vtr{2} \cr
\vdots & \vdots & \ddots & \vdots \vtr{2} \cr
0      & 0      & \ldots & 0 \cr
}
\right) \ .
$$
Relation (\ref{AUTD}) shows that $D$ is essentially equal to the
behaviour of the automaton ${\cal A}_{n-1}$ defined by

\begin{itemize}
\item the set of states of ${\cal A}_{n-1}$ is $1,2,\dots,n-1$,
\item the only edges in ${\cal A}_{n-1}$ go from $i$ to $j$ with $i < j$
and are labeled by $a_{i+1,j}$,
\item each state $i$ is equiped with an initial and a final arrow,
respectively labeled by $a_{1i}$ and by $a_{i+1,n}$.
\end{itemize}

\noindent
We give below the automaton ${\cal A}_4$ which illustrates
the general structure of ${\cal A}_{n}$.

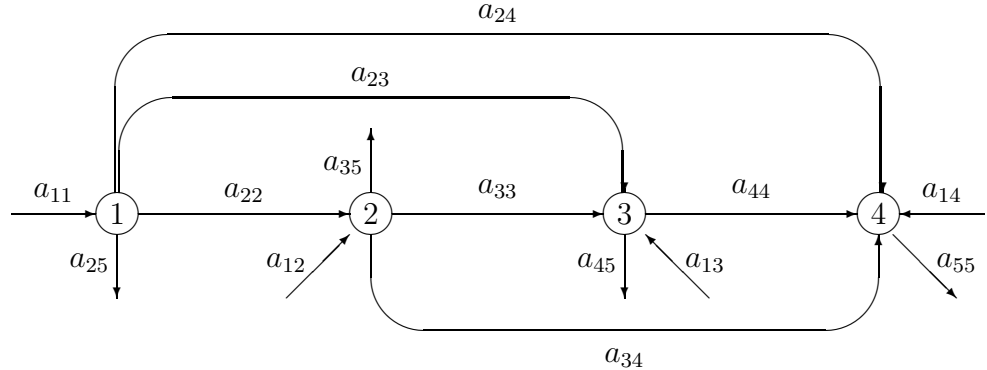
\setlength{\unitlength}{0.8pt}

\centerline{
\begin{picture}(430,210)(-60,-85)
\put(0,0){\circle{20}}
\put(120,0){\circle{20}}
\put(240,0){\circle{20}}
\put(360,0){\circle{20}}
\put(0,0){\makebox(0,0){1}}
\put(120,0){\makebox(0,0){2}}
\put(240,0){\makebox(0,0){3}}
\put(360,0){\makebox(0,0){4}}
\put(-50,0){\vector(1,0){40}}
\put(10,0){\vector(1,0){100}}
\put(130,0){\vector(1,0){100}}
\put(250,0){\vector(1,0){100}}
\put(410,0){\vector(-1,0){40}}
\put(0,-10){\vector(0,-1){30}}
\put(120,10){\vector(0,1){30}}
\put(240,-10){\vector(0,-1){30}}
\put(367,-10){\vector(1,-1){30}}
\put(80,-40){\vector(1,1){30}}
\put(280,-40){\vector(-1,1){30}}
\put(60,10){\makebox(0,0){$a_{22}$}}
\put(180,12){\makebox(0,0){$a_{33}$}}
\put(300,12){\makebox(0,0){$a_{44}$}}
\put(-30,10){\makebox(0,0){$a_{11}$}}
\put(80,-23){\makebox(0,0){$a_{12}$}}
\put(278,-23){\makebox(0,0){$a_{13}$}}
\put(390,10){\makebox(0,0){$a_{14}$}}
\put(-13,-23){\makebox(0,0){$a_{25}$}}
\put(107,23){\makebox(0,0){$a_{35}$}}
\put(227,-23){\makebox(0,0){$a_{45}$}}
\put(397,-23){\makebox(0,0){$a_{55}$}}
\put(180,10){\oval(362,150)[t]}
\put(120,10){\oval(238,90)[t]}
\put(240,-10){\oval(240,90)[b]}
\put(180,95){\makebox(0,0){$a_{24}$}}
\put(120,65){\makebox(0,0){$a_{23}$}}
\put(240,-68){\makebox(0,0){$a_{34}$}}
\put(362,12){\vector(0,-1){2}}
\put(360,-12){\vector(0,1){2}}
\put(240,12){\vector(0,-1){2}}
\end{picture}
}

\noindent
It is now clear that
$$
D
= a_{1n}
+
\ \sum_{i<j} \ a_{1i}\ (\ \sum \
\hbox{words labelling paths from $i$ to $j$ in
${\cal A}_{n-1}$} \, )
\ a_{j+1,n} \ ,
$$
which is the right-hand side of (\ref{POLENT}). \cqfd


\end{document}